\documentclass[longauth]{aa} 
\usepackage{graphicx}
\usepackage{txfonts}
\usepackage{amsmath,amsfonts,amssymb} 
\usepackage{natbib}
\usepackage{multicol,caption}
\usepackage{lipsum}
\usepackage{booktabs}
\usepackage{threeparttable}
\usepackage{multirow}
\usepackage{here}
\usepackage{lscape}

\bibpunct{(}{)}{;}{a}{}{,}
\usepackage[colorlinks=true,urlcolor=blue,linkcolor=blue,citecolor=blue]{hyperref}

\usepackage{threeparttable}
\usepackage{booktabs}
\usepackage{longtable}
\usepackage{ragged2e}

\usepackage{tikz,xcolor,hyperref}

\definecolor{lime}{HTML}{A6CE39}
\DeclareRobustCommand{\orcidicon}{%
	\hspace{-1.5mm}
	\begin{tikzpicture}
	\draw[lime, fill=lime] (0,0) 
	circle [radius=0.16] 
	node[white] {{\fontfamily{qag}\selectfont \tiny ID}};
	\draw[white, fill=white] (-0.0625,0.095) 
	circle [radius=0.007];
	\end{tikzpicture}
	\hspace{-2.5mm}
}
\foreach \x in {A, ..., Z}{%
	\expandafter\xdef\csname orcid\x\endcsname{\noexpand\href{https://orcid.org/\csname orcidauthor\x\endcsname}{\noexpand\orcidicon}}
}

\foreach \x in {a, ..., z}{%
	\expandafter\xdef\csname orcid\x\endcsname{\noexpand\href{https://orcid.org/\csname orcidauthor\x\endcsname}{\noexpand\orcidicon}}
}

\begin{document}

\title{TOI-6508\,b: A massive transiting brown dwarf orbiting a low-mass star}

\newcommand{\orcidauthorA}{0000-0003-1464-9276} 
\newcommand{\orcidauthorB}{}
\newcommand{\orcidauthorC}{0000-0001-6416-1274}
\newcommand{\orcidauthorD}{}
\newcommand{\orcidauthorE}{}
\newcommand{\orcidauthorF}{0000-0002-4909-5763}
\newcommand{\orcidauthorG}{}
\newcommand{\orcidauthorH}{}
\newcommand{\orcidauthorI}{0000-0002-0149-1302}
\newcommand{\orcidauthorJ}{}
\newcommand{\orcidauthorK}{}
\newcommand{\orcidauthorL}{}
\newcommand{\orcidauthorM}{}
\newcommand{\orcidauthorN}{}
\newcommand{\orcidauthorO}{}
\newcommand{\orcidauthorP}{}
\newcommand{\orcidauthorQ}{}
\newcommand{\orcidauthorR}{}
\newcommand{\orcidauthorS}{0000-0002-2532-2853}
\newcommand{\orcidauthorT}{0000-0003-2163-1437}
\newcommand{\orcidauthorU}{0000-0002-4265-047X}
\newcommand{\orcidauthorV}{}
\newcommand{\orcidauthorW}{}
\newcommand{\orcidauthorY}{}
\newcommand{\orcidauthorX}{}
\newcommand{\orcidauthorZ}{}
\newcommand{\orcidauthora}{}
\newcommand{\orcidauthorb}{}
\newcommand{\orcidauthorc}{}
\newcommand{\orcidauthord}{}
\newcommand{\orcidauthore}{}
\newcommand{\orcidauthorf}{}
\newcommand{\orcidauthorg}{}
\newcommand{\orcidauthorh}{}
\newcommand{\orcidauthori}{}
\newcommand{\orcidauthorj}{}
\newcommand{\orcidauthorl}{}
\newcommand{\orcidauthorm}{}
\newcommand{\orcidauthorn}{}
\newcommand{\orcidauthoro}{}
\newcommand{\orcidauthorp}{}
\newcommand{\orcidauthorq}{}
\newcommand{\orcidauthorr}{}
\newcommand{\orcidauthors}{}
\newcommand{\orcidauthort}{}
\newcommand{\orcidauthoru}{}
\newcommand{\orcidauthorv}{}
\newcommand{\orcidauthorw}{}
\newcommand{\orcidauthorx}{0000-0002-6523-9536}
\newcommand{\orcidauthory}{0000-0002-9807-5435}
\newcommand{\orcidauthorz}{0000-0002-1420-1837}
\newcommand{\orcidauthoraa}{}
\newcommand{\orcidauthorbb}{}
\newcommand{\orcidauthorcc}{}
\newcommand{\orcidauthordd}{}

\author{
K. Barkaoui\orcidA{}\inst{\ref{astro_liege},\ref{MIT},\ref{IAC_Laguna}}\thanks{E-mail: \color{blue}khalid.barkaoui@uliege.be}
\and D.~Sebastian\inst{\ref{birmingham}} 
\and S.~Z\'u\~niga-Fern\'andez\orcidZ{}\inst{\ref{astro_liege}} 
\and A.H.M.J.~Triaud\inst{\ref{birmingham}} 
\and B.V.~Rackham\inst{\ref{MIT},\ref{Kavli_MIT}} 
\and A.J.~Burgasser\inst{\ref{UCSDiego}} 
\and T.W.~Carmichael\orcidC{}\inst{\ref{Insti_Hawaii}} 
\and M.~Gillon\inst{\ref{astro_liege}}   
\and C.~Theissen\inst{\ref{UCSDiego}} 
\and E.~Softich\inst{\ref{UCSDiego}} 
\and B.~Rojas-Ayala\inst{\ref{tara}} 
\and G.~Srdoc\inst{\ref{Kotiza_Obs}}  
\and A.~Soubkiou\inst{\ref{ouka}}
\and A.~Fukui\orcidF{}\inst{\ref{Univ_tokyo},\ref{IAC_Laguna}}    
\and M. Timmermans\inst{\ref{astro_liege},\ref{birmingham}} 
\and M.~Stalport\inst{\ref{star_liege}}  
\and A.~Burdanov\inst{\ref{MIT}} 
\and D.R.~Ciardi\inst{\ref{ipac}} 
\and K.A.~Collins\inst{\ref{Harvard_USA}} 
\and Y.T.~Davis\inst{\ref{birmingham}} 
\and F.~Davoudi\inst{\ref{astro_liege}} 
\and J.~de~Wit\inst{\ref{MIT}} 
\and B.O.~Demory\inst{\ref{unibe}} 
\and S.~Deveny\inst{\ref{Bay_Area},\ref{Ames_NASA}} 
\and G.~Dransfield\inst{\ref{birmingham}} 
\and E.~Ducrot\inst{\ref{Paris_Region},\ref{cea}} 
\and L.~Florian\inst{\ref{ETH_Zur_Queloz}} 
\and T.~Gan\inst{\ref{Univ_Montreal},\ref{Tsinghua_Univ}}  
\and Y.~G\'omez~Maqueo~Chew\inst{\ref{ciudad}} 
\and M.J.~Hooton\inst{\ref{Cavendish}} 
\and S.~B.~Howell\orcidS{}\inst{\ref{Ames_NASA}} 
\and J.~M.~Jenkins\inst{\ref{Ames_NASA}} 
\and C.~Littlefield\inst{\ref{Bay_Area},\ref{Ames_NASA}} 
\and E.~L.~Martín\inst{\ref{IAC_Laguna}} 
\and F.~Murgas\inst{\ref{IAC_Laguna},\ref{Univ_LaLaguna}} 
\and P.~Niraula\inst{\ref{MIT},\ref{Kavli_MIT}} 
\and E.~Palle\inst{\ref{IAC_Laguna},\ref{Univ_LaLaguna}} 
\and P.P.~Pedersen\inst{\ref{Cavendish},\ref{ETH_Zur_Queloz}} 
\and F.J.~Pozuelos\inst{\ref{iaa}} 
\and D.~Queloz\inst{\ref{Cavendish},\ref{ETH_Zur_Queloz}} 
\and G.~Ricker\inst{\ref{Kavli_MIT}} 
\and R.P.~Schwarz\inst{\ref{Harvard_USA}} 
\and S.Seager\inst{\ref{Dep_Kavli},\ref{MIT},\ref{MIT_Aeronaut}} 
\and A.~Shporer\inst{\ref{Dep_Kavli}}  
\and M.G.~Scott\inst{\ref{birmingham}} 
\and C.~Stockdale\orcidT{}\inst{\ref{Haze_Austra}}   
\and J.~Winn\orcidU{}\inst{\ref{Astro_Prin}}   
}

\institute{
Astrobiology Research Unit, Universit\'e de Li\`ege, All\'ee du 6 Ao\^ut 19C, B-4000 Li\`ege, Belgium \label{astro_liege}
\and Department of Earth, Atmospheric and Planetary Science, Massachusetts Institute of Technology, 77 Massachusetts Avenue, Cambridge, MA 02139, USA \label{MIT}
\and Instituto de Astrof\'isica de Canarias (IAC), Calle V\'ia L\'actea s/n, 38200, La Laguna, Tenerife, Spain \label{IAC_Laguna} 
\and School of Physics \& Astronomy, University of Birmingham, Edgbaston, Birmingham B15 2TT, UK \label{birmingham}
\and Department of Physics and Kavli Institute for Astrophysics and Space Research, Massachusetts Institute of Technology, Cambridge, MA 02139, USA \label{Kavli_MIT}
\and Department of Astronomy \& Astrophysics, UC San Diego, 9500 Gilman Drive, La Jolla, CA 92093, USA \label{UCSDiego}
\and Institute for Astronomy, University of Hawai‘i, 2680 Woodlawn Drive, Honolulu, HI 96822, USA \label{Insti_Hawaii}
\and Instituto de Alta Investigaci\'on, Universidad de Tarapac\'a, Casilla 7D, Arica, Chile \label{tara}
\and Kotizarovci Observatory, Sarsoni 90, 51216 Viskovo, Croatia \label{Kotiza_Obs}
\and Oukaimeden Observatory, High Energy Physics and Astrophysics Laboratory, Faculty of sciences Semlalia, Cadi Ayyad University, Marrakech, Morocco \label{ouka}
\and Komaba Institute for Science, The University of Tokyo, 3-8-1 Komaba, Meguro, Tokyo 153-8902, Japan \label{Univ_tokyo}
\and Space Sciences, Technologies and Astrophysics Research (STAR) Institute, Universit\'e de Li\`ege, All\'ee du 6 Ao\^ut 19C, B-4000 Li\`ege, Belgium \label{star_liege}
\and NASA Exoplanet Science Institute, IPAC, California Institute of Technology, Pasadena, CA 91125 USA \label{ipac}
\and Center for Astrophysics \textbar \ Harvard \& Smithsonian, 60 Garden Street, Cambridge, MA 02138, USA \label{Harvard_USA}
\and Center for Space and Habitability, University of Bern, Gesellschaftsstrasse 6, 3012, Bern, Switzerland \label{unibe}
\and Paris Region Fellow, Marie Sklodowska-Curie Action \label{Paris_Region}
\and AIM, CEA, CNRS, Universit\'e Paris-Saclay, Universit\'e de Paris, F-91191 Gif-sur-Yvette, France \label{cea}
\and Institut Trottier de recherche sur les exoplan\`etes, D\'epartement de Physique, Universit\'e de Montr\'eal, Montr\'eal, Qu\'ebec, Canada \label{Univ_Montreal}
\and Department of Astronomy, Tsinghua University, Beijing 100084, People's Republic of China \label{Tsinghua_Univ}
\and Universidad Nacional Aut\'onoma de M\'exico, Instituto de Astronom\'ia, AP 70-264, Ciudad de M\'exico,  04510, M\'exico \label{ciudad}
\and Cavendish Laboratory, JJ Thomson Avenue, Cambridge CB3 0HE, UK \label{Cavendish}
\and Bay Area Environmental Research Institute, Moffett Field, CA 94035, USA \label{Bay_Area}
\and NASA Ames Research Center, Moffett Field, CA 94035, USA \label{Ames_NASA}
\and Departamento de Astrof\'{i}sica, Universidad de La Laguna (ULL), 38206 La Laguna, Tenerife, Spain \label{Univ_LaLaguna}
\and Institute for Particle Physics and Astrophysics , ETH Z\"urich, Wolfgang-Pauli-Strasse 2, 8093 Z\"urich, Switzerland \label{ETH_Zur_Queloz}
\and Instituto de Astrof\'isica de Andaluc\'ia (IAA-CSIC), Glorieta de la Astronom\'ia s/n, 18008 Granada, Spain \label{iaa}
\and Department of Physics and Kavli Institute for Astrophysics and Space Research, Massachusetts Institute of Technology, Cambridge, MA 02139, USA \label{Dep_Kavli}
\and Department of Aeronautics and Astronautics, MIT, 77 Massachusetts Avenue, Cambridge, MA 02139, USA \label{MIT_Aeronaut}
\and Hazelwood Observatory, Australia \label{Haze_Austra}
\and Department of Astrophysical Sciences, Princeton University, Princeton, NJ 08544, USA \label{Astro_Prin}
}

\date{Received/accepted}
\titlerunning{TOI-6508\,b}\authorrunning{K. Barkaoui et al.}	 

\abstract{ 
    We report the discovery of a transiting brown dwarf orbiting a low-mass star, TOI-6508\,b. Today, only $\sim$50 transiting brown dwarfs have been discovered. TOI-6508\,b was first detected with data from the Transiting Exoplanet Survey Satellite (\emph{TESS}) in Sectors 10, 37 and 63. Ground-based follow-up photometric data were collected with the SPECULOOS-South and LCOGT-1m telescopes, and RV measurements were obtained with the Near InfraRed Planet Searcher (NIRPS) spectrograph. We find that TOI-6508\,b has a mass of $M_p = 72.5^{+7.6}_{-5.1} M_{\rm Jup}$ and a radius of $R_p = 1.03 \pm 0.03R_{\rm Jup}$. 
    Our modeling shows that the data are consistent with an eccentric orbit of 19~day and an eccentricity of $e=0.28^{+0.09}_{-0.08}$.  
    TOI-6508\,b has a mass ratio of $M_{\rm BD}/M_\star = 0.40$, makes it the second highest mass ratio brown dwarf that transits a low-mass star. The host has a mass of $M_\star = 0.174 \pm 0.004M_\odot$, a radius of $R_\star = 0.205 \pm 0.006R_\odot$, an effective temperature of $T_{\rm eff} = 2930 \pm 70$~K, and a metallicity of $[Fe/H] = -0.22 \pm 0.08$.
    This makes TOI-6508\,b an interesting discovery that has come to light in a region still sparsely populated.
}


\maketitle

\section{Introduction}

Brown dwarfs (BDs) are traditionally defined as objects between giant planets ($\sim$13~$M_{\rm Jup}$) and stars ($\sim$80~$M_{\rm Jup}$), with radii ranging from 0.7 to 1.4~$R_{\rm Jup}$. The lower limit that separates giant planets from BDs corresponds to the ignition of deuterium fusion in the core of the BD. This limit varies within the range 11--16~$M_{\rm Jup}$  depending on the abundance of deuterium and the bulk metallicity  \citep{Spiegel_2011ApJ}. 
The upper limit that separates the BDs and stars corresponds to hydrogen fusion, and varies within the range 75--80~$M_{\rm Jup}$ depending on the stellar initial formation conditions \citep{Baraffe_2002A&A}. 
Based on this mass definition, the cores of BDs are partially degenerate. To improve the characterization and classification of the BDs, the transit method is very useful as it brings additional information, specifically the radius. However, any candidate companion close in size to $1R_{\rm J}$ can be a brown dwarf, giant planet, or a low-mass star as it is not clear solely from the radius, making the measurement of the companion's mass crucial for BD discovery.
\\

Transiting BDs orbiting low-mass stars offer us valuable opportunities to measure the radius and mass (with the combination with radial velocity technique) and orbital parameters of the system.
The relatively small size of the star leads to a large transit signal on the order of ten percent. Furthermore, the relatively low mass results in a huge radial-velocity signal on the order of several kilometers per second. This makes possible high-precision measurements of a BD's mass and radius.
The mass and radius are key for exploring the physical properties of the BDs, in order to improve our understanding of the mechanisms of formation and evolution of these mysterious sub-stellar objects. \citep{Baraffe_2002A&A, Saumon_2008ApJ,Phillips_2020A&A,Chabrier_2023A&A}. 

Our current understanding of planetary formation predicts a low probability for the existence of Jupiter-like planets and BDs around low-mass stars with $M_\star \leq 0.4 M_\odot$ , and their formation by core accretion becomes increasingly unlikely as $M_\star > 0.4 M_\odot$.
\citep{Kanodia_2022AJ,Palle_2021AA,Burn_2021AA}.
However, we have discovered only 9 transiting BDs orbiting low-mass stars to date, with host masses ranging from 0.25--0.65 M$_\odot$. Due to the small size of this sample, the occurrence rate of BDs around low-mass stars is still highly uncertain;  more detections are necessary to compare observations to theoretical expectations. 

In this work, we present a new system orbiting a low-mass M dwarf ($M_\star = 0.17 \pm 0.02 M_\odot$) in a 19-day eccentric orbit ($e = 0.28^{+0.09}_{-0.08}$), TOI-6508. This system contains a transiting BD, TOI-6508\,b with a mass of $M_{\rm BD} = 72.5~M_{\rm Jup}$ and a radius of $R_{\rm BD} = 1.03~R_{\rm Jup}$ around an M5 star. 

The paper is organized as follows. We present TESS data and ground-based photometric and spectroscopic observations in Section~\ref{section_observation}. Stellar characterization of TOI-6508 (spectroscopic and spectral energy distribution analysis) is presented in Section~\ref{sec:stellar_charac}. Section~\ref{sec:mcmc_fit} describes the global modeling of photometric and radial velocity data. Finally, a discussion and our conclusions are presented in Section~\ref{discu_conclusion}.

\section{Observations and data reduction} \label{section_observation}

\subsection{TESS photometric observation}
\label{sec:tess_photometry}

\begin{figure}[!]
	\centering
	\includegraphics[scale=0.5]{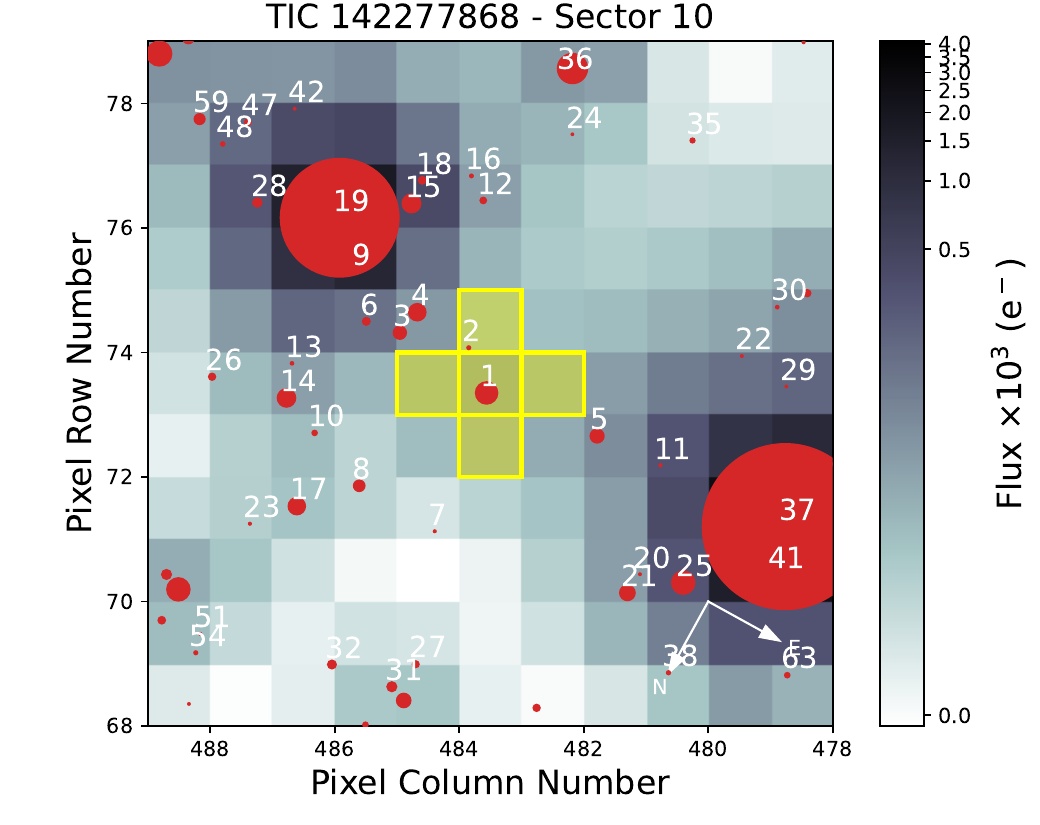}\\
     \includegraphics[scale=0.5]{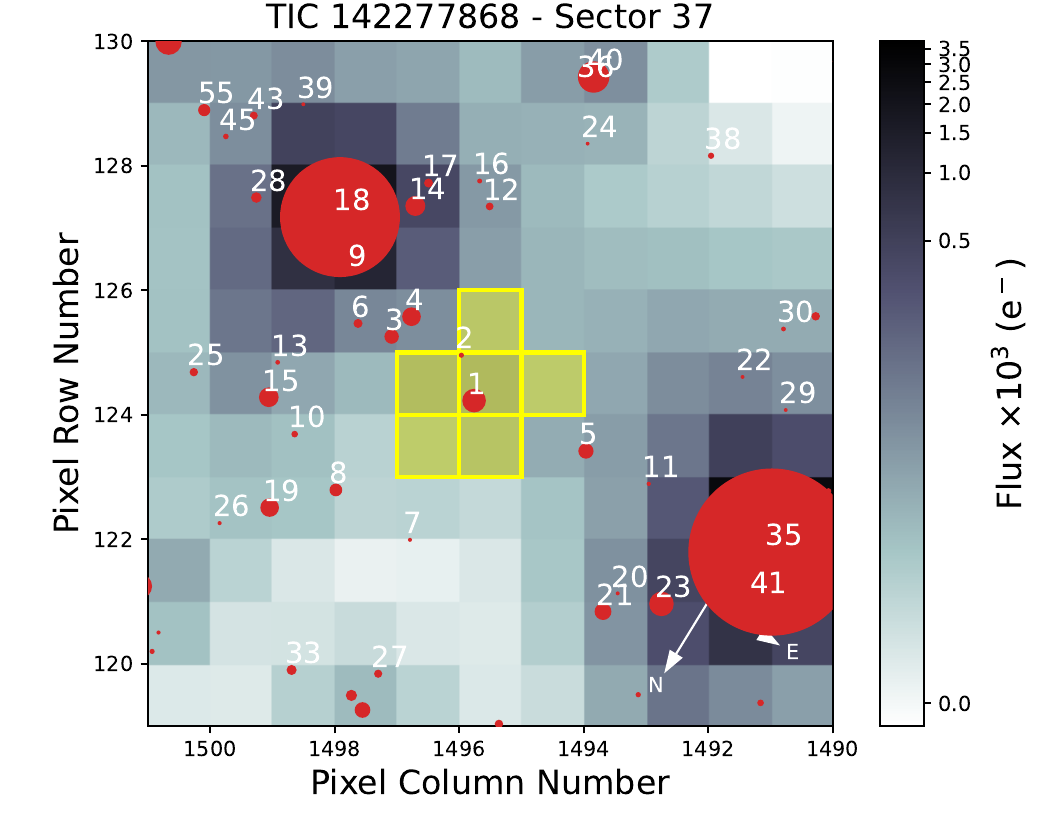}\\
     \includegraphics[scale=0.5]{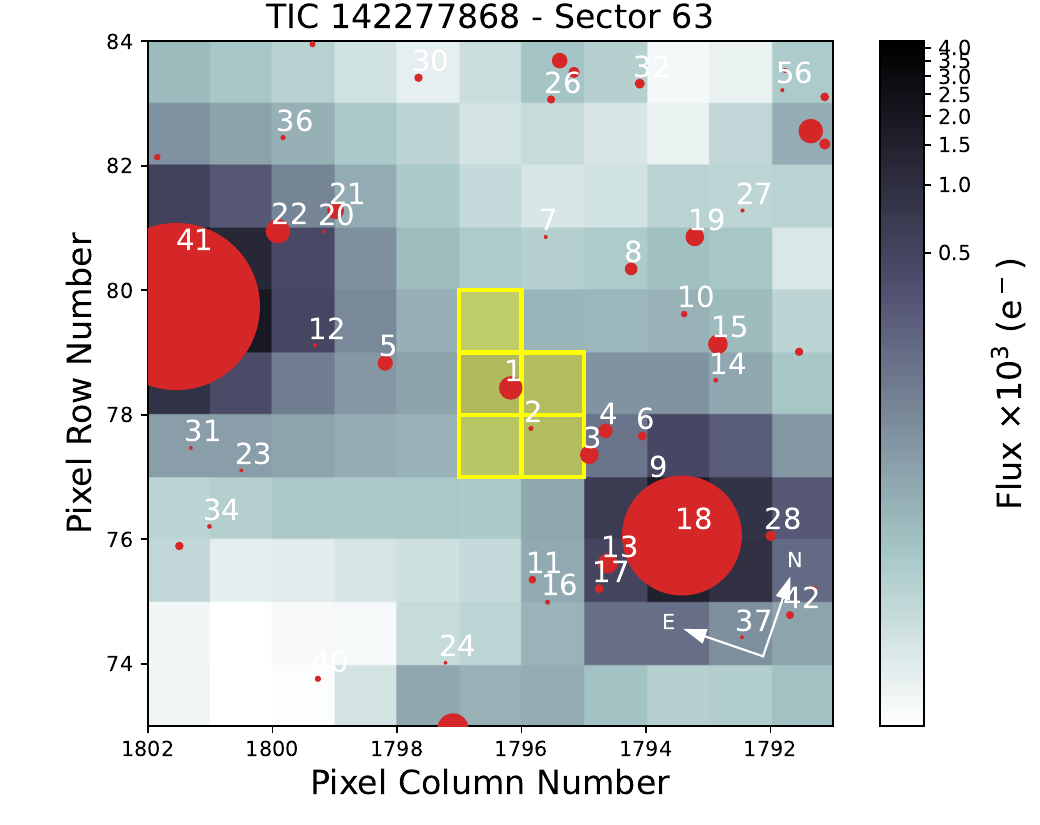}
	\caption{TESS target pixel file images for TOI-6508 observed in Sectors 10 ({\it top}), 37 ({\it middle}) and 63 ({\it bottom}). The plots are made with the {\tt tpfplotter} \citep{Aller_2020AandA} package. The pixels highlighted in yellow show the TESS apertures. The red dots show the positions of Gaia DR3 sources, and their sizes correspond to their TESS magnitudes.}
	\label{TESS_FOV}
\end{figure}

\begin{figure*}[!]
	\centering
	\includegraphics[scale=0.33]{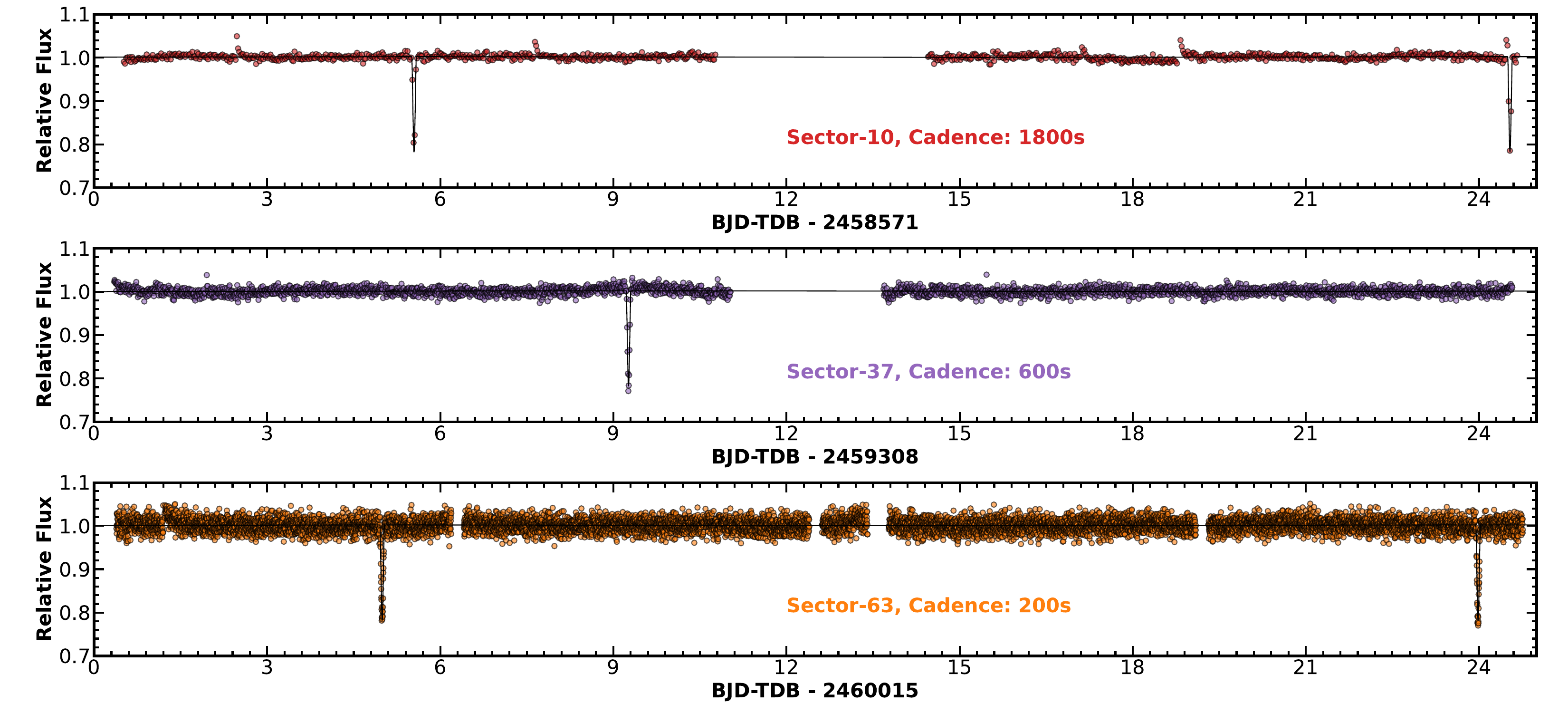}
	\caption{TESS PDC-SAP flux of TOI-6508 extracted from the full frame images (FFIs). The target was observed in Sectors 10 ({\it top}) at 1800-second, 37 ({\it middle}) at 600-second, and 63 ({\it bottom}) at 200-second cadence. The solid line is the best-fitting model of the transit.}
	\label{TESS_LCs_BJD}
\end{figure*}

The host star TIC~142277868 (TOI-6508) was observed by the \emph{TESS} mission \citep{Ricker_2015JATIS_TESS} in Sector 10, from March 26 to 22 April 2019.
The relevant data are available within the full-frame images (FFIs) with a cadence of 1800 seconds. The target was re-observed during the extended mission in Sectors 37 and 63, and FFI data are available with a cadence of 600 seconds and 200 seconds, respectively (see Table~\ref{TESS_obs_table}).
We used the Pre-search Data Conditioning Simple Aperture Photometry flux (PDC-SAP; \citealt{Stumpe_2012PASP,Smith_2012PASP,Stumpe_2014}), constructed by the TESS Science Processing Operations Center (SPOC; \citealt{SPOC_Jenkins_2016SPIE}) at the Ames Research Center, from the Mikulski Archive for Space Telescopes \footnote{\url{https://archive.stsci.edu/missions-and-data/tess}}. The PDC-SAP light curves were calibrated for any instrument systematics and crowding effects. The TOI-6508 light curves are normalized using the {\tt lightkurve} \citep{Lightkurve_2018ascl} Python package.  Figure~\ref{TESS_FOV} shows the TOI-6508 field of view in the TESS data, as well as the photometric apertures that were used to construct the
light curves. The locations of nearby Gaia DR3 sources are also marked \citep{Gaia_Collaboration_2021AandA}. Figure~\ref{TESS_LCs_BJD} shows the TESS photometric data.

\begin{table}[!]
 \begin{center}
 {\renewcommand{\arraystretch}{1.4}
 \resizebox{0.45\textwidth}{!}{
 \begin{tabular}{l c c c c}
 \toprule
Sector  &  Exptime [s] & Camera & CCD &  Observation date\\ 
 \hline
 10 & 1800 & 2 & 4 & 2019 Mar-26 -- Apr-22 \\
 37 & 600 &  2 & 4 & 2021 Apr-02 -- Apr-28 \\
 63 & 200 &  1 & 2 & 2023 Mar-10 -- Apr-06 \\
\hline
 \end{tabular}}}
 \caption{TESS observations log for TOI-6508.}
 \label{TESS_obs_table}
 \end{center}
\end{table}

\subsection{Ground-based photometric observation}
\label{sec:gb_photometry}
\begin{figure*}[!]
	\centering
	\includegraphics[scale=0.4]{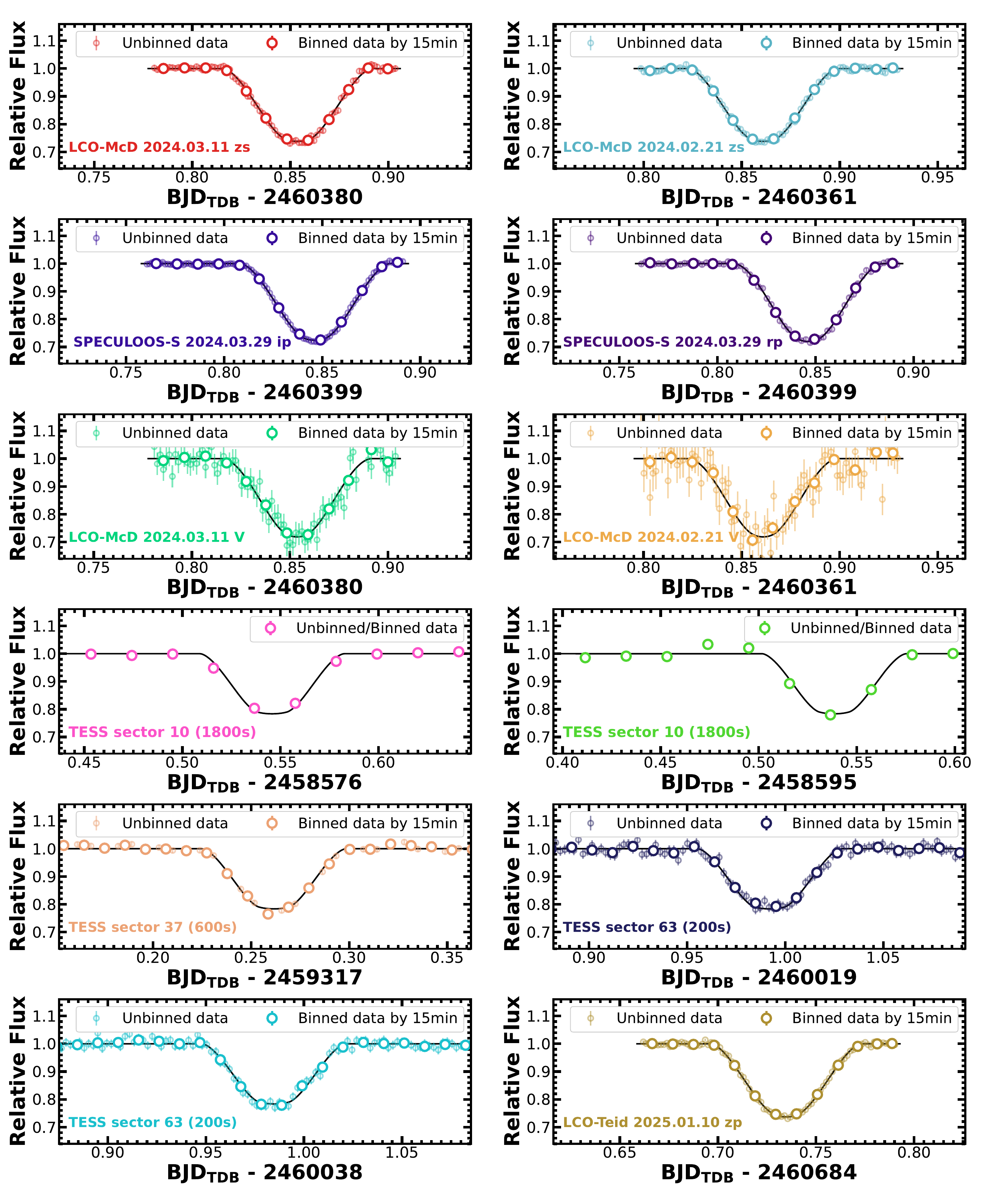}
	\caption{Individual TESS (Sectors 10, 37 and 63) and ground-based (collected with SPECULOOS-S-1.0m and LCOGT-1.0m) light curves of TOI-6508\,b. The colored data points show the relative flux measurements (unbinned and binned data). The black solid lines show the best-fitting transit model.}
	\label{fig:LCs_tess_gb}
\end{figure*}

\begin{figure*}[!]
	\centering
	\includegraphics[scale=0.35]{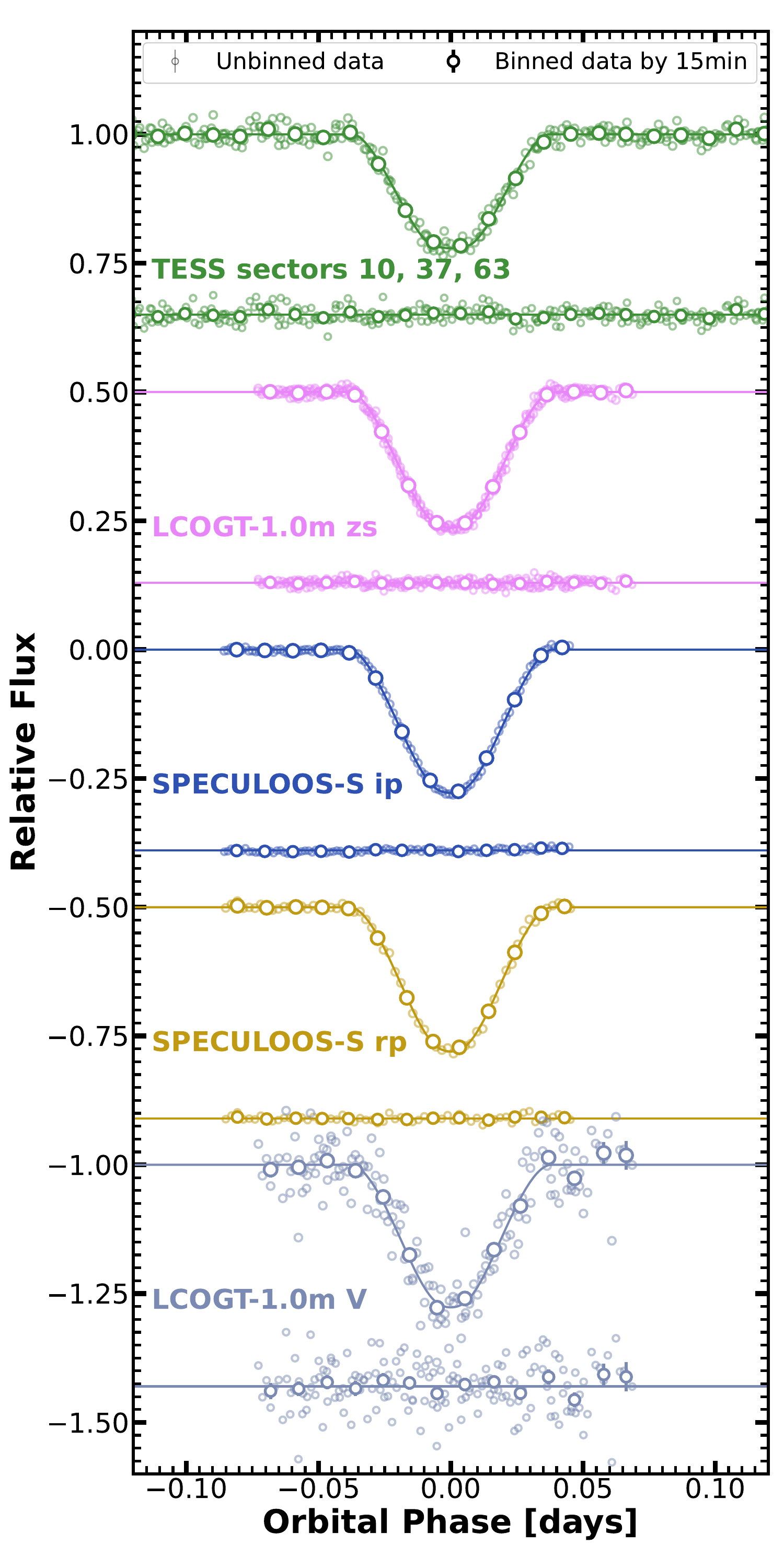}
    \includegraphics[scale=0.35]{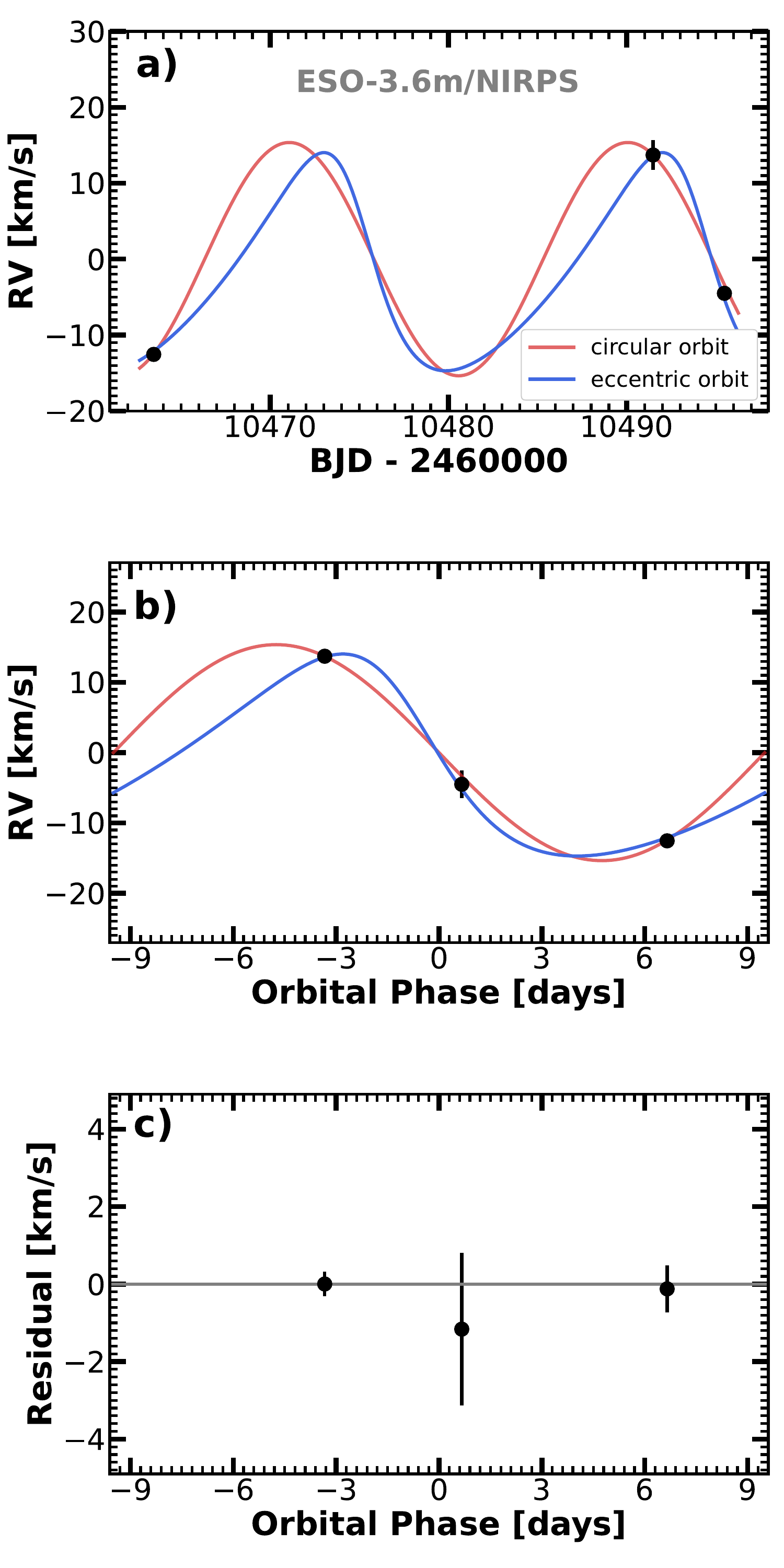}
	\caption{Photometric and radial velocity observations for TOI-6508. The {\it left panel} shows TESS and ground-based phase-folded light curves (unbinned and binned by 15min data) of TOI-6508\,b. The light curves are shifted along y-axis for the visibility. The residuals are also presented bellow each light curve.
    The {\it right panels} show radial velocity measurements collected by the NIRPS spectrograph. {\it a)}, RV observations vs. time. {\it b)}, Phase-folded RV measurements. {\it c)}, RV residuals ($O - C$). The red data points and red solid line show RV measurements and best-fitting assuming a circular orbit. While, the blue data points and blue solid line show RVs measurements and best-fitting assuming an eccentric orbit. }
	\label{fig:lcs_and_rv}
\end{figure*}

We performed ground-based follow-up observations of TOI-6508 as part of the TESS follow-up observing program (TFOP) in order to confirm the transit event on the target star, rule out nearby eclipsing binaries (NEBs) as the source of the transit signal, measure the transit depth, as well as refine the transit ephemerides. 
To schedule time-series observations, we used the {\tt TESS transit finder} tool \citep{jensen2013}, which is a customized version of the {\tt Tapir} software package.
The ground-based photometric observations are summarized in Table~\ref{GB_obs_table}.  The observed transit light curves are presented in Figures~\ref{fig:LCs_tess_gb} and \ref{fig:lcs_and_rv}.

\subsubsection{SPECULOOS-South}

We used the SPECULOOS-South (Search for habitable Planets EClipsing ULtra-cOOl Stars, \citet{Jehin2018Msngr,Delrez2018,Sebastian_2021AA}) facilities to observe the transits of TOI-6508\,b simultaneously in the Sloan-$r'$ and -$i'$ filters on UTC March 29 2024  with an exposure time of 180\,seconds and 105\,seconds, respectively. Each 1.0\,m  telescope is equipped with a 2K$\times$2K CCD camera with a pixel scale of 0.35$\arcsec$ and a total FOV of 12$\arcmin$$\times$12$\arcmin$.
Data reduction and aperture photometry were performed using the {\tt PROSE}\footnote{{\tt Prose:} \url{https://github.com/lgrcia/prose}} pipeline \citep{prose}.

\subsubsection{LCOGT-1.0m}

We used the Las Cumbres Observatory Global Telescope (LCOGT; \citealt{Brown_2013}) 1.0m facilities to observe simultaneously four full transits of TOI-6508\,b in the Pan-STARRS-$z_\mathrm{s}$ and V filters. Two transits were observed on UTC Feb 21 2024, and two others were observed on UTC March 11 2024.
An additional transit of TOI-6508\,b was observed with LCO-Teid-1m0 at the Teide Observatory on UTC Jan 10 2025 in the Pan-STARRS-$z_\mathrm{s}$. The LCOGT telescopes are equipped with $4096 \times 4096$ SINISTRO camera with a pixel scale of $0.389\arcsec$ per pixel and a total FOV of $26' \times 26'$. 

\begin{figure}[!]
	\centering
	\includegraphics[scale=0.24]{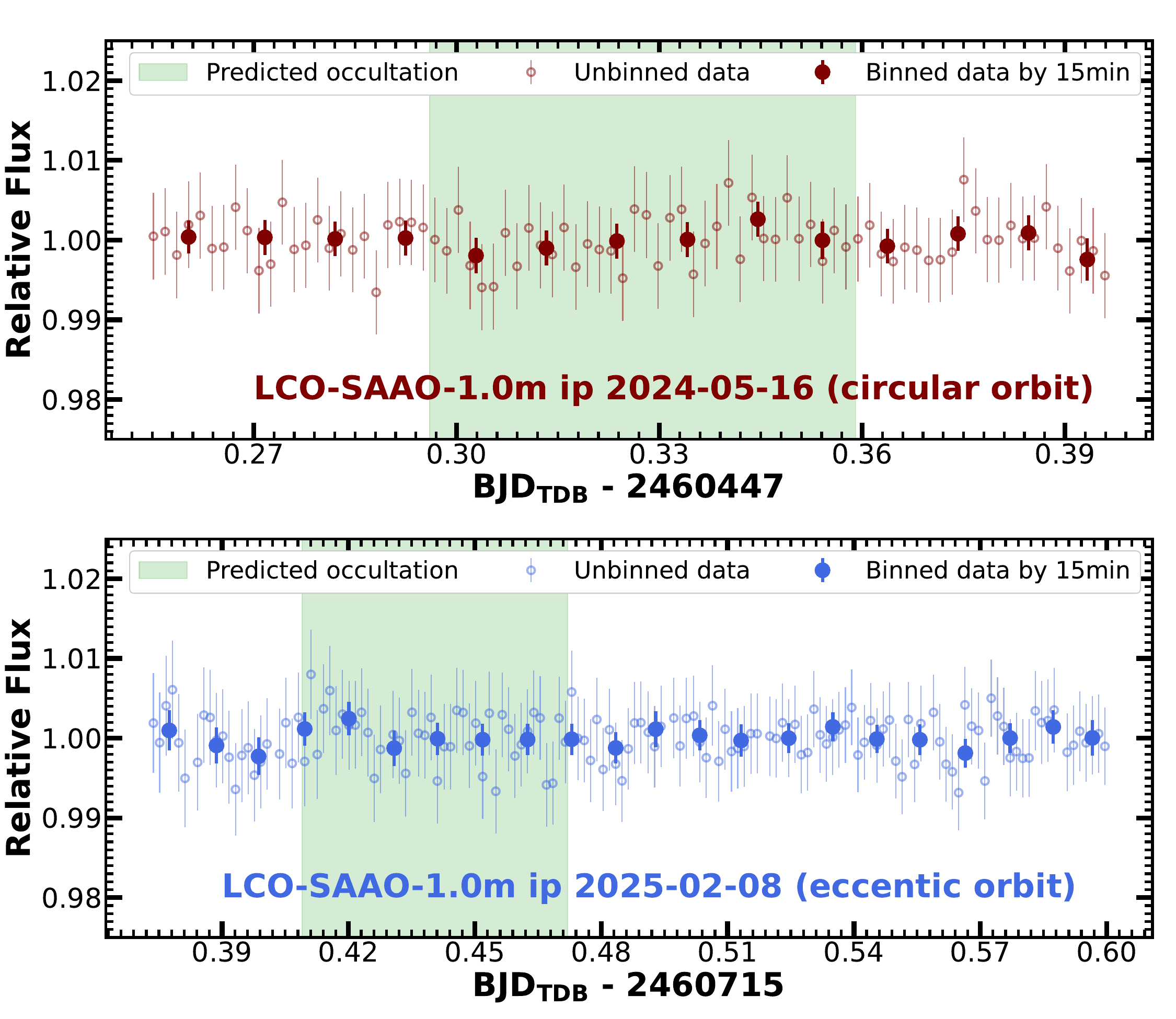}
	\caption{The secondary eclipse observations of TOI-6508\,b from LCO-SAAO-1.0m in the Sloan-$i'$. {\it Top panel} shows the data collected on  UTC May 16 2024 assuming a circular orbit and {\it bottom panel} shows the data collected on UTC Feb 8 2025 assuming an eccentric orbit ($e= 0.28$ constrained from
our global MCMC analysis).  No significant secondary eclipse is observed. The green region shows the predicted secondary eclipse ingress and egress.}
	\label{fig:LCs_tess_gb_eclipse}
\end{figure}

TOI-6508\,b was also observed during an occulation at the Southern African Astronomical Observatory (SAAO) in the Sloan-$i'$ filter. The first observation was carried out on UTC May 16 2024 assuming a circular orbit. The Second observation was carried out on UTC February 8 2025 assuming an eccentric orbit. Figure~\ref{fig:LCs_tess_gb_eclipse} shows the secondary eclipse light curves.

The science data processing was performed using the standard LCOGT {\tt BANZAI} pipeline \citep{McCully_2018SPIE10707E}, and aperture \& differential photometric were performed using {\tt AstroImageJ}\footnote{{\tt AstroImageJ:}\url{https://www.astro.louisville.edu/software/astroimagej/}} \citep{Collins_2017}.

\begin{table*}[!]
\caption{Observational  log for TOI-6508\,b: Telescope,  date of the observation, filter, exposure time(s), and FWHM of the point-spread function and photometric aperture are tabulated.}
 \begin{center}
 {\renewcommand{\arraystretch}{1.4}
 \begin{tabular}{l c c c c c c cccccc}
 \toprule
Telescope & Date (UT) & Filter &  Exptime  &  FWHM & Aperture & Comment \\ 
       &           &        &     [second] & [arcsec] & [arcsec]  &  \\
 \hline
LCO-Teid-1.0m & Jan 9 2025 & Pan-STARRS-$z_\mathrm{s}$ & 100 & 1.8 & 4.0 & Full transit \\
SPECULOOS-South-1.0m & March 29 2024 & Sloan-$r'$ & 180 & 2.0 & 2.1 & Full transit \\
SPECULOOS-South-1.0m & March 29 2024 & Sloan-$i'$ & 105 & 2.1 & 3.5 & Full transit \\
LCO-McD-1.0m & March 11 2024 & Pan-STARRS-$z_\mathrm{s}$ & 100 & 2.2 & 4.7 & Full transit \\
LCO-McD-1.0m & March 11 2024 & V & 100 & 2.9 & 5.4 & Full transit \\
LCO-McD-1.0m & Feb 21 2024 & Pan-STARRS-$z_\mathrm{s}$ & 100 & 1.8 & 4.7 & Full transit \\
LCO-McD-1.0m & Feb 21 2024 & V & 100 & 3.1 & 5.4 & Full transit \\
\hline
LCO-SAAO-1.0m & May 16 2024 & Sloan-$i'$ & 120 & 1.9 & 4.3 & Full occultation \\
LCO-SAAO-1.0m & Feb 8 2025 & Sloan-$i'$ & 100 & 2.1 & 4.0 & Full occultation \\
\hline
 \end{tabular}}
 \label{GB_obs_table}
 \end{center}
\end{table*}

\subsection{Spectroscopic observation for TOI-6508}

\subsubsection{NIRPS observation}
\label{Nirps_obs}

TOI-6508 was observed with the Near-InfraRed Planet Searcher (NIRPS; \citet{Bouchy_NIRPS_2017Msngr,Wildi_NIRPS_2022SPIE12184E}) spectrograph installed on the ESO-3.6m telescope at La Silla observatory in Chile. NIRPS is fiber-fed, stabilized high-resolution ($R \approx 70,000$) echelle spectrograph operating in the near-infrared covering the range from 950nm to 1800nm, under an adaptive optics system. The observations were carried out as a DDT program (ID:113.27QV, Cycle: 113, PI: K. Barkaoui).
Over 3 individual nights, we collected two spectra of TOI-6508 per night with NIRPS with an exposure time of 900\,s. The data were collected on June 1, 9 and July 3 2024 with an average signal-to-noise-ratio (SNR) of 12 at $1.6\,{\rm \mu m}$.\\ 
The NIRPS data were reduced with the nominal pipeline for NIRPS data reduction for the ESO science archive through the VLT Data Flow System\footnote{\url{https://www.eso.org/sci/software/pipelines/}}. This pipeline is based on the publicly available ESPRESSO pipeline, which utilizes recipes adapted from software originally developed for the ESPRESSO instrument \citep{Pepe2021} and specifically refined for near-infrared spectroscopy\footnote{{\tt Nirps pipeline:}\url{https://www.eso.org/sci/software/pipelines/nirps/nirps-pipe-recipes.html}} \citep{NIRPS2024}.\\
Radial velocities (RVs) were extracted using a cross-correlation with the line-mask for M4-type stars, which is implemented in the NIRPS pipeline. Due to the relatively low SNR we only use orders between 1.4-1.87\,${\rm \mu m}$. The RV measurements are presented in Table~\ref{table_rvs_nirps}.

\begin{table}[!]
\centering
	{\renewcommand{\arraystretch}{1.2}
 \resizebox{0.4\textwidth}{!}{
		\begin{tabular}{l|c|c}
		\hline
     BJD$_{\rm TDB}$     & RV [km/s] & $\sigma_{\rm RV}$ [km/s]      \\
        \hline
        2460463.468286  &  24.384 & 0.203  \\
        2460491.481160  &  51.406  & 0.159\\
        2460495.471592  &  30.840   & 0.790 \\
		\hline
  
		\end{tabular}} }
		\caption{RV measurements for TOI-6508 obtained with the NIRPS spectrograph.}
		\label{table_rvs_nirps}
\end{table}

\subsubsection{Shane/Kast optical spectroscopy} 
\label{Shane_Kast_obs}

We observed TOI-6508 with the Kast double spectrograph \citep{kastspectrograph} mounted on the Lick Observatory Shane 3m Telescope on UTC April 2024. Conditions were clear and windy with 1$\farcs$8 at the southern declination of the source. 
We used the 2$\farcs$5 (6~pixel) aligned with the parallactic angle and 
the 600/7500 grating in the Kast red channel to acquire 5800--9000~{\AA} spectra at a resolution of $\lambda/\Delta\lambda$ $\approx$ 900. 
Two exposures of 1200\,s each were acquired at an average airmass of 3.08, 
followed by observation of the nearby G2~V star HD 113207 ($V$ = 7.62) at a similar airmass.
The flux standard Feige~34 ($V$ = 7.6; \citealt{1990ApJ...358..344M,1990AJ.....99.1621O}) was observed earlier in the night. HeNeAr arclamps, quartz flat-field lamps, and bias frames were obtained at the start of the night for wavelength and pixel response calibration.
Data were reduced using {\tt kastredux}\footnote{{\tt kastredux:} \url{https://github.com/aburgasser/kastredux}} following standard procedures for optical spectroscopic data reduction (cf.\ \citealt{Barkaoui_2024A&A}). Our final spectrum has a signal-to-noise (S/N) of 76 at $\lambda$ $\approx$ 7500~{\AA}.

\subsubsection{IRTF/SpeX spectroscopy}
\label{IRTF_SpeX_obs}

We collected a medium-resolution near-infrared spectrum of TOI-6508 on 10 May 2024 (UT) using the SpeX spectrograph \citep{Rayner2003} on the 3.2-m NASA Infrared Telescope Facility (IRTF) during clear conditions with 0$\farcs$5 seeing.
We used the short-wavelength cross-dispersed (SXD) mode and the $0\farcs3 \times 15''$ slit aligned to the parallactic angle, yielding $R{\approx}2000$ spectra covering 0.80--2.42\,$\mu$m.
We collected 12 integrations of 300\,s on the target, nodding in an ABBA pattern.
Afterwards we gathered a set of standard SXD flat-field and arc-lamp calibrations and six, 30\,s integrations of the A0\,V standard HD\,100330.
We reduced the data with \texttt{Spextool v4.1} \citep{Cushing2004}, following the standard approach \citep[e.g.,][]{Delrez2022, Barkaoui2023A&A, Barkaoui_2024A&A}.
The reduced spectrum has a median S/N of 107 per pixel and 2.5\,pixels per resolution element.

\subsubsection{High-resolution imaging}

\begin{figure}[!]
	\centering
	\includegraphics[scale=0.6]{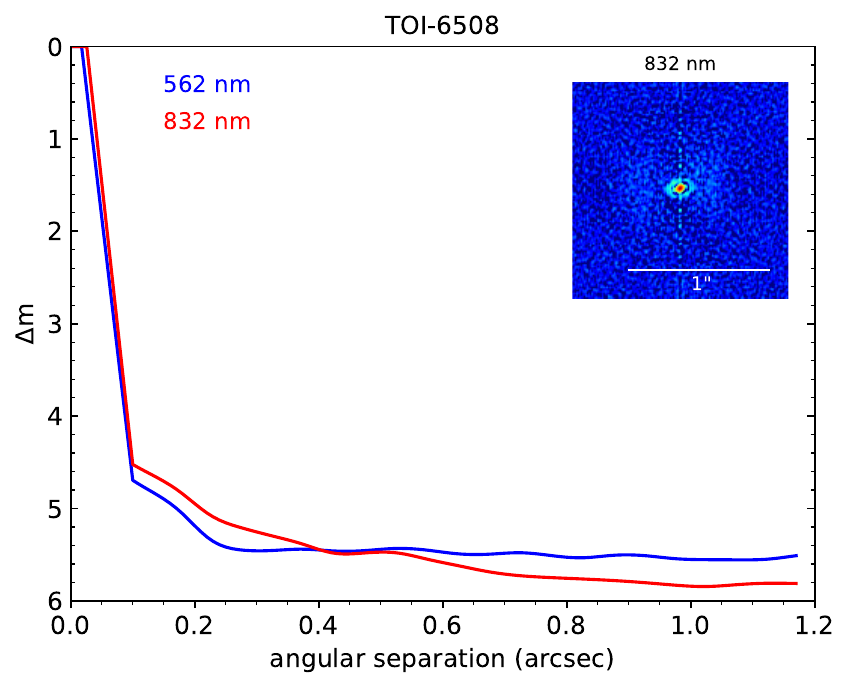}
	\caption{High-resolution imaging for TOI-6508 with 5$\sigma$ magnitude contrast curves in both filters as a function of the angular separation out to 1.2$\arcsec$. The inset shows the reconstructed 832 nm image of TOI-6508 with a 1$\arcsec$ scale bar. TOI-6508 was found to have no close companions from the diffraction limit (0.02$\arcsec$) out to 1.2$\arcsec$ to within the contrast levels achieved.}
	\label{fig:high_res}
\end{figure}

As part of the validation and confirmation process for a transiting exoplanet observation, high-resolution imaging is one of the critical assets required. The presence of a close companion star, whether truly bound or line of sight, will provide ''third-light'' contamination of the observed transit, leading to derived properties for the exoplanet and host star that are incorrect \citep{Ciardi_2015,Furlan_2017AJ,Furlan_2020}. In addition, it has been shown that the presence of a close companion dilutes small planet transits ($<1.2 R_\oplus$) to the point of non-detection \citep{Lester_2021}. Given that nearly one-half of FGK stars are in binary or multiple star systems \citep{Matson_2018} high-resolution imaging yields crucial information toward our understanding of each discovered exoplanet as well as more global information on its formation, dynamics and evolution.\\
TOI-6508 was observed on UTC 2025 January 10 using the Zorro speckle instrument on Gemini South \citep{Scott_2021FrASS}.  Zorro provides simultaneous speckle imaging in two bands (562 nm and 832 nm) with output data products including a reconstructed image and robust contrast limits on companion detections. Nine sets of 1000 60-ms frames were obtained for TOI-6508 simultaneously in each channel. The data was reduced using our standard software pipeline \citep{Howell_2011AJ}. Figure~\ref{fig:high_res} shows the 5-sigma magnitude contrast curves obtained and our 832 nm reconstructed speckle image. We find that TOI-6508 is a single star with no close companion brighter than about 5-6 magnitudes from the diffraction limit (0.02$\arcsec$) out to 1.2$\arcsec$.  At the distance to TOI-6508 ($d=48.5$~pc)  TOI-700, these angular limits correspond to spatial limits of 0.97 to 58 au.

\section{Stellar properties for TOI-6508}
\label{sec:stellar_charac}
\begin{table}[!]
\caption{Astrometry, photometry, and spectroscopy stellar properties of TOI-6508.}
\centering
	{\renewcommand{\arraystretch}{1.6}
        \resizebox{0.5\textwidth}{!}{
		\begin{tabular}{lcc}
			\hline
			\hline
			\multicolumn{3}{c}{  Star information}   \\
			\hline
			{\it Target designations:}  & & \\
			            & \multicolumn{2}{l}{TOI 6508 }  \\
                        & \multicolumn{2}{l}{TIC 142277868 }  \\
			              & \multicolumn{2}{l}{GAIA DR3 3465796341653768192 } \\
			            & \multicolumn{2}{l}{2MASS J12030396-3339552 }   \\
                        & \multicolumn{2}{l}{SIPS J1203-3339} \\
   			Parameter & Value &  Source   \\
			\hline
            \hline
			{\it Parallax  and distance:} &   \\
			RA [J2000]     &  12:03:03.42  &  (1) \\
			Dec [J2000]    & -33:39:55.8   &  (1)\\
			Plx [$mas$] & $20.64 \pm 0.05$ &   (1)\\
                $\mu_{RA}$ [mas yr$^{-1}$] & $-413.59 \pm 0.06$ & (1) \\
                $\mu_{Dec}$ [mas yr$^{-1}$] & $-35.14 \pm 0.03$ & (1) \\
			Distance [pc]  & $48.44 \pm 0.12$ & (1)\\
			\hline
			{\it Photometric properties:} & \\
			TESS$_{\rm mag}$           &  $14.309 \pm 0.008$  & (2)  \\
			$V_{\rm mag}$ [UCAC4]       & $17.18 \pm 0.20$  & (3) \\
            $B_{\rm mag}$ [UCAC4]       & $18.3$  & (3) \\
            $R_{\rm mag}$ [UCAC4]       & $16.7$  & (3) \\
			$J_{\rm mag}$ [2MASS]       & $12.39 \pm 0.02$  &  (4) \\
			$H_{\rm mag}$ [2MASS]       & $11.83 \pm 0.03$  &    (4) \\
			$K_{\rm mag}$ [2MASS]       & $11.49 \pm 0.02$  &   (4) \\			
			$G_{\rm mag}$ [Gaia DR3]    & $15.79 \pm 0.001$  &  (1)  \\
			$W1_{\rm mag}$ [WISE]       & $11.310 \pm 0.022$ &  (5) \\
			$W2_{\rm mag}$ [WISE]       & $11.098 \pm 0.021$  &   (5) \\
			$W3_{\rm mag}$ [WISE]       & $10.89 \pm 0.09$  &   (5)\\
            $W4_{\rm mag}$ [WISE]       & $8.995$  &   (5)\\
			\hline
   			\multicolumn{2}{l}{\it Spectroscopic and derived parameters}  \\
			$T_{\rm eff}$ [K]              &  $ 2930 \pm 70 $  &   this work\\
			$\log g_\star$ [dex]           &  $ 5.05 \pm 0.02 $  &  this work\\
			$[Fe/H]$ [dex]                 & $ -0.22 \pm 0.08 $  &  this work\\
			$M_\star$  [$M_\odot$]         & $ 0.174 \pm 0.004 $  &  $^a$ this work\\
			$R_\star$  [$R_\odot$]         & $ 0.205 \pm 0.006 $ & $^a$ this work\\
			$F_{\rm bol}$  [erg s$^{-1}$ cm$^{-2}$]   &  $ (4.48 \pm 0.21) \times 10^{-11} $  &  this work\\
			$Av$ [mag]    &  $0.1 \pm 0.1$ & this work\\
			$\rho_\star$  [$\rho_\odot$]   & $20.2 \pm 1.8$  & this work \\
			$Age$  [Gyr]                   & $\lesssim$7 & this work \\
			Spectral type                  & M6$\pm$1 & [Shane/Kast] \\
			Spectral type                  &  M5.0 $\pm$ 0.5  & this work [IRTF/SpeX] \\
            \hline
	\end{tabular} }}
	\tablefoot{Astrometry, photometry, and spectroscopy stellar properties of TOI-6508. 
	{\bf (1):} Gaia EDR3 \cite{Gaia_Collaboration_2021AandA}; 
	{\bf (2)} \emph{TESS} Input Catalog \cite{Stassun_2018AJ_TESS_Catalog}; 
	{\bf (3)} UCAC4 \cite{Zacharias_2012yCat.1322};
	{\bf (4)} 2MASS \cite{Skrutskie_2006AJ_2MASS};
	{\bf (5)} WISE \cite{Cutri_2014yCat.2328}. $^a$ Stellar mass and radius values are computed from \cite{Mann_2015, Mann:2019}
	\label{stellarpar}}
\end{table}

\subsection{Shane/Kast}
\label{sec:Shane_Kast_charac}

\begin{figure}
    \centering
    \includegraphics[width=0.52\textwidth]{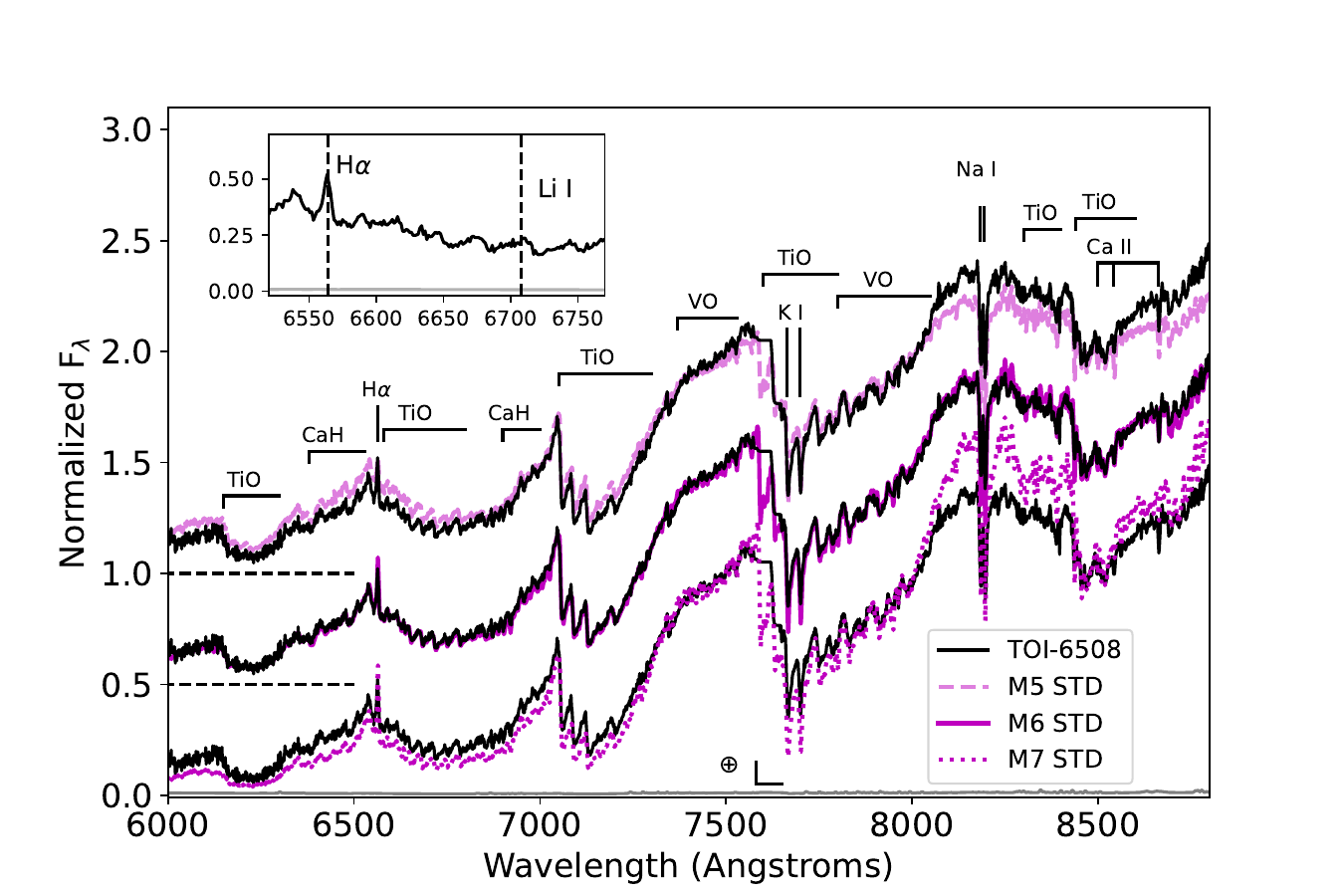}
    \caption{Kast spectrum of TOI-6508 (black lines) compared to M5, M6, and M7 standards from \citet[magenta lines]{2007AJ....133..531B}. All spectra are normalized at 7500~{\AA} and for clarity (zeropoints are indicated by dashed lines). Key atomic and molecular spectral features are labeled, as are regions of strong telluric absorption ($\oplus$). The inset box highlights the 6520--6770{\AA} region containing the H$\alpha$ emission (detected) and Li~I absorption lines (absent).}
    \label{fig:kast}
\end{figure}

Figure~\ref{fig:kast} shows the Kast optical spectrum of TOI-6508.
Characteristic spectral features for mid- to late- M dwarfs are present, including CaH, TiO, and VO molecular features and line absorption from Na~I, K~I and Ca~II. The spectrum is an excellent match to the M6 SDSS spectral template data from \citet{2007AJ....133..531B}, implying an optical classification of M6$\pm$1. H$\alpha$ is seen in clear emission with an equivalent width of $-$2.95$\pm$0.18~{\AA}, which corresponds to a relative emission luminosity of $\log_{10}L_{H\alpha}/L_{bol}$ = $-$4.36$\pm$0.10 using the $\chi$-factor calibration of \citet{2014ApJ...795..161D}.
The presence of magnetic emission is consistent with an activity age $\lesssim$7~Gyr \citep{2008AJ....135..785W,2023MNRAS.526.4787R}.
The relative strength of TiO and CaH absorption in the 7000~{\AA} region yields a $\zeta$ value of 0.85$\pm$0.02 \citep{2013AJ....145..102L}, near the boundary between dwarf and subdwarf classes, and the \citet{2013AJ....145...52M} metallicity-$\zeta$ relation yields [Fe/H]$ = -$0.17$\pm$0.20.

\subsection{IRTF/SpeX}
\label{sec:IRTF_SpeX_charac}

\begin{figure}
    \centering
    \includegraphics[width=\linewidth]{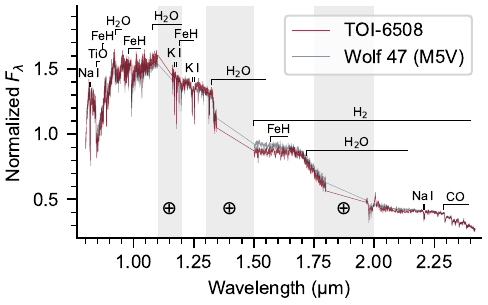}
    \caption{
    SpeX SXD spectrum of TOI-6508 (red) alongside the spectrum of the M5 dwarf Wolf 47 (grey) for comparison.
    Atomic and molecular features of M dwarfs are annotated, and regions of strong telluric absorption are shaded.
    }
    \label{fig:spex}
\end{figure}

Figure\,\ref{fig:spex} shows the SpeX SXD spectrum of TOI-6508.
Following previous SpeX analyses \citep[e.g.,][]{Triaud2023, Gillon2024, Timmermans2024}, we used the SpeX Prism Library Analysis Toolkit \citep[SPLAT, ][]{splat} to assign a spectral type and estimate a stellar metallicity.
We compared the spectrum to those of single-star spectral standards in the IRTF Spectral Library \citep{Cushing2005, Rayner2009}.
Finding the best match to the M5 dwarf Wolf\,47, we adopt an infrared spectral type of M5.0 $\pm$ 0.5, slightly earlier than but consistent with the optical classification.
Using the H2O--K2 index \citep{Rojas-Ayala2012} and \citet{Mann2014} relation, we estimate a sub-solar stellar iron abundance of $\mathrm{[Fe/H]} = -0.22 \pm 0.08$ for TOI-6508, consistent with the optical metallicity.

\subsection{SED analysis}
\begin{figure}[!]
	\centering
	\includegraphics[scale=0.3]{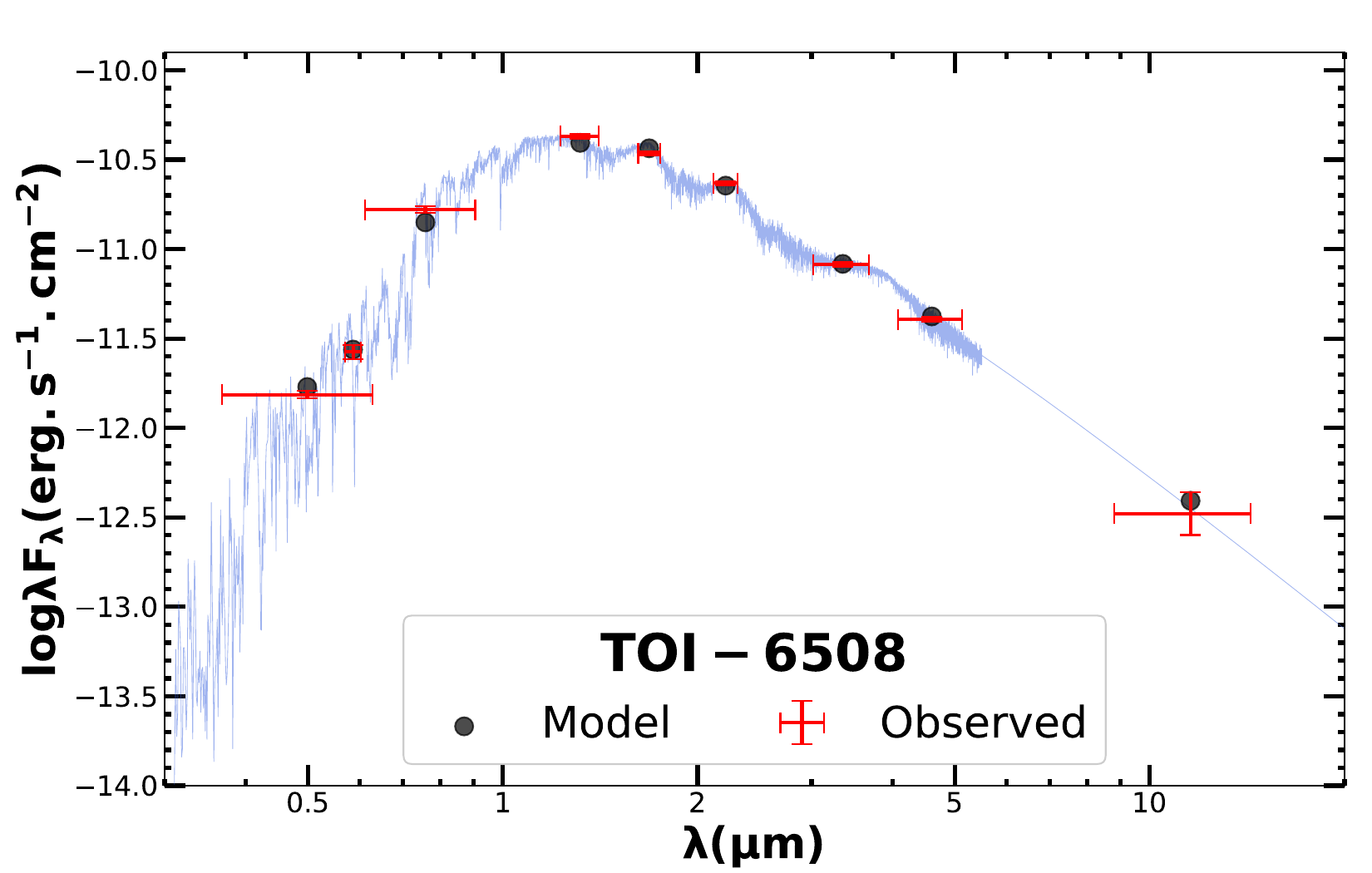}
	\caption{SED fit of TOI-6508. The red symbols with error-bars are the observed fluxes, and black dots are the PHOENIX model fluxes. The best-fitting NextGen atmosphere model is presented in blue.}
	\label{fig:sed}
\end{figure}

As an independent determination of the basic stellar parameters, we performed an analysis of the broadband spectral energy distribution (SED) of the star together with the {\it Gaia\/} DR3 parallax \citep[with no systematic offset applied; see, e.g.,][]{StassunTorres:2021}, as described in \citet{Stassun:2016,Stassun:2017,Stassun:2018}. We used the the 2MASS $JHK_S$ magnitudes, the {\it WISE} W1--W3, and the {\it Gaia} $G_{\rm BP} G_{\rm RP}$ magnitudes. Together, the available photometry spans the full stellar SED over the wavelength range 0.4--10~$\mu$m (see Figure~\ref{fig:sed}).  
 
We performed a fit using PHOENIX stellar atmosphere models \citep{Husser:2013}, with the free parameter being the effective temperature ($T_{\rm eff}$), and the extinction $A_V$ which we limited to maximum line-of-sight value from the Galactic dust maps of \citet{Schlegel:1998}; we used the metallicity ([Fe/H]) value derived from spectroscopic observations (see Sections~\ref{sec:Shane_Kast_charac} and \ref{sec:IRTF_SpeX_charac}). The resulting fit (Figure~\ref{fig:sed}) has a best-fit $A_V = 0.1 \pm 0.1$ and $T_{\rm eff} = 2930 \pm 70$~K, with a reduced $\chi^2$ of 3.8. Integrating the (unreddened) model SED gives the bolometric flux at Earth, $F_{\rm bol} = 4.48 \pm 0.21 \times 10^{-11}$ erg~s$^{-1}$~cm$^{-2}$. Taking the $F_{\rm bol}$ and {\it Gaia\/} parallax gives directly the stellar bolometric luminosity, $L_{\rm bol} = 0.00327 \pm 0.00015$~L$_\odot$. The stellar radius follows from the Stefan-Boltzmann relation, giving $R_\star = 0.222 \pm 0.013 $~R$_\odot$. In addition, we estimated the stellar mass from the empirical $M_K$ relations of \citet{Mann:2019}, giving $M_\star = 0.20 \pm 0.01$~M$_\odot$. Finally, we used the measured chromospheric activity $\log H\alpha / \log L_{\rm bol}$ with the empirical relations from \citet{Stassun:2012} to predict the degree of radius inflation, which in this case is predicted to be $\approx$7\%, roughly consistent with the apparent inflation of $\sim$10\%.

\subsection{Archival imaging for TOI-6508}

\begin{figure*}[!]
	\centering
	\includegraphics[scale=0.7]{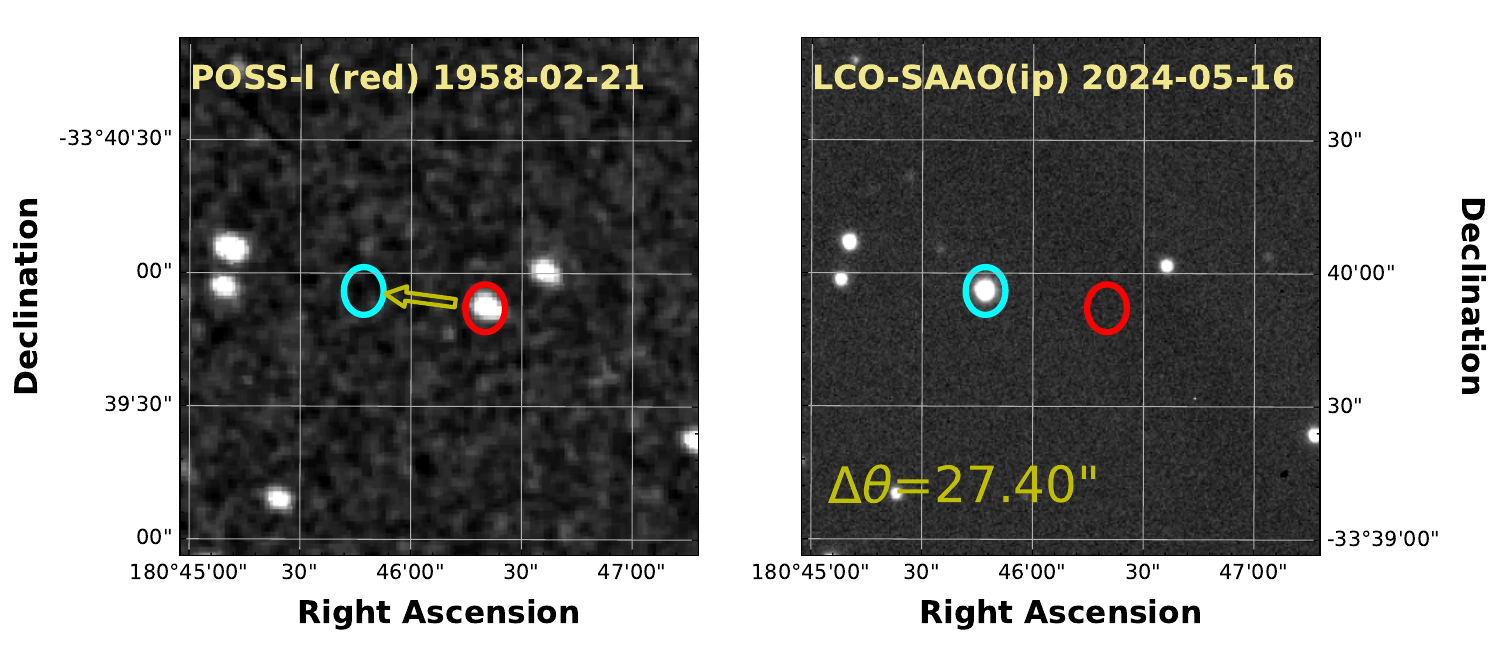}
	\caption{Evolution of TOI-6508's position over time.  {\it Left panel} shows archival image of TOI-6508 taken using a photographic plate on the Palomar Schmidt Telescope in the red filter. {\it Right panel} shows the Sloan-$i'$ image from LCO-SAAO-1.0m taken in 2024.}
	\label{archival_images}
\end{figure*}

We used the archival images of TOI-6508 to exclude the possibility that there are background stars blended with the target at its current position.
TOI-6508 has a high proper motion of 414~mas/yr. We used the POSS-I/red data taken in 1958, and the LCO-SAAO-1.0m/Sloan-$i'$ data taken in 2024, spanning 72 years with our new observations. 
The target has been shifted   by 27\farcs40 from 1958 to 2024. There is no bright background source in the current day position of the target (see Figure~\ref{archival_images}).

\subsection{Stellar rotation}

\begin{figure}[!]
	\centering
	\includegraphics[scale=0.36]{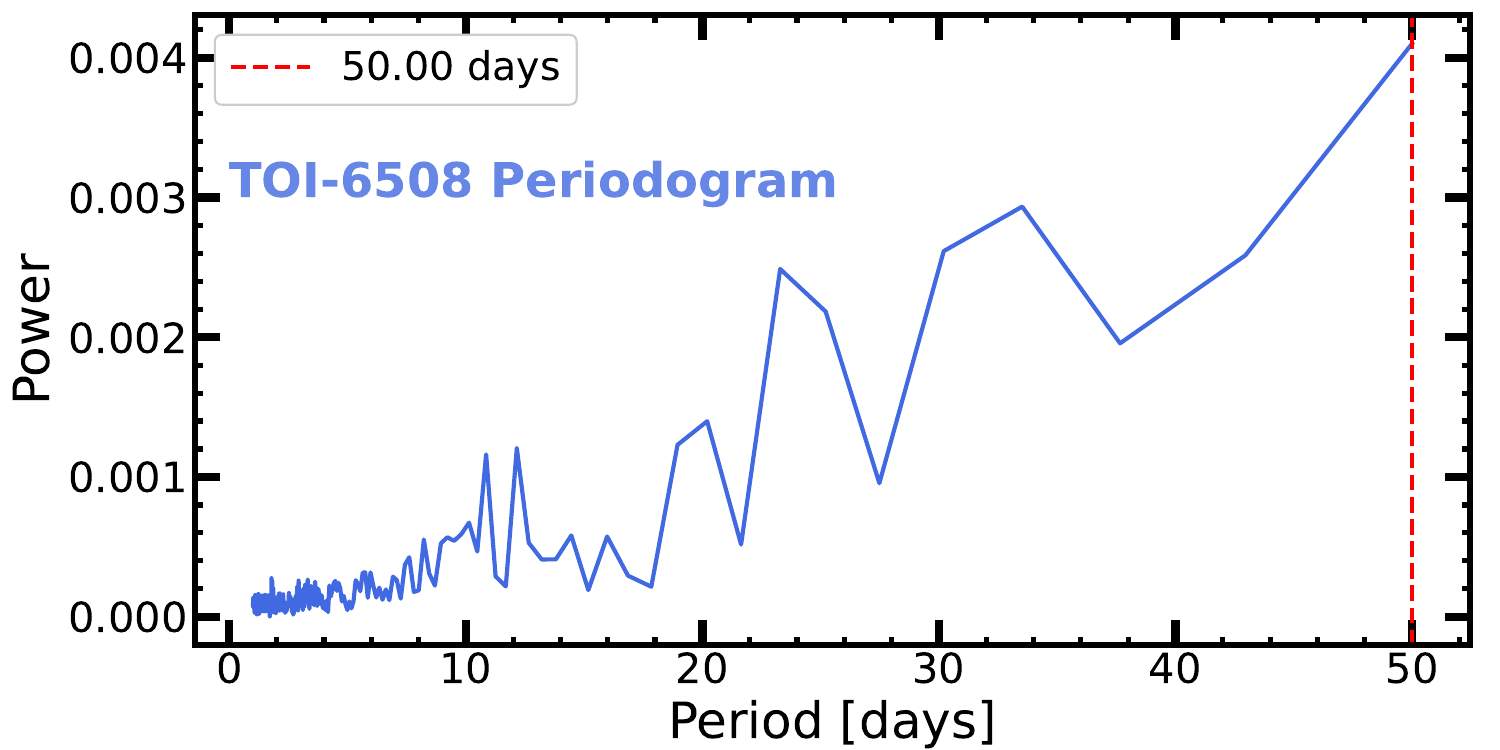}
	\caption{TESS Systematics-Insensitive Periodogram ({\tt TESS-SIP}) of TOI-6508 using  the TESS data from Sectors 10, 37 and 63 (blue line).}
	\label{fig:SIP}
\end{figure}

We searched for photometric modulation in TESS observations using the {\tt TESS-SIP} \citep{Hedges_2020} package. {\tt TESS-SIP} returns two outputs simulateneousely, which are  a Lomb–Scargle periodogram \citep{Lomb_1976ApSS,Scargle_1982ApJ} and detrend systematics.
In our case, we used the available TESS photometric data from Sectors 10, 37 and 63. We limit our search to a rotation period range 1--50~days. Our results showed that no indications stellar modulation in the TESS data of TOI-6508 (see Figure~\ref{fig:SIP}). This implies that the rotational periods of the target star is probably longer than the TESS observation window for a single sector.

\section{Photometric and RVs analysis}
\label{sec:mcmc_fit}

We performed a global modeling of transit light curves obtained from  the TESS mission (described Section~\ref{sec:tess_photometry}), and SPECULOOS-South-1.0m and LCO-SAAO-1.0m telescopes (described in Section~\ref{sec:gb_photometry}), together with the radial velocity measurements collected by the ESO-3.6m/NIRPS spectrograph (described in Section~\ref{Nirps_obs}), using the Metropolis-Hastings \citep{Metropolis_1953,Hastings_1970} method implemented in {\tt TRAFIT}, a revised version of the Markov chain Monte Carlo (MCMC) code described in \cite{Gillon2010AA,Gillon2012,Gillon2014AA}. We followed the same strategy as described in \citet{Barkaoui2023A&A,Barkaoui_2024A&A}.
The photometric data were modeled using the \cite{Mandel2002} quadratic limb-darkening model, multiplied by a transit baseline, in order to correct for external systematic effects related to the time and FWHM of the PSF as well as the airmass, and background light level. The radial velocity data were modeled with the two-body Keplerian model \citep{Murray_2010exopbook}.

The baseline model for each transit was selected by minimizing the Bayesian information criterion (BIC; \citet{schwarz1978}).
The error bars of TOI-6508 radial velocity measurements were quadratically rescaled using the `jitter' noise, while the
photometric error bars were rescaled using the correction factor $CF = \beta_{w} \times \beta_{r}$, where $\beta_{r}$ is the red noise and $\beta_{w}$ is the white noise \citep{Gillon2012}.  

For the joint fit, the global free parameters for transit modeling are the mid-transit time at a reference epoch ($T_0$), the orbital period ($P$), the transit depth ($dF$), the impact parameter ($b$), the stellar density $\rho_\star$, as well as the total transit duration ($W$). We applied a Gaussian prior distribution on the stellar quadratic limb-darkening coefficients ($u_1$ and $u_2$), stellar mass $M_\star$, radius $R_\star$ and effective temperature $T_{\rm eff}$ (computed from SED analysis), as well as the stellar atmospheric parameters computed from spectroscopic analysis (metallicity $T_{\rm eff}$ and surface gravity $\log g_\star$).
The quadratic limb-darkening coefficients $u_1$ and $u_2$ for the TESS, Pan-STARRS-$z_\mathrm{s}$, Sloan-$i'$, Sloan-$r'$, and Johnson-$V$ filters were computed using the stellar parameters ($T_{\rm eff}$, $[Fe/H]$ and $\log g_\star$) and Tables from \cite{Claret_2012AA,Claret_2018AandA}. During our analysis, we converted the quadratic limb-darkening coefficients $u_1$ and $u_2$ into the combination $q_1 = (u_1 + u_2)^2$ and $q_2 = 0.5u_1(u_1 + u_2)^{-1}$ proposed by \cite{Kipping_2013MNRAS.435.2152K}.

We performed two MCMC fits. The first fit assumed an eccentric orbit (i.e. free eccentricity) and the second assumed a perfectly circular orbit (i.e., $e=0$). Our results favored an eccentric orbit solution based on the Bayes factor $BC = \exp{(-\Delta BIC/2)}>1000$. 
For each transit, a preliminary analysis was performed composed of one Markov chain with $5\times 10^5$ steps to compute the $CF$ (correction factor; \citet{Gillon2012}). Then, we performed a final MCMC fit  composed of five Markov chains with one million steps to infer the physical properties of the system. The convergence for each Markov chain has been checked based on the \cite{Gelman1992} statistical test. Our final results for the eccentric orbit solution are presented in Table~\ref{tab:mcmc_results}.


\begin{table}[!]
\caption{Derived physical parameters of the TOI-6508\,b system with 1-$\sigma$ for the eccentric orbit solution.}
	\begin{center}
		{\renewcommand{\arraystretch}{1.4}
				\resizebox{0.48\textwidth}{!}{
			\begin{tabular}{ll}
		\hline
            \multicolumn{2}{c}{TOI-6508}  \\
        \hline
				Parameter &  Value    \\   
	    \hline
            \multicolumn{2}{l}{\it Quadratic Limb-Darkening coefficients}  \\
        \hline
			    $u_{\rm 1,TESS}$ &  $0.32 \pm 0.02$   \\
				$u_{\rm 2,TESS}$ &  $0.23 \pm 0.03$   \\
				$u_{\rm 1,{\rm Pan-STARRS-z_s}}$ &  $0.29 \pm 0.02$   \\
				$u_{\rm 2,{\rm Pan-STARRS-z_s}}$ &  $0.19 \pm 0.04$  \\
				$u_{\rm 1,{\rm Sloan}-i'}$ &  $0.43 \pm 0.01$   \\
				$u_{\rm 2,{\rm Sloan}-i'}$ &  $0.30 \pm 0.01$   \\
				$u_{\rm 1,{\rm Sloan}-r'}$ &  $0.70 \pm 0.01$  \\
				$u_{\rm 2,{\rm Sloan}-r'}$ &  $0.19 \pm 0.01$   \\
                $u_{\rm 1,{\rm Johnson}-V}$ &  $0.73 \pm 0.01$  \\
				$u_{\rm 2,{\rm Johnson}-V}$ &  $0.19 \pm 0.01$   \\
        \hline
            \multicolumn{2}{l}{\it Derived stellar parameters}  \\
        \hline
				Stellar mass, $M_\star$ [$M_\odot$ ]      & $0.1744 ^{+0.0203}_{-0.0198}$        \\
				Stellar radius,  $R_\star$ [$R_\odot$]      & $0.2041^{+0.0061}_{-0.0061}$       \\
    			Mean density,  $\rho_\star$ [$\rho_\odot$]   &  $20.38^{+3.36}_{-2.65} $      \\
				Luminosity, $L_\star$ [$L_\odot$]       & $0.003044 _{-0.000319}^{+0.000353}$    \\
                Effective temperature, $T_{\rm eff}$ [K] & $ 3003^{+71}_{-69} $  \\ 
        \hline
        \multicolumn{2}{l}{\it Derived BD parameters}  \\
        \hline
            Radius ratio $R_p/R_\star$               &  $ 0.5036^{+0.0037}_{-0.0030} $   \\
            Orbital period $P$ [days]                &  $18.99265922 ^{+0.0000688}_{-0.0000681}$   \\ 
            Transit-timing $T_0$                     &  $ 10399.8462781 \pm 0.0000685$     \\
            $[{\rm BJD}_{\rm TDB} - 2450000]$              &     \\
            Orbital semi-major axis $a$ [AU]         &  $ 0.08659 \pm 0.00296$    \\
            Impact parameter $b$ [$R_\star$]         &  $0.631^{+0.039}_{-0.044} $    \\
			Transit duration $W$ [min]               &  $ 109.2 \pm 0.5 $    \\
			Scaled semi-major axis  $a/R_\star$      &  $ 90.96^{+4.53}_{-3.65}$     \\
			Orbital inclination $i$ [deg]            &  $ 89.60 \pm 0.04$    \\
            Eccentricity $e$                         &  $ 0.28^{+0.09}_{-0.08} $ \\
            $\sqrt{e} \cos(w)$                       &  $ 0.35_{-0.20}^{+0.17} $ \\
            $\sqrt{e} \sin(w)$                       &  $ 0.375_{-0.093}^{+0.068} $ \\
            RV semi-amplitude $K$ [$km/s$]           &  $ 14.88^{+1.05}_{-0.69} $     \\
            Mass ratio $M_{\rm BD}/M_\star$          &  $ 0.397^{+0.053}_{-0.062} $    \\
            BD Radius $R_{\rm BD}$ [$R_{\rm Jupiter} $]     &  $ 1.026_{-0.032}^{+0.031} $   \\
            BD Mass $M_{\rm BD}$ [$M_{\rm Jupiter}$ ]       &  $ 72.53_{-5.09}^{+7.61} $      \\
            BD density $\rho_{\rm BD}$  [$g/cm^3$]   &    $ 89.0_{-8.2}^{+13.7} $ \\
            Surface gravity $\log g_{\rm BD}$        &   $ 5.2469_{-0.0338}^{+0.0535} $ \\
            Incident flux $<F>$  [$<F_\oplus>$]                    & $ 0.407 ^{+0.055}_{-0.051} $ \\
   \hline
		\end{tabular}}}
	\end{center}
	\label{tab:mcmc_results}
\end{table}

\section{Discussion and Conclusion} \label{discu_conclusion}

Transiting BDs around M dwarf stars are rare, and they are helpful for understanding the formation and evolution of such systems. Only $\sim 10$ M dwarf/BD systems are known, and more detections are required to probe the formation and evolution paths.

In this paper, we present a BD orbiting a low-mass star, TOI-6508\,b. 
The target was observed with the TESS mission during Sectors 10, 37 and 73 with long-cadence of 1800\,s, 600\,s and 200\,s, respectively (Section~\ref{sec:tess_photometry}). The candidate was first identified by TESS. 
Ground-based photometric follow-up observations were performed with the SPECULOOS-South-1.0m and LCOGT-McD-1.0m telescopes (Section~\ref{sec:gb_photometry}). Radial velocity measurements were collected using the NIRPS spectrograph as described in Section~\ref{Nirps_obs}.
The host star was characterized by combining optical spectra collected by IRTF/SpeX and Shane/Kast instruments, the spectral energy distribution (SED), and stellar evolutionary models (Section~\ref{sec:stellar_charac}). TOI-6508 is a $K_{\rm mag} = 11.5$ M5.5 sub-solar star with metallicity of $[Fe/H] = -0.22\pm0.08$, a mass of $M_\star = 0.174\pm 0.004~M_\odot$, a radius of $R_\star = 0.205\pm 0.006~R_\odot$ and an effective temperature of $T_{\rm eff} = 2930\pm 70~K$.

We performed a global analysis of the TESS observations together with ground-based photometric and radial velocity observations in order to derive the physical parameters of the system (Section~\ref{sec:mcmc_fit}).
Table~\ref{stellarpar} shows the stellar physical characteristics of the host star TOI-6508. 
The derived physical parameters of the system are presented in Table \ref{tab:mcmc_results}. The posterior distribution parameters of TOI-6508\,b are presented in Figure~\ref{corner_TOI6508}.
We find that TOI-6508\,b is a massive brown dwarf with a mass of $M_{\rm BD} = 72.53^{+7.61}_{-5.09}M_{\rm Jup}$ and a radius of $R_{\rm BD} = 1.026^{+0.031}_{-0.032}R_{\rm Jup}$. It is the second highest mass ratio BD transiting a low-mass star. 

\begin{figure}[!]
	\centering
	\includegraphics[scale=0.3]{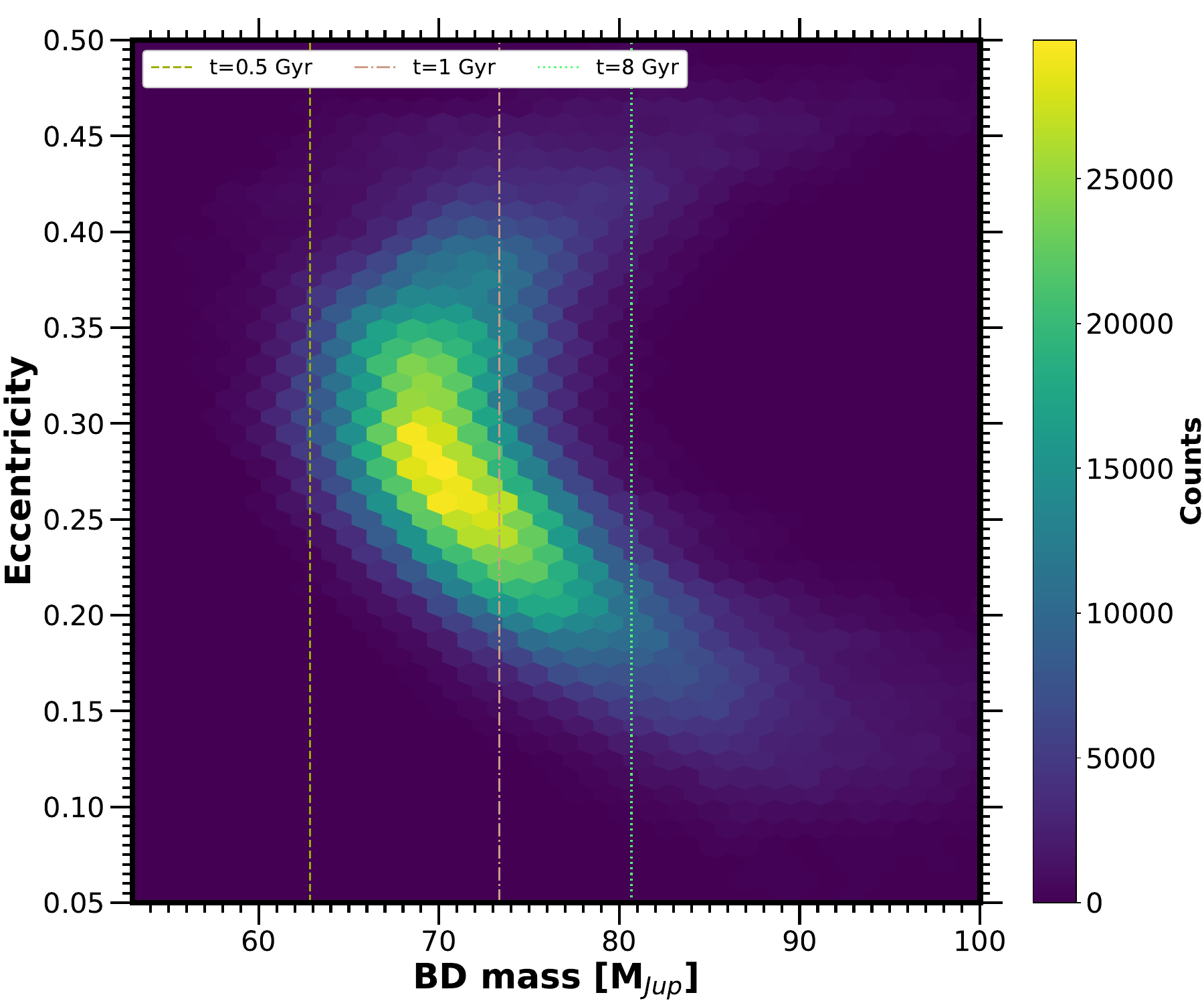}
	\caption{Posterior probability distribution of the eccentricity and the mass of TOI-6508\,b. Vertical colored lines show the evolutionary models from \citet{Baraffe_2003A&A}.}
	\label{diagram:ecc_mass}
\end{figure}

\begin{figure}[!]
	\centering
	\includegraphics[scale=0.26]{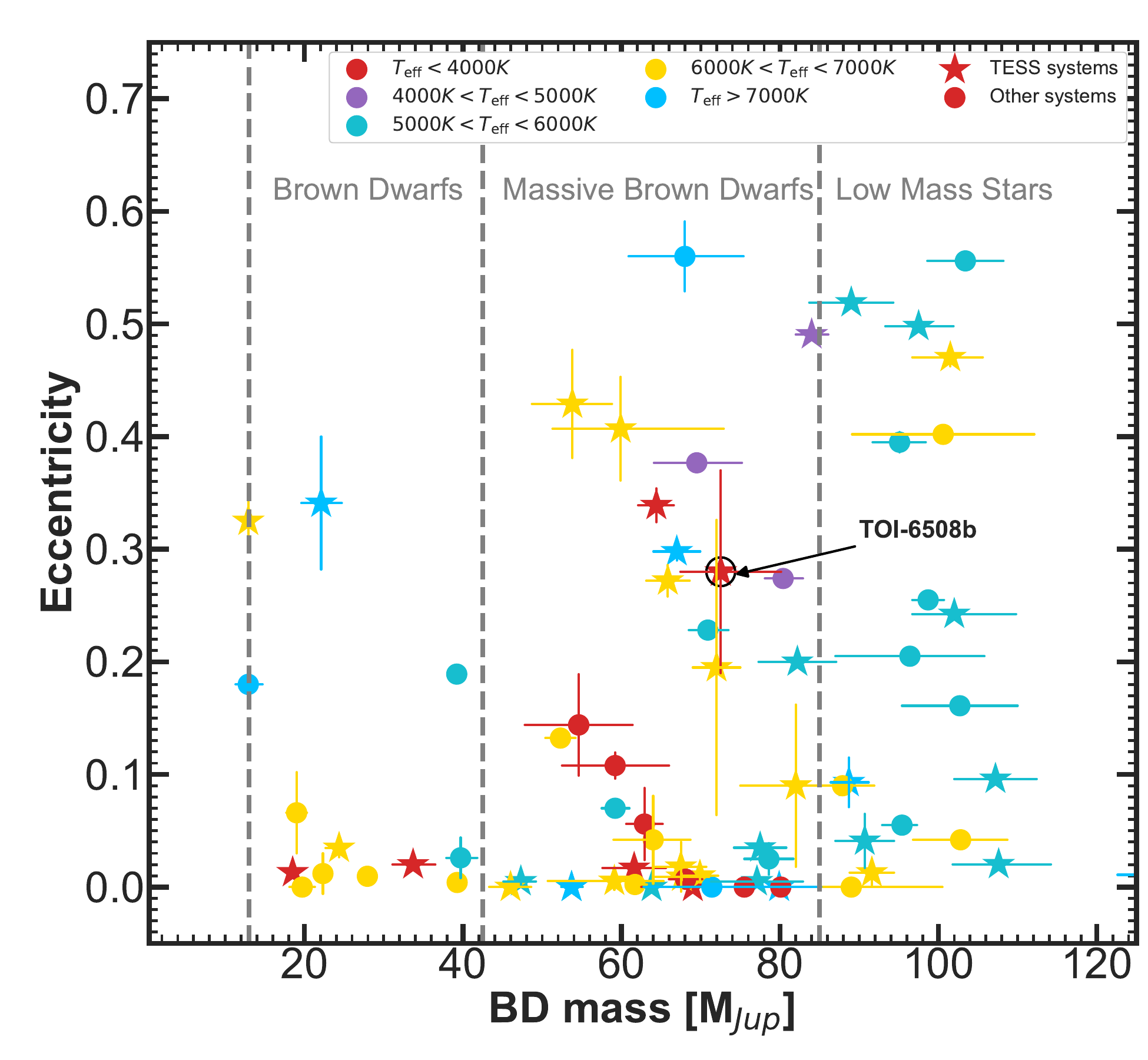}
	\caption{The eccentricity as a function of the mass of transiting BDs from Table~\ref{tab:full_BDs_lists}. The color of each point indicates the effective temperature of the host star. TESS BD systems are highlighted by the stars and other systems by dots. TOI-6508\,b is highlighted by the black circle. }
	\label{diagram:Mass_ecc}
\end{figure}

\begin{figure}[!]
	\centering
	\includegraphics[scale=0.27]{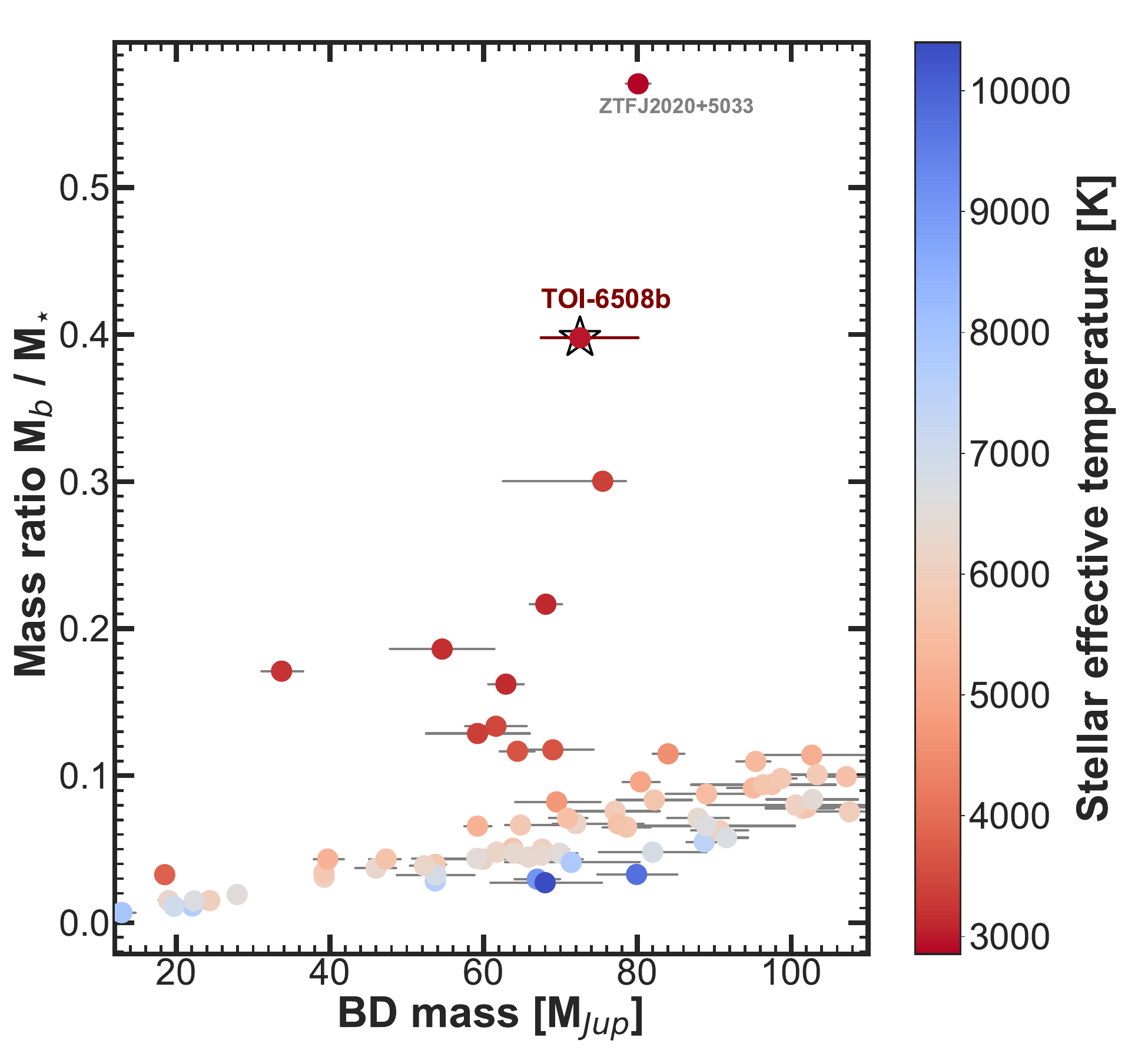}
	\caption{Comparison of TOI-6508\,b to other transiting BD systems from Table~\ref{tab:full_BDs_lists}. TOI-6508\,b is the second highest mass ratio transiting BD after ZTF J2020+5033 \citep{ZTFJ2020}. Dot are colored according to the stellar effective temperature.}
	\label{diagram:Mass_ratio}
\end{figure}

During our modeling, we performed two MCMC fits. The first assuming a circular orbit and the second assuming an eccentric orbit. The best solution is compatible with an eccentric orbit, based on the Bayes factor $BC$.
TOI-6508\,b  orbits its host star with an orbital period of $P = 18.99265922 ^{+0.0000688}_{-0.0000681}$~days and an eccentricity of $e = 0.28^{+0.09}_{-0.08}$.
Figure~\ref{diagram:ecc_mass} shows the posterior probability distribution of the orbital eccentricity and the mass of TOI-6508\,b, including the evolutionary models from \citet{Baraffe_2003A&A}.
Additional observations of radial velocity are required to improve the orbital eccentricity and mass measurements of TOI-6508\,b (see Figure~\ref{diagram:Mass_ecc}).
Figure~\ref{diagram:Mass_ratio} presents the mass ratio $M_{\rm BD}/M_\star$ as a function of the BD mass. TOI-6508\,b has the second highest mass ratio among all known transiting BDs.

\begin{figure*}[!]
	\centering
     \includegraphics[scale=0.32]{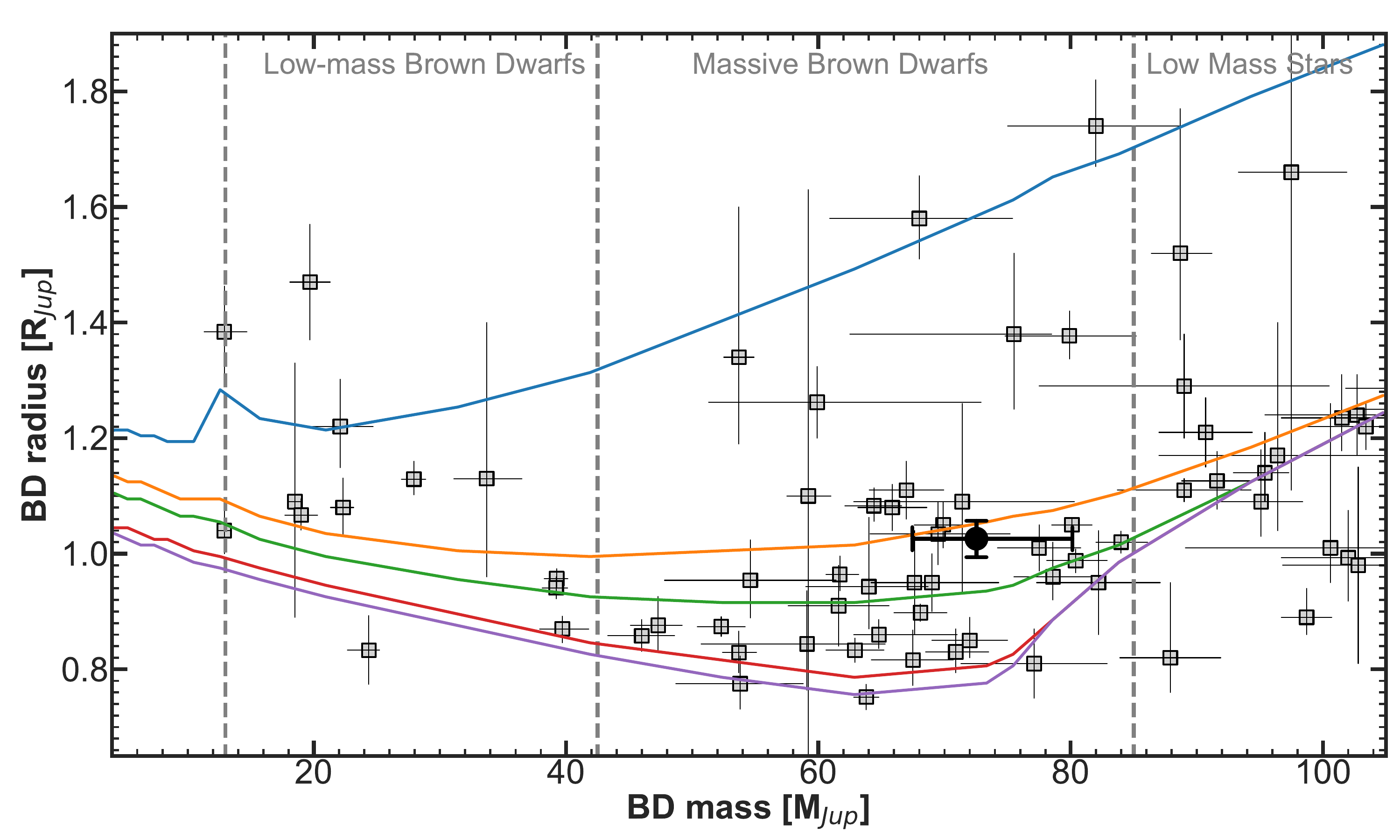}
	\caption{The Radius against the mass of transiting BDs from Table~\ref{tab:full_BDs_lists}. The colored solid lines indicate the evolutionary models from \cite{Baraffe_2003A&A} with different ages from 0.1 to 10 Gyrs. TOI-6508\,b is highlighted by the black dot with error bars. }
	\label{diagram:Mass_Radius}
\end{figure*}

\begin{figure*}[!]
	\centering
     \includegraphics[scale=0.32]{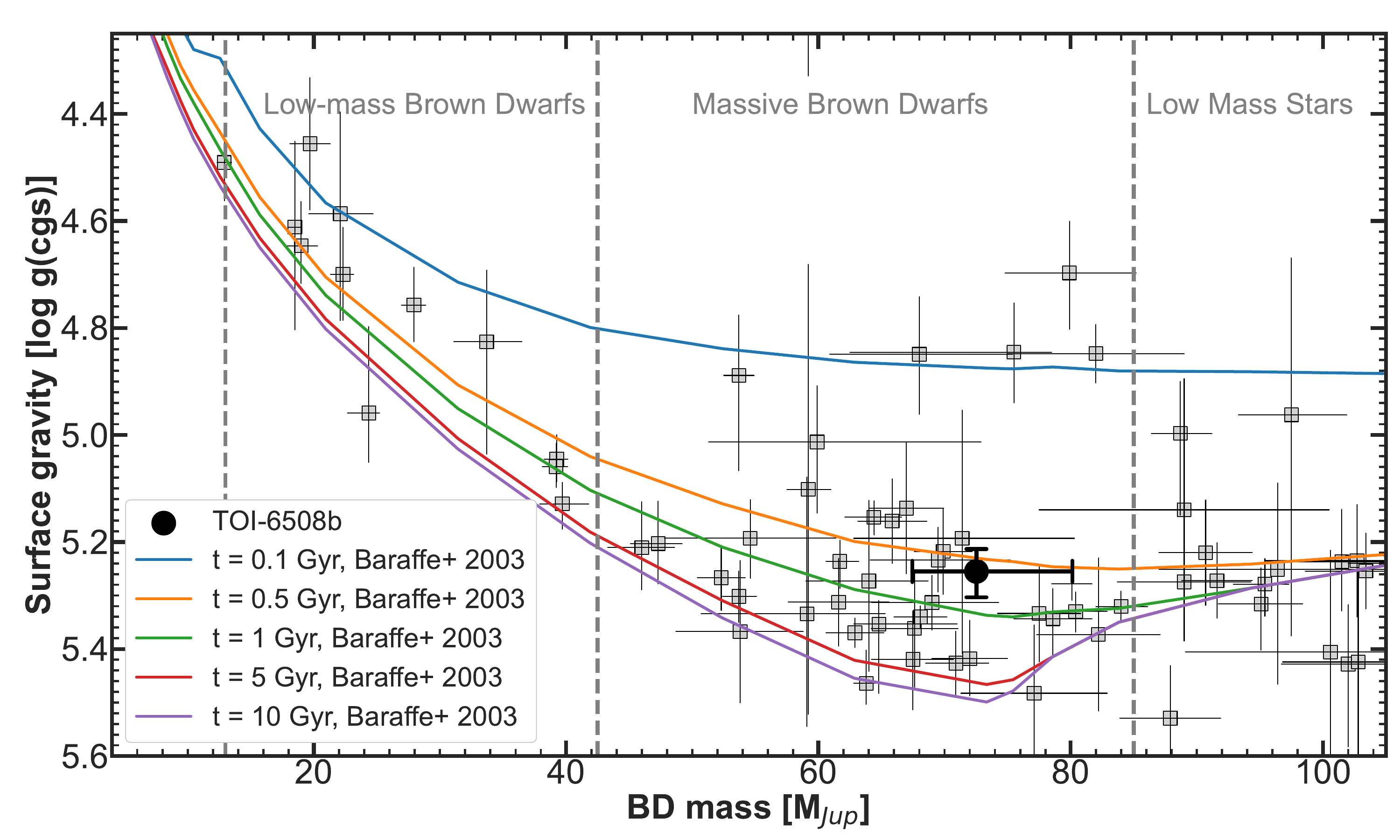}
	\caption{The surface gravity measurements $\log g_{\rm BD}$ as a function of the mass of transiting BDs from Table~\ref{tab:full_BDs_lists}. The colored solid lines indicate the evolutionary models from \cite{Baraffe_2003A&A} with different ages from 0.1 to 10 Gyrs. TOI-6508\,b is highlighted by the black dot with error bars. }
	\label{diagram:Mass_logg}
\end{figure*}

The surface gravity of the transiting BDs $g_{\rm BD}$, can be derived directly  from the transit observations and radial velocity measurements, and is given by,
\begin{equation}
    g_{\rm BD} = \frac{G M_{\rm BD}}{R^2_{\rm BD}} = \frac{2\pi K_{\rm RV}\sqrt{(1 - e^2)}}{P\sin(i)r^2_{\rm BD}},
\end{equation}
where, the semi-amplitude $K_{\rm RV}$ and the orbital eccentricity $e$ are derived from the radial velocity fit. The orbital period $P$, the scaled BD radius $r_{\rm BD} = a/R_{\rm BD}$, and the orbital inclination $i$ are derived from the fit of transit light curves.  The BD's surface gravity is related directly to the observable parameters independent of those of the host star.
\\
Figure~\ref{diagram:Mass_Radius} presents the radius--mass diagram for all known transiting objects with masses ranging 10 and 120~$M_{\rm Jup}$. Figure~\ref{diagram:Mass_logg} presents the surface gravity as a function of the mass of known transiting objects. Table~\ref{tab:full_BDs_lists} shows the updated list of transiting BDs from
\citet{Carmichael_2023} and \citet{Henderson_2024}. Some new objects have been included from \cite{Vowell_2025arXiv250109795V}. As a preliminary comparison, TOI-6508\,b is well placed within the edge of the brown dwarf regime, means, near the hydrogen burning limit \citep{Baraffe_2002A&A}. We also presented the tabulated isochrones models for substellar objects derived by \cite{Baraffe_2003A&A} (colored solid lines), with different ages of 0.1, 0.5, 1, 5 and 10 Gyr \footnote{Isochrones Models: \url{http://perso.ens-lyon.fr/isabelle.baraffe/}}. \\

\begin{figure}[!]
	\includegraphics[scale=0.23]{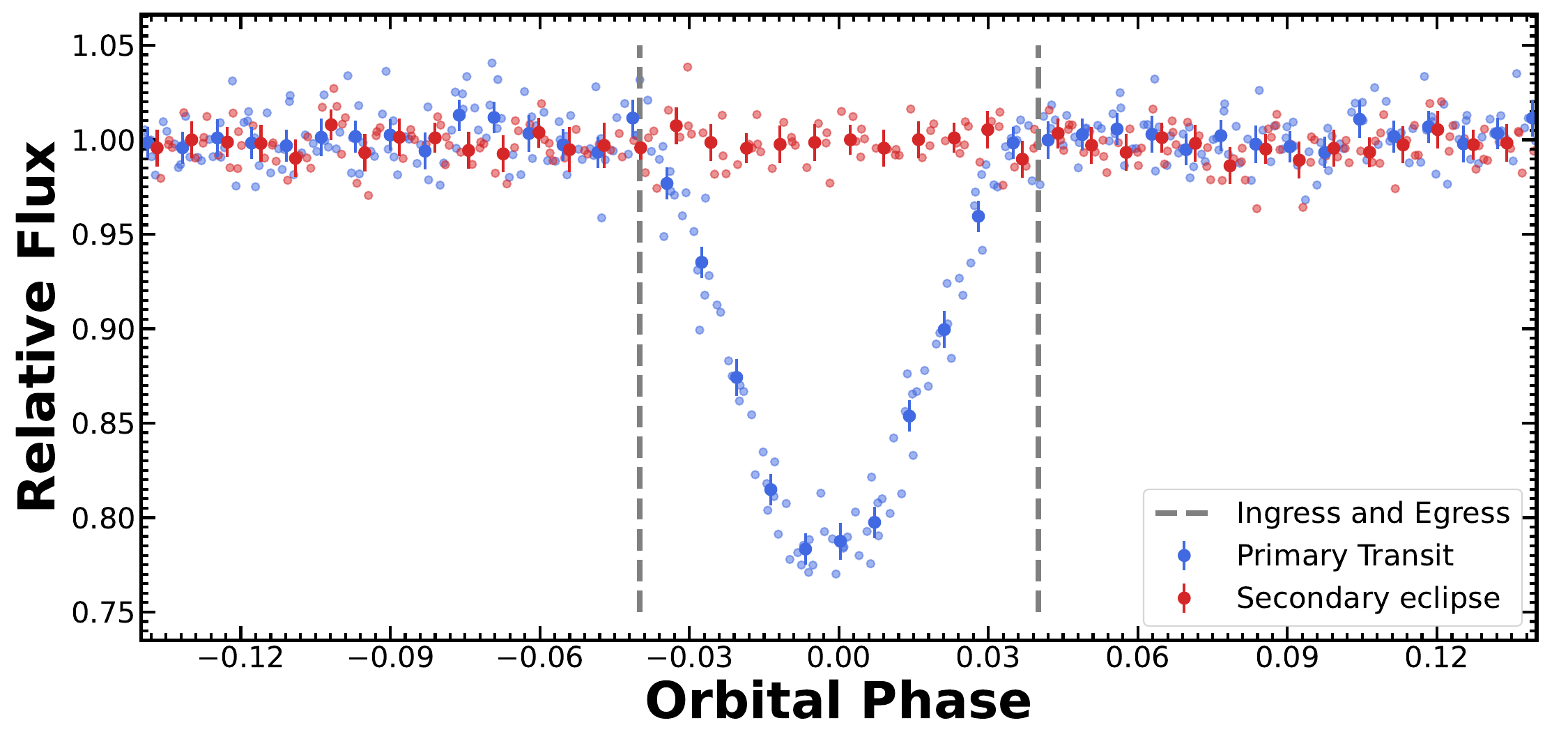}
	\caption{TESS PDC-SAP flux of TOI-6508 extracted from the full frame images (FFIs). The blue data points show the TESS folded-phased transit light curves of TOI-650b\,b. While, the red data points show the TESS folded-phased secondary eclipse light curves assuming an eccentricity of $e = 0.28$ (constrained from our global MCMC analysis). Based on the TESS data, we might rule out a secondary eclipse of $\delta_{\rm occult}\approx10$~ppt.}
	\label{tess_phased_folded_lcs}
\end{figure}

TOI-6508\,b shows a deeper primary eclipse of 250~ppt (parts per thousand), but no detectable secondary eclipse. This implies that the secondary component has a lower surface brightness than the primary component. Phase-folded TESS observations are shown in Figure~\ref{tess_phased_folded_lcs}.
The absence of a detectable secondary eclipse in the data suggested that the companion is a brown dwarf. Based on TESS data, we might rule out a secondary eclipse of $\delta_{\rm occult}\approx10$~ppt. We observed two full occultations of TOI-6508\,b from LCO-SAAO-1m0 in the Sloan-$i'$ on UTC May 16 2024 (assuming a circular orbit) and UTC Feb 8 2025 (assuming an eccentric orbit of $e=0.28$ constrained from our global MCMC analysis). Based on these observations, we might rule out a secondary eclipse of $\delta_{\rm occult}\approx3$~ppt (see Figure~\ref{fig:LCs_tess_gb_eclipse}).
Moreover, the effective temperature $T_{\rm BD}$ for the companion can be computed by combining the companion and stellar radius ratio $R_{\rm BD}/R_\star$ with the Planck function for blackbody via the formula: 
\begin{equation}
    \delta_{\rm occult}= \left(\frac{R_{\rm BD}}{R_\star}\right)^2 \frac{B_{\rm BD}(\lambda, T_{\rm BD})}{B_\star(\lambda, T_{\rm eff})},
\end{equation}
where $B_{\rm BD}(\lambda, T_{\rm BD})$ and $B_\star(\lambda, T_{\rm eff})$ are the Planck
distribution functions for the companion and host star, respectively. This resulted in an effective temperature of the companion of $T_{\rm BD} < 1800$K, indicative of a brown dwarf.
Since the luminosity of a BD is mainly emitted at infrared wavelengths, the secondary eclipse should be deeper when observed at infrared wavelengths. Moreover, a new observation in the infrared is required to confirm or not the secondary eclipse of the companion. If a secondary eclipse is present, this will allow an independent determination of the effective temperature of the brown dwarf TOI-6508\,b. 
Moreover, the combination of low mass and low luminosity of the host star, and low incident flux of the companion, make TOI-6508\,b a favorable target for upcoming secondary eclipse observation with the \emph{JWST}, in order to measure its luminosity, its Albedo, and its effective temperature.

\section{Acknowledgments}
The postdoctoral fellowship of KB is funded by F.R.S.-FNRS grant T.0109.20 and by the Francqui Foundation.
This publication benefits from the support of the French Community of Belgium in the context of the FRIA Doctoral Grant awarded to MT.
MG is F.R.S.-FNRS Research Director. 
Author F.J.P acknowledges financial support from the Severo Ochoa grant CEX2021-001131-S funded by MCIN/AEI/10.13039/501100011033 and 
Ministerio de Ciencia e Innovación through the project PID2022-137241NB-C43.
This material is based upon work supported by the National Aeronautics and Space Administration under Agreement No.\ 80NSSC21K0593 for the program ``Alien Earths''.
The results reported herein benefited from collaborations and/or information exchange within NASA’s Nexus for Exoplanet System Science (NExSS) research coordination network sponsored by NASA’s Science Mission Directorate.
Based on observations collected at the European Southern Observatory under ESO programme 113.27QV.001.
Visiting Astronomer at the Infrared Telescope Facility, which is operated by the University of Hawaii under contract 80HQTR24DA010 with the National Aeronautics and Space Administration.
Funding for the TESS mission is provided by NASA's Science Mission Directorate. KAC acknowledges support from the TESS mission via subaward s3449 from MIT.
This paper made use of data collected by the TESS mission, obtained from the Mikulski Archive for Space Telescopes MAST data archive at the Space Telescope Science Institute (STScI). Funding for the TESS mission is provided by the NASA Explorer Program. STScI is operated by the Association of Universities for Research in Astronomy, Inc., under NASA contract NAS 5–26555. We acknowledge the use of public TESS data from pipelines at the TESS Science Office and at the TESS Science Processing Operations Center.  Resources supporting this work were provided by the NASA High-End Computing (HEC) Program through the NASA Advanced Supercomputing (NAS) Division at Ames Research Center for the production of the SPOC data products.
This research has made use of the Exoplanet Follow-up Observation Program (ExoFOP; DOI: 10.26134/ExoFOP5) website, which is operated by the California Institute of Technology, under contract with the National Aeronautics and Space Administration under the Exoplanet Exploration Program.
Based on data collected by the SPECULOOS-South Observatory at the ESO Paranal Observatory in Chile. The ULiege's contribution to SPECULOOS has received funding from the European Research Council under the European Union's Seventh Framework Programme (FP/2007-2013) (grant Agreement n$^\circ$ 336480/SPECULOOS), from the Balzan Prize and Francqui Foundations, from the Belgian Scientific Research Foundation (F.R.S.-FNRS; grant n$^\circ$ T.0109.20), from the University of Liege, and from the ARC grant for Concerted Research Actions financed by the Wallonia-Brussels Federation. 
The Birmingham contribution is in part funded by the European Union's Horizon 2020 research and innovation programme (grant's agreement n$^{\circ}$ 803193/BEBOP), and from the Science and Technology Facilities Council (STFC; grant n$^\circ$ ST/S00193X/1, ST/W000385/1 and ST/Y001710/1).
The Cambridge contribution is supported by a grant from the Simons Foundation (PI Queloz, grant number 327127).
This work makes use of observations from the LCOGT network. Part of the LCOGT telescope time was granted by NOIRLab through the Mid-Scale Innovations Program (MSIP). MSIP is funded by NSF.
Some of the observations in this paper made use of the High-Resolution Imaging instrument Zorro and were obtained under Gemini LLP Proposal Number: GN/S-2021A-LP-105. Zorro was funded by the NASA Exoplanet Exploration Program and built at the NASA Ames Research Center by Steve B. Howell, Nic Scott, Elliott P. Horch, and Emmett Quigley. Zorro was mounted on the Gemini South telescope of the international Gemini Observatory, a program of NSF’s OIR Lab, which is managed by the Association of Universities for Research in Astronomy (AURA) under a cooperative agreement with the National Science Foundation. on behalf of the Gemini partnership: the National Science Foundation (United States), National Research Council (Canada), Agencia Nacional de Investigación y Desarrollo (Chile), Ministerio de Ciencia, Tecnología e Innovación (Argentina), Ministério da Ciência, Tecnologia, Inovações e Comunicações (Brazil), and Korea Astronomy and Space Science Institute (Republic of Korea).


\bibliographystyle{aa}
\bibliography{aa.bib}

\begin{thebibliography}{137}
\expandafter\ifx\csname natexlab\endcsname\relax\def\natexlab#1{#1}\fi

\bibitem[{Acton {et~al.}(2021)Acton, Goad, Burleigh, Casewell, Breytenbach, Nielsen, Smith, Anderson, Battley, Bayliss, Bouchy, Bryant, Csizmadia, Eigmüller, Gill, Gillen, Grieves, Günther, Henderson, Hodgkin, Jackman, Jenkins, Lendl, McCormac, Moyano, Nelson, Sefako, Smith, Stalport, Thomas, Tilbrook, Udry, West, Wheatley, Worters, Vines, \& Alves}]{19b}
Acton, J.~S., Goad, M.~R., Burleigh, M.~R., {et~al.} 2021, \mnras, 505, 2741–2752

\bibitem[{{Aller} {et~al.}(2020){Aller}, {Lillo-Box}, {Jones}, {Miranda}, \& {Barcel{\'o} Forteza}}]{Aller_2020AandA}
{Aller}, A., {Lillo-Box}, J., {Jones}, D., {Miranda}, L.~F., \& {Barcel{\'o} Forteza}, S. 2020, \aap, 635, A128

\bibitem[{Artigau {et~al.}(2024)Artigau, Bouchy, Doyon, Baron, Malo, Wildi, Pepe, Cook, Thibault, Reshetov, Dumusque, Lovis, Sosnowska, Martins, Medeiros, Delfosse, Santos, Rebolo, Abreu, Allain, Allart, Auger, Barros, Bazinet, Blind, Boisse, Bonfils, Bourrier, Bovay, Broeg, Brousseau, Bruniquel, Cabral, Cadieux, Carmona, Carteret, Challita, Chazelas, Cloutier, Coelho, Cointepas, Conod, Cowan, Cristo, da~Silva, Dauplaise, Gomes, Delgado-Mena, Ehrenreich, Faria, Figueira, Forveille, Frensch, Gagn{\'e}, Genest, Genolet, Hern{\'a}ndez, Temich, Grieves, Hernandez, Hobson, Hoeijmakers, Kerley, Krishnamurthy, Lafreni{\`e}re, Lamontagne, Larue, Leaf, Le{\~a}o, Lim, Curto, Martins, Melo, Messias, Mignon, Moranta, Mordasini, Moulla, Mounzer, L'Heureux, Nari, Nielsen, Osborn, Parc, Pasquini, Passegger, Pelletier, Peroux, Piaulet, Plotnykov, Poulin-Girard, Rasilla, Saint-Antoine, Sarajic, Segovia, Seidel, S{\'e}gransan, Silva, Srivastava, Stefanov, Mascare{\~n}o, Sordet, Teixeira, Udry, Valencia, Vall{\'e}e, Vandal,
  Vaulato, Wade, Wardenier, Wehb{\'e}, Weisserman, Wevers, \& Zins}]{NIRPS2024}
Artigau, {\'E}., Bouchy, F., Doyon, R., {et~al.} 2024, in Ground-based and Airborne Instrumentation for Astronomy X, ed. J.~J. Bryant, K.~Motohara, \& J.~R.~D. Vernet, Vol. 13096, International Society for Optics and Photonics (SPIE), 130960C

\bibitem[{{Artigau} {et~al.}(2021){Artigau}, {H{\'e}brard}, {Cadieux}, {Vandal}, {Cook}, {Doyon}, {Gagn{\'e}}, {Moutou}, {Martioli}, {Frasca}, {Jahandar}, {Lafreni{\`e}re}, {Malo}, {Donati}, {Cort{\'e}s-Zuleta}, {Boisse}, {Delfosse}, {Carmona}, {Fouqu{\'e}}, {Morin}, {Rowe}, {Marino}, {Papini}, {Ciardi}, {Lund}, {Martins}, {Pelletier}, {Arnold}, {Bouchy}, {Forveille}, {Santos}, {Bonfils}, {Figueira}, {Fausnaugh}, {Ricker}, {Latham}, {Seager}, {Winn}, {Jenkins}, {Ting}, {Torres}, \& {Gomes da Silva}}]{toi1278b}
{Artigau}, {\'E}., {H{\'e}brard}, G., {Cadieux}, C., {et~al.} 2021, \aj, 162, 144

\bibitem[{{Baraffe} {et~al.}(2002){Baraffe}, {Chabrier}, {Allard}, \& {Hauschildt}}]{Baraffe_2002A&A}
{Baraffe}, I., {Chabrier}, G., {Allard}, F., \& {Hauschildt}, P.~H. 2002, \aap, 382, 563

\bibitem[{{Baraffe} {et~al.}(2003){Baraffe}, {Chabrier}, {Barman}, {Allard}, \& {Hauschildt}}]{Baraffe_2003A&A}
{Baraffe}, I., {Chabrier}, G., {Barman}, T.~S., {Allard}, F., \& {Hauschildt}, P.~H. 2003, \aap, 402, 701

\bibitem[{{Barkaoui} {et~al.}(2024){Barkaoui}, {Schwarz}, {Narita}, {Mistry}, {Magliano}, {Hirano}, {Maity}, {Burgasser}, {Rackham}, {Murgas}, {Pozuelos}, {Stassun}, {Everett}, {Ciardi}, {Lamman}, {Pass}, {Bieryla}, {Aganze}, {Esparza-Borges}, {Collins}, {Covone}, {de Leon}, {D{\'e}vora-Pajares}, {de Wit}, {Fukuda}, {Fukui}, {Gerasimov}, {Gillon}, {Hayashi}, {Howell}, {Ikoma}, {Ikuta}, {Jenkins}, {Karpoor}, {Kawai}, {Kimura}, {Kotani}, {Latham}, {Mori}, {Pall{\'e}}, {Parviainen}, {Patel}, {Ricker}, {Relles}, {Shporer}, {Seager}, {Softich}, {Srdoc}, {Tamura}, {Theissen}, {Twicken}, {Vanderspek}, {Watanabe}, {Watkins}, {Winn}, \& {Wohler}}]{Barkaoui_2024A&A}
{Barkaoui}, K., {Schwarz}, R.~P., {Narita}, N., {et~al.} 2024, \aap, 687, A264

\bibitem[{{Barkaoui} {et~al.}(2023){Barkaoui}, {Timmermans}, {Soubkiou}, {Rackham}, {Burgasser}, {Chouqar}, {Pozuelos}, {Collins}, {Howell}, {Simcoe}, {Melis}, {Stassun}, {Tregloan-Reed}, {Cointepas}, {Gillon}, {Bonfils}, {Furlan}, {Gnilka}, {Almenara}, {Alonso}, {Benkhaldoun}, {Bonavita}, {Bouchy}, {Burdanov}, {Chinchilla}, {Davoudi}, {Delrez}, {Demangeon}, {Dominik}, {Demory}, {de Wit}, {Dransfield}, {Ducrot}, {Fukui}, {Hinse}, {Hooton}, {Jehin}, {Jenkins}, {J{\o}rgensen}, {Latham}, {Garcia}, {Carrazco-Gaxiola}, {Ghachoui}, {G{\'o}mez Maqueo Chew}, {G{\"u}nther}, {McCormac}, {Murgas}, {Murray}, {Narita}, {Niraula}, {Pedersen}, {Queloz}, {Rebolo-L{\'o}pez}, {Ricker}, {Sabin}, {Sajadian}, {Schanche}, {Schwarz}, {Seager}, {Sebastian}, {Sefako}, {Sohy}, {Southworth}, {Srdoc}, {Thompson}, {Triaud}, {Vanderspek}, {Wells}, {Winn}, \& {Z{\'u}{\~n}iga-Fern{\'a}ndez}}]{Barkaoui2023A&A}
{Barkaoui}, K., {Timmermans}, M., {Soubkiou}, A., {et~al.} 2023, \aap, 677, A38

\bibitem[{{Bayliss} {et~al.}(2017){Bayliss}, {Hojjatpanah}, {Santerne}, {Dragomir}, {Zhou}, {Shporer}, {Col{\'o}n}, {Almenara}, {Armstrong}, {Barrado}, {Barros}, {Bento}, {Boisse}, {Bouchy}, {Brown}, {Brown}, {Cameron}, {Cochran}, {Demangeon}, {Deleuil}, {D{\'\i}az}, {Fulton}, {Horne}, {H{\'e}brard}, {Lillo-Box}, {Lovis}, {Mawet}, {Ngo}, {Osborn}, {Palle}, {Petigura}, {Pollacco}, {Santos}, {Sefako}, {Siverd}, {Sousa}, \& {Tsantaki}}]{epic201702477b}
{Bayliss}, D., {Hojjatpanah}, S., {Santerne}, A., {et~al.} 2017, \aj, 153, 15

\bibitem[{{Benni} {et~al.}(2021){Benni}, {Burdanov}, {Krushinsky}, {Bonfanti}, {H{\'e}brard}, {Almenara}, {Dalal}, {Demangeon}, {Tsantaki}, {Pepper}, {Stassun}, {Vanderburg}, {Belinski}, {Kashaev}, {Barkaoui}, {Kim}, {Kang}, {Antonyuk}, {Dyachenko}, {Rastegaev}, {Beskakotov}, {Mitrofanova}, {Pozuelos}, {Kuznetsov}, {Popov}, {Kiefer}, {Wilson}, {Ricker}, {Vanderspek}, {Latham}, {Seager}, {Jenkins}, {Sokov}, {Sokova}, {Marchini}, {Papini}, {Salvaggio}, {Banfi}, {Ba{\c{s}}t{\"u}rk}, {Torun}, {Yal{\c{c}}{\i}nkaya}, {Ivanov}, {Valyavin}, {Jehin}, {Gillon}, {Pak{\v{s}}tien{\.{e}}}, {Hentunen}, {Shadick}, {Bretton}, {W{\"u}nsche}, {Garlitz}, {Jongen}, {Molina}, {Girardin}, {Grau Horta}, {Naves}, {Benkhaldoun}, {Joner}, {Spencer}, {Bieryla}, {Stevens}, {Jensen}, {Collins}, {Charbonneau}, {Quintana}, {Mullally}, \& {Henze}}]{gpx1b}
{Benni}, P., {Burdanov}, A.~Y., {Krushinsky}, V.~V., {et~al.} 2021, \mnras, 505, 4956

\bibitem[{{Bochanski} {et~al.}(2007){Bochanski}, {West}, {Hawley}, \& {Covey}}]{2007AJ....133..531B}
{Bochanski}, J.~J., {West}, A.~A., {Hawley}, S.~L., \& {Covey}, K.~R. 2007, \aj, 133, 531

\bibitem[{{Bonomo} {et~al.}(2015){Bonomo}, {Sozzetti}, {Santerne}, {Deleuil}, {Almenara}, {Bruno}, {D{\'\i}az}, {H{\'e}brard}, \& {Moutou}}]{kepler39b}
{Bonomo}, A.~S., {Sozzetti}, A., {Santerne}, A., {et~al.} 2015, \aap, 575, A85

\bibitem[{{Bouchy} {et~al.}(2011){Bouchy}, {Deleuil}, {Guillot}, {Aigrain}, {Carone}, {Cochran}, {Almenara}, {Alonso}, {Auvergne}, {Baglin}, {Barge}, {Bonomo}, {Bord{\'e}}, {Csizmadia}, {de Bondt}, {Deeg}, {D{\'\i}az}, {Dvorak}, {Endl}, {Erikson}, {Ferraz-Mello}, {Fridlund}, {Gandolfi}, {Gazzano}, {Gibson}, {Gillon}, {Guenther}, {Hatzes}, {Havel}, {H{\'e}brard}, {Jorda}, {L{\'e}ger}, {Lovis}, {Llebaria}, {Lammer}, {MacQueen}, {Mazeh}, {Moutou}, {Ofir}, {Ollivier}, {Parviainen}, {P{\"a}tzold}, {Queloz}, {Rauer}, {Rouan}, {Santerne}, {Schneider}, {Tingley}, \& {Wuchterl}}]{corot15b}
{Bouchy}, F., {Deleuil}, M., {Guillot}, T., {et~al.} 2011, \aap, 525, A68

\bibitem[{{Bouchy} {et~al.}(2017){Bouchy}, {Doyon}, {Artigau}, {Melo}, {Hernandez}, {Wildi}, {Delfosse}, {Lovis}, {Figueira}, {Canto Martins}, {Gonz{\'a}lez Hern{\'a}ndez}, {Thibault}, {Reshetov}, {Pepe}, {Santos}, {de Medeiros}, {Rebolo}, {Abreu}, {Adibekyan}, {Bandy}, {Benz}, {Blind}, {Bohlender}, {Boisse}, {Bovay}, {Broeg}, {Brousseau}, {Cabral}, {Chazelas}, {Cloutier}, {Coelho}, {Conod}, {Cumming}, {Delabre}, {Genolet}, {Hagelberg}, {Jayawardhana}, {K{\"a}ufl}, {Lafreni{\`e}re}, {de Castro Le{\~a}o}, {Malo}, {de Medeiros Martins}, {Matthews}, {Metchev}, {Oshagh}, {Ouellet}, {Parro}, {Rasilla Pi{\~n}eiro}, {Santos}, {Sarajlic}, {Segovia}, {Sordet}, {Udry}, {Valencia}, {Vall{\'e}e}, {Venn}, {Wade}, \& {Saddlemyer}}]{Bouchy_NIRPS_2017Msngr}
{Bouchy}, F., {Doyon}, R., {Artigau}, {\'E}., {et~al.} 2017, The Messenger, 169, 21

\bibitem[{Brown {et~al.}(2013)Brown, Baliber, Bianco, Bowman, Burleson, Conway, Crellin, Depagne, Vera, Dilday, Dragomir, Dubberley, Eastman, Elphick, Falarski, Foale, Ford, Fulton, Garza, Gomez, Graham, Greene, Haldeman, Hawkins, Haworth, Haynes, Hidas, Hjelstrom, Howell, Hygelund, Lister, Lobdill, Martinez, Mullins, Norbury, Parrent, Paulson, Petry, Pickles, Posner, Rosing, Ross, Sand, Saunders, Shobbrook, Shporer, Street, Thomas, Tsapras, Tufts, Valenti, Horst, Walker, White, \& Willis}]{Brown_2013}
Brown, T.~M., Baliber, N., Bianco, F.~B., {et~al.} 2013, Publications of the Astronomical Society of the Pacific, 125, 1031

\bibitem[{{Burgasser} \& {Splat Development Team}(2017)}]{splat}
{Burgasser}, A.~J. \& {Splat Development Team}. 2017, in Astronomical Society of India Conference Series, Vol.~14, Astronomical Society of India Conference Series, 7--12

\bibitem[{{Burn} {et~al.}(2021){Burn}, {Schlecker}, {Mordasini}, {Emsenhuber}, {Alibert}, {Henning}, {Klahr}, \& {Benz}}]{Burn_2021AA}
{Burn}, R., {Schlecker}, M., {Mordasini}, C., {et~al.} 2021, \aap, 656, A72

\bibitem[{{Ca{\~n}as} {et~al.}(2018){Ca{\~n}as}, {Bender}, {Mahadevan}, {Fleming}, {Beatty}, {Covey}, {De Lee}, {Hearty}, {Garc{\'\i}a-Hern{\'a}ndez}, {Majewski}, {Schneider}, {Stassun}, \& {Wilson}}]{kepler503b}
{Ca{\~n}as}, C.~I., {Bender}, C.~F., {Mahadevan}, S., {et~al.} 2018, \apjl, 861, L4

\bibitem[{Carmichael(2022)}]{Carmichael_2023}
Carmichael, T.~W. 2022, Monthly Notices of the Royal Astronomical Society, 519, 5177

\bibitem[{{Carmichael} {et~al.}(2022){Carmichael}, {Irwin}, {Murgas}, {Pall{\'e}}, {Stassun}, {Bartnik}, {Collins}, {de Leon}, {Esparza-Borges}, {Fedewa}, {Fong}, {Fukui}, {Jenkins}, {Kagetani}, {Latham}, {Lund}, {Mann}, {Moldovan}, {Morgan}, {Narita}, {Painter}, {Parviainen}, {Quintana}, {Ricker}, {Schulte}, {Schwarz}, {Seager}, {Sokolovsky}, {Twicken}, \& {Winn}}]{carmichael22}
{Carmichael}, T.~W., {Irwin}, J.~M., {Murgas}, F., {et~al.} 2022, \mnras, 514, 4944

\bibitem[{{Carmichael} {et~al.}(2019){Carmichael}, {Latham}, \& {Vanderburg}}]{carmichael19}
{Carmichael}, T.~W., {Latham}, D.~W., \& {Vanderburg}, A.~M. 2019, \aj, 158, 38

\bibitem[{{Carmichael} {et~al.}(2020){Carmichael}, {Quinn}, {Mustill}, {Huang}, {Zhou}, {Persson}, {Nielsen}, {Collins}, {Ziegler}, {Collins}, {Rodriguez}, {Shporer}, {Brahm}, {Mann}, {Bouchy}, {Fridlund}, {Stassun}, {Hellier}, {Seidel}, {Stalport}, {Udry}, {Pepe}, {Ireland}, {{\v{Z}}erjal}, {Brice{\~n}o}, {Law}, {Jord{\'a}n}, {Espinoza}, {Henning}, {Sarkis}, \& {Latham}}]{carmichael20}
{Carmichael}, T.~W., {Quinn}, S.~N., {Mustill}, A.~J., {et~al.} 2020, \aj, 160, 53

\bibitem[{Carmichael {et~al.}(2021)Carmichael, Quinn, Zhou, Grieves, Irwin, Stassun, Vanderburg, Winn, Bouchy, Brasseur, Briceño, Caldwell, Charbonneau, Collins, Colon, Eastman, Fausnaugh, Fong, Fűrész, Huang, Jenkins, Kielkopf, Latham, Law, Lund, Mann, Ricker, Rodriguez, Schwarz, Shporer, Tenenbaum, Wood, \& Ziegler}]{carmichael21}
Carmichael, T.~W., Quinn, S.~N., Zhou, G., {et~al.} 2021, \aj, 161, 97

\bibitem[{{Chabrier} {et~al.}(2023){Chabrier}, {Baraffe}, {Phillips}, \& {Debras}}]{Chabrier_2023A&A}
{Chabrier}, G., {Baraffe}, I., {Phillips}, M., \& {Debras}, F. 2023, \aap, 671, A119

\bibitem[{{Chaturvedi} {et~al.}(2016){Chaturvedi}, {Chakraborty}, {Anandarao}, {Roy}, \& {Mahadevan}}]{j2343}
{Chaturvedi}, P., {Chakraborty}, A., {Anandarao}, B.~G., {Roy}, A., \& {Mahadevan}, S. 2016, \mnras, 462, 554

\bibitem[{Ciardi {et~al.}(2015)Ciardi, Beichman, Horch, \& Howell}]{Ciardi_2015}
Ciardi, D.~R., Beichman, C.~A., Horch, E.~P., \& Howell, S.~B. 2015, The Astrophysical Journal, 805, 16

\bibitem[{{Claret}(2018)}]{Claret_2018AandA}
{Claret}, A. 2018, \aap, 618, A20

\bibitem[{{Claret} {et~al.}(2012){Claret}, {Hauschildt}, \& {Witte}}]{Claret_2012AA}
{Claret}, A., {Hauschildt}, P.~H., \& {Witte}, S. 2012, \aap, 546, A14

\bibitem[{Collins {et~al.}(2017)Collins, Kielkopf, Stassun, \& Hessman}]{Collins_2017}
Collins, K.~A., Kielkopf, J.~F., Stassun, K.~G., \& Hessman, F.~V. 2017, The Astronomical Journal, 153, 77

\bibitem[{Csizmadia(2016)}]{NLTTrad}
Csizmadia, S. 2016, in The {CoRoT} Legacy Book ({EDP} Sciences), 143

\bibitem[{{Csizmadia} {et~al.}(2015){Csizmadia}, {Hatzes}, {Gandolfi}, {Deleuil}, {Bouchy}, {Fridlund}, {Szabados}, {Parviainen}, {Cabrera}, {Aigrain}, {Alonso}, {Almenara}, {Baglin}, {Bord{\'e}}, {Bonomo}, {Deeg}, {D{\'\i}az}, {Erikson}, {Ferraz-Mello}, {Tadeu dos Santos}, {Guenther}, {Guillot}, {Grziwa}, {H{\'e}brard}, {Klagyivik}, {Ollivier}, {P{\"a}tzold}, {Rauer}, {Rouan}, {Santerne}, {Schneider}, {Mazeh}, {Wuchterl}, {Carpano}, \& {Ofir}}]{corot33b}
{Csizmadia}, S., {Hatzes}, A., {Gandolfi}, D., {et~al.} 2015, \aap, 584, A13

\bibitem[{{Cushing} {et~al.}(2005){Cushing}, {Rayner}, \& {Vacca}}]{Cushing2005}
{Cushing}, M.~C., {Rayner}, J.~T., \& {Vacca}, W.~D. 2005, \apj, 623, 1115

\bibitem[{{Cushing} {et~al.}(2004){Cushing}, {Vacca}, \& {Rayner}}]{Cushing2004}
{Cushing}, M.~C., {Vacca}, W.~D., \& {Rayner}, J.~T. 2004, \pasp, 116, 362

\bibitem[{{Cutri} {et~al.}(2021){Cutri}, {Wright}, {Conrow}, {Fowler}, {Eisenhardt}, {Grillmair}, {Kirkpatrick}, {Masci}, {McCallon}, {Wheelock}, {Fajardo-Acosta}, {Yan}, {Benford}, {Harbut}, {Jarrett}, {Lake}, {Leisawitz}, {Ressler}, {Stanford}, {Tsai}, {Liu}, {Helou}, {Mainzer}, {Gettngs}, {Gonzalez}, {Hoffman}, {Marsh}, {Padgett}, {Skrutskie}, {Beck}, {Papin}, \& {Wittman}}]{Cutri_2014yCat.2328}
{Cutri}, R.~M., {Wright}, E.~L., {Conrow}, T., {et~al.} 2021, VizieR Online Data Catalog, II/328

\bibitem[{{David} {et~al.}(2019){David}, {Hillenbrand}, {Gillen}, {Cody}, {Howell}, {Isaacson}, \& {Livingston}}]{rik72b}
{David}, T.~J., {Hillenbrand}, L.~A., {Gillen}, E., {et~al.} 2019, \apj, 872, 161

\bibitem[{{Deleuil} {et~al.}(2008){Deleuil}, {Deeg}, {Alonso}, {Bouchy}, {Rouan}, {Auvergne}, {Baglin}, {Aigrain}, {Almenara}, {Barbieri}, {Barge}, {Bruntt}, {Bord{\'e}}, {Collier Cameron}, {Csizmadia}, {de La Reza}, {Dvorak}, {Erikson}, {Fridlund}, {Gandolfi}, {Gillon}, {Guenther}, {Guillot}, {Hatzes}, {H{\'e}brard}, {Jorda}, {Lammer}, {L{\'e}ger}, {Llebaria}, {Loeillet}, {Mayor}, {Mazeh}, {Moutou}, {Ollivier}, {P{\"a}tzold}, {Pont}, {Queloz}, {Rauer}, {Schneider}, {Shporer}, {Wuchterl}, \& {Zucker}}]{corot3b}
{Deleuil}, M., {Deeg}, H.~J., {Alonso}, R., {et~al.} 2008, \aap, 491, 889

\bibitem[{{Delrez} {et~al.}(2018){Delrez}, {Gillon}, {Queloz}, {Demory}, {Almleaky}, {de Wit}, {Jehin}, {Triaud}, {Barkaoui}, {Burdanov}, {Burgasser}, {Ducrot}, {McCormac}, {Murray}, {Silva Fernandes}, {Sohy}, {Thompson}, {Van Grootel}, {Alonso}, {Benkhaldoun}, \& {Rebolo}}]{Delrez2018}
{Delrez}, L., {Gillon}, M., {Queloz}, D., {et~al.} 2018, in Society of Photo-Optical Instrumentation Engineers (SPIE) Conference Series, Vol. 10700, Ground-based and Airborne Telescopes VII, ed. H.~K. {Marshall} \& J.~{Spyromilio}, 107001I

\bibitem[{{Delrez} {et~al.}(2022){Delrez}, {Murray}, {Pozuelos}, {Narita}, {Ducrot}, {Timmermans}, {Watanabe}, {Burgasser}, {Hirano}, {Rackham}, {Stassun}, {Van Grootel}, {Aganze}, {Cointepas}, {Howell}, {Kaltenegger}, {Niraula}, {Sebastian}, {Almenara}, {Barkaoui}, {Baycroft}, {Bonfils}, {Bouchy}, {Burdanov}, {Caldwell}, {Charbonneau}, {Ciardi}, {Collins}, {Daylan}, {Demory}, {de Wit}, {Dransfield}, {Fajardo-Acosta}, {Fausnaugh}, {Fukui}, {Furlan}, {Garcia}, {Gnilka}, {G{\'o}mez Maqueo Chew}, {G{\'o}mez-Mu{\~n}oz}, {G{\"u}nther}, {Harakawa}, {Heng}, {Hooton}, {Hori}, {Ikoma}, {Jehin}, {Jenkins}, {Kagetani}, {Kawauchi}, {Kimura}, {Kodama}, {Kotani}, {Krishnamurthy}, {Kudo}, {Kunovac}, {Kusakabe}, {Latham}, {Littlefield}, {McCormac}, {Melis}, {Mori}, {Murgas}, {Palle}, {Pedersen}, {Queloz}, {Ricker}, {Sabin}, {Schanche}, {Schroffenegger}, {Seager}, {Shiao}, {Sohy}, {Standing}, {Tamura}, {Theissen}, {Thompson}, {Triaud}, {Vanderspek}, {Vievard}, {Wells}, {Winn}, {Zou}, {Z{\'u}{\~n}iga-Fern{\'a}ndez}, \&
  {Gillon}}]{Delrez2022}
{Delrez}, L., {Murray}, C.~A., {Pozuelos}, F.~J., {et~al.} 2022, \aap, 667, A59

\bibitem[{{D{\'\i}az} {et~al.}(2013){D{\'\i}az}, {Damiani}, {Deleuil}, {Almenara}, {Moutou}, {Barros}, {Bonomo}, {Bouchy}, {Bruno}, {H{\'e}brard}, {Montagnier}, \& {Santerne}}]{koi205b}
{D{\'\i}az}, R.~F., {Damiani}, C., {Deleuil}, M., {et~al.} 2013, \aap, 551, L9

\bibitem[{{D{\'\i}az} {et~al.}(2014){D{\'\i}az}, {Montagnier}, {Leconte}, {Bonomo}, {Deleuil}, {Almenara}, {Barros}, {Bouchy}, {Bruno}, {Damiani}, {H{\'e}brard}, {Moutou}, \& {Santerne}}]{koi189b}
{D{\'\i}az}, R.~F., {Montagnier}, G., {Leconte}, J., {et~al.} 2014, \aap, 572, A109

\bibitem[{{Douglas} {et~al.}(2014){Douglas}, {Ag{\"u}eros}, {Covey}, {Bowsher}, {Bochanski}, {Cargile}, {Kraus}, {Law}, {Lemonias}, {Arce}, {Fierroz}, \& {Kundert}}]{2014ApJ...795..161D}
{Douglas}, S.~T., {Ag{\"u}eros}, M.~A., {Covey}, K.~R., {et~al.} 2014, \apj, 795, 161

\bibitem[{{El-Badry} {et~al.}(2023){El-Badry}, {Burdge}, {van Roestel}, \& {Rodriguez}}]{ZTFJ2020}
{El-Badry}, K., {Burdge}, K.~B., {van Roestel}, J., \& {Rodriguez}, A.~C. 2023, The Open Journal of Astrophysics, 6, 33

\bibitem[{{Ferreira dos Santos} {et~al.}(2024){Ferreira dos Santos}, {Rice}, {Wang}, \& {Wang}}]{toi2533b}
{Ferreira dos Santos}, T., {Rice}, M., {Wang}, X.-Y., \& {Wang}, S. 2024, \aj, 168, 145

\bibitem[{{Furlan} \& {Howell}(2017)}]{Furlan_2017AJ}
{Furlan}, E. \& {Howell}, S.~B. 2017, \aj, 154, 66

\bibitem[{Furlan \& Howell(2020)}]{Furlan_2020}
Furlan, E. \& Howell, S.~B. 2020, The Astrophysical Journal, 898, 47

\bibitem[{{Gaia Collaboration} {et~al.}(2021){Gaia Collaboration}, {Brown}, {Vallenari}, {Prusti}, {de Bruijne}, {Babusiaux}, {Biermann}, {Creevey}, {Evans}, {Eyer}, {Hutton}, {Jansen}, {Jordi}, {Klioner}, {Lammers}, {Lindegren}, {Luri}, {Mignard}, {Panem}, {Pourbaix}, {Randich}, {Sartoretti}, {Soubiran}, {Walton}, {Arenou}, {Bailer-Jones}, {Bastian}, {Cropper}, {Drimmel}, {Katz}, {Lattanzi}, {van Leeuwen}, {Bakker}, {Cacciari}, {Casta{\~n}eda}, {De Angeli}, {Ducourant}, {Fabricius}, {Fouesneau}, {Fr{\'e}mat}, {Guerra}, {Guerrier}, {Guiraud}, {Jean-Antoine Piccolo}, {Masana}, {Messineo}, {Mowlavi}, {Nicolas}, {Nienartowicz}, {Pailler}, {Panuzzo}, {Riclet}, {Roux}, {Seabroke}, {Sordo}, {Tanga}, {Th{\'e}venin}, {Gracia-Abril}, {Portell}, {Teyssier}, {Altmann}, {Andrae}, {Bellas-Velidis}, {Benson}, {Berthier}, {Blomme}, {Brugaletta}, {Burgess}, {Busso}, {Carry}, {Cellino}, {Cheek}, {Clementini}, {Damerdji}, {Davidson}, {Delchambre}, {Dell'Oro}, {Fern{\'a}ndez-Hern{\'a}ndez}, {Galluccio}, {Garc{\'\i}a-Lario},
  {Garcia-Reinaldos}, {Gonz{\'a}lez-N{\'u}{\~n}ez}, {Gosset}, {Haigron}, {Halbwachs}, {Hambly}, {Harrison}, {Hatzidimitriou}, {Heiter}, {Hern{\'a}ndez}, {Hestroffer}, {Hodgkin}, {Holl}, {Jan{\ss}en}, {Jevardat de Fombelle}, {Jordan}, {Krone-Martins}, {Lanzafame}, {L{\"o}ffler}, {Lorca}, {Manteiga}, {Marchal}, {Marrese}, {Moitinho}, {Mora}, {Muinonen}, {Osborne}, {Pancino}, {Pauwels}, {Petit}, {Recio-Blanco}, {Richards}, {Riello}, {Rimoldini}, {Robin}, {Roegiers}, {Rybizki}, {Sarro}, {Siopis}, {Smith}, {Sozzetti}, {Ulla}, {Utrilla}, {van Leeuwen}, {van Reeven}, {Abbas}, {Abreu Aramburu}, {Accart}, {Aerts}, {Aguado}, {Ajaj}, {Altavilla}, {{\'A}lvarez}, {{\'A}lvarez Cid-Fuentes}, {Alves}, {Anderson}, {Anglada Varela}, {Antoja}, {Audard}, {Baines}, {Baker}, {Balaguer-N{\'u}{\~n}ez}, {Balbinot}, {Balog}, {Barache}, {Barbato}, {Barros}, {Barstow}, {Bartolom{\'e}}, {Bassilana}, {Bauchet}, {Baudesson-Stella}, {Becciani}, {Bellazzini}, {Bernet}, {Bertone}, {Bianchi}, {Blanco-Cuaresma}, {Boch}, {Bombrun}, {Bossini},
  {Bouquillon}, {Bragaglia}, {Bramante}, {Breedt}, {Bressan}, {Brouillet}, {Bucciarelli}, {Burlacu}, {Busonero}, {Butkevich}, {Buzzi}, {Caffau}, {Cancelliere}, {C{\'a}novas}, {Cantat-Gaudin}, {Carballo}, {Carlucci}, {Carnerero}, {Carrasco}, {Casamiquela}, {Castellani}, {Castro-Ginard}, {Castro Sampol}, {Chaoul}, {Charlot}, {Chemin}, {Chiavassa}, {Cioni}, {Comoretto}, {Cooper}, {Cornez}, {Cowell}, {Crifo}, {Crosta}, {Crowley}, {Dafonte}, {Dapergolas}, {David}, {David}, {de Laverny}, {De Luise}, {De March}, {De Ridder}, {de Souza}, {de Teodoro}, {de Torres}, {del Peloso}, {del Pozo}, {Delbo}, {Delgado}, {Delgado}, {Delisle}, {Di Matteo}, {Diakite}, {Diener}, {Distefano}, {Dolding}, {Eappachen}, {Edvardsson}, {Enke}, {Esquej}, {Fabre}, {Fabrizio}, {Faigler}, {Fedorets}, {Fernique}, {Fienga}, {Figueras}, {Fouron}, {Fragkoudi}, {Fraile}, {Franke}, {Gai}, {Garabato}, {Garcia-Gutierrez}, {Garc{\'\i}a-Torres}, {Garofalo}, {Gavras}, {Gerlach}, {Geyer}, {Giacobbe}, {Gilmore}, {Girona}, {Giuffrida}, {Gomel}, {Gomez},
  {Gonzalez-Santamaria}, {Gonz{\'a}lez-Vidal}, {Granvik}, {Guti{\'e}rrez-S{\'a}nchez}, {Guy}, {Hauser}, {Haywood}, {Helmi}, {Hidalgo}, {Hilger}, {H{\l}adczuk}, {Hobbs}, {Holland}, {Huckle}, {Jasniewicz}, {Jonker}, {Juaristi Campillo}, {Julbe}, {Karbevska}, {Kervella}, {Khanna}, {Kochoska}, {Kontizas}, {Kordopatis}, {Korn}, {Kostrzewa-Rutkowska}, {Kruszy{\'n}ska}, {Lambert}, {Lanza}, {Lasne}, {Le Campion}, {Le Fustec}, {Lebreton}, {Lebzelter}, {Leccia}, {Leclerc}, {Lecoeur-Taibi}, {Liao}, {Licata}, {Lindstr{\o}m}, {Lister}, {Livanou}, {Lobel}, {Madrero Pardo}, {Managau}, {Mann}, {Marchant}, {Marconi}, {Marcos Santos}, {Marinoni}, {Marocco}, {Marshall}, {Martin Polo}, {Mart{\'\i}n-Fleitas}, {Masip}, {Massari}, {Mastrobuono-Battisti}, {Mazeh}, {McMillan}, {Messina}, {Michalik}, {Millar}, {Mints}, {Molina}, {Molinaro}, {Moln{\'a}r}, {Montegriffo}, {Mor}, {Morbidelli}, {Morel}, {Morris}, {Mulone}, {Munoz}, {Muraveva}, {Murphy}, {Musella}, {Noval}, {Ord{\'e}novic}, {Orr{\`u}}, {Osinde}, {Pagani}, {Pagano},
  {Palaversa}, {Palicio}, {Panahi}, {Pawlak}, {Pe{\~n}alosa Esteller}, {Penttil{\"a}}, {Piersimoni}, {Pineau}, {Plachy}, {Plum}, {Poggio}, {Poretti}, {Poujoulet}, {Pr{\v{s}}a}, {Pulone}, {Racero}, {Ragaini}, {Rainer}, {Raiteri}, {Rambaux}, {Ramos}, {Ramos-Lerate}, {Re Fiorentin}, {Regibo}, {Reyl{\'e}}, {Ripepi}, {Riva}, {Rixon}, {Robichon}, {Robin}, {Roelens}, {Rohrbasser}, {Romero-G{\'o}mez}, {Rowell}, {Royer}, {Rybicki}, {Sadowski}, {Sagrist{\`a} Sell{\'e}s}, {Sahlmann}, {Salgado}, {Salguero}, {Samaras}, {Sanchez Gimenez}, {Sanna}, {Santove{\~n}a}, {Sarasso}, {Schultheis}, {Sciacca}, {Segol}, {Segovia}, {S{\'e}gransan}, {Semeux}, {Shahaf}, {Siddiqui}, {Siebert}, {Siltala}, {Slezak}, {Smart}, {Solano}, {Solitro}, {Souami}, {Souchay}, {Spagna}, {Spoto}, {Steele}, {Steidelm{\"u}ller}, {Stephenson}, {S{\"u}veges}, {Szabados}, {Szegedi-Elek}, {Taris}, {Tauran}, {Taylor}, {Teixeira}, {Thuillot}, {Tonello}, {Torra}, {Torra}, {Turon}, {Unger}, {Vaillant}, {van Dillen}, {Vanel}, {Vecchiato}, {Viala}, {Vicente},
  {Voutsinas}, {Weiler}, {Wevers}, {Wyrzykowski}, {Yoldas}, {Yvard}, {Zhao}, {Zorec}, {Zucker}, {Zurbach}, \& {Zwitter}}]{Gaia_Collaboration_2021AandA}
{Gaia Collaboration}, {Brown}, A.~G.~A., {Vallenari}, A., {et~al.} 2021, \aap, 650, C3

\bibitem[{{Garcia} {et~al.}(2022){Garcia}, {Timmermans}, {Pozuelos}, {Ducrot}, {Gillon}, {Delrez}, {Wells}, \& {Jehin}}]{prose}
{Garcia}, L.~J., {Timmermans}, M., {Pozuelos}, F.~J., {et~al.} 2022, \mnras, 509, 4817

\bibitem[{{Gelman} \& {Rubin}(1992)}]{Gelman1992}
{Gelman}, A. \& {Rubin}, D.~B. 1992, Statistical Science, 7, 457

\bibitem[{{Gill} {et~al.}(2022){Gill}, {Ulmer-Moll}, {Wheatley}, {Bayliss}, {Burleigh}, {Acton}, {Casewell}, {Watson}, {Lendl}, {Worters}, {Sefako}, {Anderson}, {Alves}, {Bouchy}, {Bryant}, {Eigm{\"u}ller}, {Gillen}, {Goad}, {Grieves}, {G{\"u}nther}, {Henderson}, {Jenkins}, {Mishra}, {Moyano}, {Osborn}, {Tilbrook}, {Udry}, {Vines}, \& {West}}]{tic320687387b}
{Gill}, S., {Ulmer-Moll}, S., {Wheatley}, P.~J., {et~al.} 2022, \mnras, 513, 1785

\bibitem[{{Gillen} {et~al.}(2017){Gillen}, {Hillenbrand}, {David}, {Aigrain}, {Rebull}, {Stauffer}, {Cody}, \& {Queloz}}]{ad3116b}
{Gillen}, E., {Hillenbrand}, L.~A., {David}, T.~J., {et~al.} 2017, \apj, 849, 11

\bibitem[{{Gillon} {et~al.}(2010){Gillon}, {Deming}, {Demory}, {Lovis}, {Seager}, {Mayor}, {Pepe}, {Queloz}, {Segransan}, {Udry}, {Delmelle}, \& {Magain}}]{Gillon2010AA}
{Gillon}, M., {Deming}, D., {Demory}, B.~O., {et~al.} 2010, \aap, 518, A25

\bibitem[{{Gillon} {et~al.}(2014){Gillon}, {Demory}, {Madhusudhan}, {Deming}, {Seager}, {Zsom}, {Knutson}, {Lanotte}, {Bonfils}, {D{\'e}sert}, {Delrez}, {Jehin}, {Fraine}, {Magain}, \& {Triaud}}]{Gillon2014AA}
{Gillon}, M., {Demory}, B.~O., {Madhusudhan}, N., {et~al.} 2014, \aap, 563, A21

\bibitem[{{Gillon} {et~al.}(2024){Gillon}, {Pedersen}, {Rackham}, {Dransfield}, {Ducrot}, {Barkaoui}, {Burdanov}, {Schroffenegger}, {G{\'o}mez Maqueo Chew}, {Lederer}, {Alonso}, {Burgasser}, {Howell}, {Narita}, {de Wit}, {Demory}, {Queloz}, {Triaud}, {Delrez}, {Jehin}, {Hooton}, {Garcia}, {Jano Mu{\~n}oz}, {Murray}, {Pozuelos}, {Sebastian}, {Timmermans}, {Thompson}, {Z{\'u}{\~n}iga-Fern{\'a}ndez}, {Aceituno}, {Aganze}, {Amado}, {Baycroft}, {Benkhaldoun}, {Berardo}, {Bolmont}, {Clark}, {Davis}, {Davoudi}, {de Beurs}, {de Leon}, {Ikoma}, {Ikuta}, {Isogai}, {Fukuda}, {Fukui}, {Gerasimov}, {Ghachoui}, {G{\"u}nther}, {Hasler}, {Hayashi}, {Heng}, {Hu}, {Kagetani}, {Kawai}, {Kawauchi}, {Kitzmann}, {Koll}, {Lendl}, {Livingston}, {Lyu}, {Meier Vald{\'e}s}, {Mori}, {McCormac}, {Murgas}, {Niraula}, {Pall{\'e}}, {Plauchu-Frayn}, {Rebolo}, {Sabin}, {Schackey}, {Schanche}, {Selsis}, {Sota}, {Stalport}, {Standing}, {Stassun}, {Tamura}, {Terada}, {Theissen}, {Turbet}, {Van Grootel}, {Varas}, {Watanabe}, \& {Zong
  Lang}}]{Gillon2024}
{Gillon}, M., {Pedersen}, P.~P., {Rackham}, B.~V., {et~al.} 2024, Nature Astronomy, 8, 865

\bibitem[{{Gillon} {et~al.}(2012){Gillon}, {Triaud}, {Fortney}, {Demory}, {Jehin}, {Lendl}, {Magain}, {Kabath}, {Queloz}, {Alonso}, {Anderson}, {Collier Cameron}, {Fumel}, {Hebb}, {Hellier}, {Lanotte}, {Maxted}, {Mowlavi}, \& {Smalley}}]{Gillon2012}
{Gillon}, M., {Triaud}, A.~H.~M.~J., {Fortney}, J.~J., {et~al.} 2012, \aap, 542, A4

\bibitem[{Grieves {et~al.}(2021)Grieves, Bouchy, Lendl, Carmichael, Mireles, Shporer, McLeod, Collins, Brahm, Stassun, Gill, Bouma, Guillot, Cointepas, Dos~Santos, Casewell, Jenkins, Henning, Nielsen, Psaridi, Udry, Ségransan, Eastman, Zhou, Abe, Agabi, Bakos, Charbonneau, Collins, Colon, Crouzet, Dransfield, Evans, Goeke, Hart, Irwin, Jensen, Jordán, Kielkopf, Latham, Marie-Sainte, Mékarnia, Nelson, Quinn, Radford, Rodriguez, Rowden, Schmider, Schwarz, Smith, Stockdale, Suarez, Tan, Triaud, Waalkes, \& Wingham}]{grieves21}
Grieves, N., Bouchy, F., Lendl, M., {et~al.} 2021, \aap, 652, A127

\bibitem[{Hastings(1970)}]{Hastings_1970}
Hastings, W.~K. 1970, Biometrika, 57, 97

\bibitem[{Hedges {et~al.}(2020)Hedges, Angus, Barentsen, Saunders, Montet, \& Gully-Santiago}]{Hedges_2020}
Hedges, C., Angus, R., Barentsen, G., {et~al.} 2020, Research Notes of the AAS, 4, 220

\bibitem[{Henderson {et~al.}(2024)Henderson, Casewell, Goad, Acton, Günther, Nielsen, Burleigh, Belardi, Tilbrook, Turner, Howell, Clark, Littlefield, Barkaoui, Alves, Anderson, Bayliss, Bouchy, Bryant, Dransfield, Ducrot, Eigmüller, Gill, Gillen, Gillon, Hawthorn, Hooton, Jackman, Jehin, Jenkins, Kendall, Lendl, McCormac, Moyano, Pedersen, Pozuelos, Ramsay, Sefako, Timmermans, Triaud, Udry, Vines, Watson, West, Wheatley, \& Zúñiga-Fernández}]{Henderson_2024}
Henderson, B.~A., Casewell, S.~L., Goad, M.~R., {et~al.} 2024, Monthly Notices of the Royal Astronomical Society, 530, 318

\bibitem[{Hodžić {et~al.}(2018)Hodžić, Triaud, Anderson, Bouchy, Cameron, Delrez, Gillon, Hellier, Jehin, Lendl, Maxted, Pepe, Pollacco, Queloz, Ségransan, Smalley, Udry, \& West}]{wasp128b}
Hodžić, V., Triaud, A. H. M.~J., Anderson, D.~R., {et~al.} 2018, \mnras, 481, 5091

\bibitem[{{Howell} {et~al.}(2011){Howell}, {Everett}, {Sherry}, {Horch}, \& {Ciardi}}]{Howell_2011AJ}
{Howell}, S.~B., {Everett}, M.~E., {Sherry}, W., {Horch}, E., \& {Ciardi}, D.~R. 2011, \aj, 142, 19

\bibitem[{{Husser} {et~al.}(2013){Husser}, {Wende-von Berg}, {Dreizler}, {Homeier}, {Reiners}, {Barman}, \& {Hauschildt}}]{Husser:2013}
{Husser}, T.~O., {Wende-von Berg}, S., {Dreizler}, S., {et~al.} 2013, \aap, 553, A6

\bibitem[{{Irwin} {et~al.}(2018){Irwin}, {Charbonneau}, {Esquerdo}, {Latham}, {Winters}, {Dittmann}, {Newton}, {Berta-Thompson}, {Berlind}, \& {Calkins}}]{lp261}
{Irwin}, J.~M., {Charbonneau}, D., {Esquerdo}, G.~A., {et~al.} 2018, \aj, 156, 140

\bibitem[{{Jackman} {et~al.}(2019){Jackman}, {Wheatley}, {Bayliss}, {Gill}, {Hodgkin}, {Burleigh}, {Braker}, {G{\"u}nther}, {Louden}, {Turner}, {Anderson}, {Belardi}, {Bouchy}, {Briegal}, {Bryant}, {Cabrera}, {Casewell}, {Chaushev}, {Costes}, {Csizmadia}, {Eigm{\"u}ller}, {Erikson}, {G{\"a}nsicke}, {Gillen}, {Goad}, {Jenkins}, {McCormac}, {Moyano}, {Nielsen}, {Pollacco}, {Poppenhaeger}, {Queloz}, {Rauer}, {Raynard}, {Smith}, {Udry}, {Vines}, {Watson}, \& {West}}]{ngts7ab}
{Jackman}, J. A.~G., {Wheatley}, P.~J., {Bayliss}, D., {et~al.} 2019, \mnras, 489, 5146

\bibitem[{{Jehin} {et~al.}(2018){Jehin}, {Gillon}, {Queloz}, {Delrez}, {Burdanov}, {Murray}, {Sohy}, {Ducrot}, {Sebastian}, {Thompson}, {McCormac}, {Almleaky}, {Burgasser}, {Demory}, {de Wit}, {Barkaoui}, {Pozuelos}, {Triaud}, \& {Grootel}}]{Jehin2018Msngr}
{Jehin}, E., {Gillon}, M., {Queloz}, D., {et~al.} 2018, The Messenger, 174, 2

\bibitem[{{Jenkins} {et~al.}(2016){Jenkins}, {Twicken}, {McCauliff}, {Campbell}, {Sanderfer}, {Lung}, {Mansouri-Samani}, {Girouard}, {Tenenbaum}, {Klaus}, {Smith}, {Caldwell}, {Chacon}, {Henze}, {Heiges}, {Latham}, {Morgan}, {Swade}, {Rinehart}, \& {Vanderspek}}]{SPOC_Jenkins_2016SPIE}
{Jenkins}, J.~M., {Twicken}, J.~D., {McCauliff}, S., {et~al.} 2016, in Society of Photo-Optical Instrumentation Engineers (SPIE) Conference Series, Vol. 9913, Software and Cyberinfrastructure for Astronomy IV, ed. G.~{Chiozzi} \& J.~C. {Guzman}, 99133E

\bibitem[{{Jensen}(2013)}]{jensen2013}
{Jensen}, E. 2013, {Tapir: A web interface for transit/eclipse observability}, Astrophysics Source Code Library, record ascl:1306.007

\bibitem[{{Johnson} {et~al.}(2011){Johnson}, {Apps}, {Gazak}, {Crepp}, {Crossfield}, {Howard}, {Marcy}, {Morton}, {Chubak}, \& {Isaacson}}]{lhs6343}
{Johnson}, J.~A., {Apps}, K., {Gazak}, J.~Z., {et~al.} 2011, \apj, 730, 79

\bibitem[{{Kanodia} {et~al.}(2022){Kanodia}, {Libby-Roberts}, {Ca{\~n}as}, {Ninan}, {Mahadevan}, {Stefansson}, {Lin}, {Jones}, {Monson}, {Parker}, {Kobulnicky}, {Swaby}, {Powers}, {Beard}, {Bender}, {Blake}, {Cochran}, {Dong}, {Diddams}, {Fredrick}, {Gupta}, {Halverson}, {Hearty}, {Logsdon}, {Metcalf}, {McElwain}, {Morley}, {Rajagopal}, {Ramsey}, {Robertson}, {Roy}, {Schwab}, {Terrien}, {Wisniewski}, \& {Wright}}]{Kanodia_2022AJ}
{Kanodia}, S., {Libby-Roberts}, J., {Ca{\~n}as}, C.~I., {et~al.} 2022, \aj, 164, 81

\bibitem[{{Khandelwal, A.} {et~al.}(2023){Khandelwal, A.}, {Sharma, Rishikesh}, {Chakraborty, Abhijit}, {Chaturvedi, Priyanka}, {Ulmer-Moll, Solène}, {Ciardi, David R.}, {Boyle, Andrew W.}, {Baliwal, Sanjay}, {Bieryla, Allyson}, {Latham, David W.}, {Prasad, Neelam J. S. S. V.}, {Nayak, Ashirbad}, {Lendl, Monika}, \& {Mordasini, Christoph}}]{toi4603b}
{Khandelwal, A.}, {Sharma, Rishikesh}, {Chakraborty, Abhijit}, {et~al.} 2023, A\&A, 672, L7

\bibitem[{{Kipping}(2013)}]{Kipping_2013MNRAS.435.2152K}
{Kipping}, D.~M. 2013, \mnras, 435, 2152

\bibitem[{{L{\'e}pine} {et~al.}(2013){L{\'e}pine}, {Hilton}, {Mann}, {Wilde}, {Rojas-Ayala}, {Cruz}, \& {Gaidos}}]{2013AJ....145..102L}
{L{\'e}pine}, S., {Hilton}, E.~J., {Mann}, A.~W., {et~al.} 2013, \aj, 145, 102

\bibitem[{Lester {et~al.}(2021)Lester, Matson, Howell, Furlan, Gnilka, Scott, Ciardi, Everett, Hartman, \& Hirsch}]{Lester_2021}
Lester, K.~V., Matson, R.~A., Howell, S.~B., {et~al.} 2021, The Astronomical Journal, 162, 75

\bibitem[{{Lightkurve Collaboration} {et~al.}(2018){Lightkurve Collaboration}, {Cardoso}, {Hedges}, {Gully-Santiago}, {Saunders}, {Cody}, {Barclay}, {Hall}, {Sagear}, {Turtelboom}, {Zhang}, {Tzanidakis}, {Mighell}, {Coughlin}, {Bell}, {Berta-Thompson}, {Williams}, {Dotson}, \& {Barentsen}}]{Lightkurve_2018ascl}
{Lightkurve Collaboration}, {Cardoso}, J. V. d.~M., {Hedges}, C., {et~al.} 2018, {Lightkurve: Kepler and TESS time series analysis in Python}, Astrophysics Source Code Library, record ascl:1812.013

\bibitem[{{Lin} {et~al.}(2023){Lin}, {Gan}, {Wang}, {Shporer}, {Rabus}, {Zhou}, {Psaridi}, {Bouchy}, {Bieryla}, {Latham}, {Mao}, {Stassun}, {Hellier}, {Howell}, {Ziegler}, {Caldwell}, {Clark}, {Collins}, {Curtis}, {Faherty}, {Gnilka}, {Grunblatt}, {Jenkins}, {Johnson}, {Law}, {Lendl}, {Littlefield}, {Lund}, {Lund}, {Mann}, {McDermott}, {Mishra}, {Mounzer}, {Paegert}, {Pritchard}, {Ricker}, {Seager}, {Srdoc}, {Sun}, {Tang}, {Udry}, {Vanderspek}, {Watanabe}, {Winn}, \& {Yu}}]{toi1608b}
{Lin}, Z., {Gan}, T., {Wang}, S.~X., {et~al.} 2023, \mnras, 523, 6162

\bibitem[{{Lomb}(1976)}]{Lomb_1976ApSS}
{Lomb}, N.~R. 1976, \apss, 39, 447

\bibitem[{{Mandel} \& {Agol}(2002)}]{Mandel2002}
{Mandel}, K. \& {Agol}, E. 2002, \apjl, 580, L171

\bibitem[{{Mann} {et~al.}(2013){Mann}, {Brewer}, {Gaidos}, {L{\'e}pine}, \& {Hilton}}]{2013AJ....145...52M}
{Mann}, A.~W., {Brewer}, J.~M., {Gaidos}, E., {L{\'e}pine}, S., \& {Hilton}, E.~J. 2013, \aj, 145, 52

\bibitem[{{Mann} {et~al.}(2014){Mann}, {Deacon}, {Gaidos}, {Ansdell}, {Brewer}, {Liu}, {Magnier}, \& {Aller}}]{Mann2014}
{Mann}, A.~W., {Deacon}, N.~R., {Gaidos}, E., {et~al.} 2014, \aj, 147, 160

\bibitem[{{Mann} {et~al.}(2019){Mann}, {Dupuy}, {Kraus}, {Gaidos}, {Ansdell}, {Ireland}, {Rizzuto}, {Hung}, {Dittmann}, {Factor}, {Feiden}, {Martinez}, {Ru{\'\i}z-Rodr{\'\i}guez}, \& {Thao}}]{Mann:2019}
{Mann}, A.~W., {Dupuy}, T., {Kraus}, A.~L., {et~al.} 2019, \apj, 871, 63

\bibitem[{Mann {et~al.}(2015)Mann, Feiden, Gaidos, Boyajian, \& von Braun}]{Mann_2015}
Mann, A.~W., Feiden, G.~A., Gaidos, E., Boyajian, T., \& von Braun, K. 2015, The Astrophysical Journal, 804, 64

\bibitem[{{Massey} \& {Gronwall}(1990)}]{1990ApJ...358..344M}
{Massey}, P. \& {Gronwall}, C. 1990, \apj, 358, 344

\bibitem[{Matson {et~al.}(2018)Matson, Howell, Horch, \& Everett}]{Matson_2018}
Matson, R.~A., Howell, S.~B., Horch, E.~P., \& Everett, M.~E. 2018, The Astronomical Journal, 156, 31

\bibitem[{{McCully} {et~al.}(2018){McCully}, {Volgenau}, {Harbeck}, {Lister}, {Saunders}, {Turner}, {Siiverd}, \& {Bowman}}]{McCully_2018SPIE10707E}
{McCully}, C., {Volgenau}, N.~H., {Harbeck}, D.-R., {et~al.} 2018, in Society of Photo-Optical Instrumentation Engineers (SPIE) Conference Series, Vol. 10707, Software and Cyberinfrastructure for Astronomy V, ed. J.~C. {Guzman} \& J.~{Ibsen}, 107070K

\bibitem[{{Metropolis} {et~al.}(1953){Metropolis}, {Rosenbluth}, {Rosenbluth}, {Teller}, \& {Teller}}]{Metropolis_1953}
{Metropolis}, N., {Rosenbluth}, A.~W., {Rosenbluth}, M.~N., {Teller}, A.~H., \& {Teller}, E. 1953, \jcp, 21, 1087

\bibitem[{{Miller} \& {Stone}(1994)}]{kastspectrograph}
{Miller}, J.~S. \& {Stone}, R.~P.~S. 1994, The Kast Double Spectrograph, Tech. Rep.~66, University of California Lick Observatory Technical Reports

\bibitem[{{Mireles} {et~al.}(2020){Mireles}, {Shporer}, {Grieves}, {Zhou}, {G{\"u}nther}, {Brahm}, {Ziegler}, {Stassun}, {Huang}, {Nielsen}, {dos Santos}, {Udry}, {Bouchy}, {Ireland}, {Wallace}, {Sarkis}, {Henning}, {Jord{\'a}n}, {Law}, {Mann}, {Paredes}, {James}, {Jao}, {Henry}, {Butler}, {Rodriguez}, {Yu}, {Flowers}, {Ricker}, {Latham}, {Vanderspek}, {Seager}, {Winn}, {Jenkins}, {Furesz}, {Hesse}, {Quintana}, {Rose}, {Smith}, {Tenenbaum}, {Vezie}, {Yahalomi}, \& {Zhan}}]{toi694}
{Mireles}, I., {Shporer}, A., {Grieves}, N., {et~al.} 2020, \aj, 160, 133

\bibitem[{{Moutou} {et~al.}(2013){Moutou}, {Bonomo}, {Bruno}, {Montagnier}, {Bouchy}, {Almenara}, {Barros}, {Deleuil}, {D{\'\i}az}, {H{\'e}brard}, \& {Santerne}}]{koi415b}
{Moutou}, C., {Bonomo}, A.~S., {Bruno}, G., {et~al.} 2013, \aap, 558, L6

\bibitem[{{Murray} \& {Correia}(2010)}]{Murray_2010exopbook}
{Murray}, C.~D. \& {Correia}, A.~C.~M. 2010, in Exoplanets, ed. S.~{Seager} (University of Arizona Press), 15--23

\bibitem[{{Nowak} {et~al.}(2017){Nowak}, {Palle}, {Gandolfi}, {Dai}, {Lanza}, {Hirano}, {Barrag{\'a}n}, {Fukui}, {Bruntt}, {Endl}, {Cochran}, {Prada Moroni}, {Prieto-Arranz}, {Kiilerich}, {Nespral}, {Hatzes}, {Albrecht}, {Deeg}, {Winn}, {Yu}, {Kuzuhara}, {Grziwa}, {Smith}, {Guenther}, {Van Eylen}, {Csizmadia}, {Fridlund}, {Cabrera}, {Eigm{\"u}ller}, {Erikson}, {Korth}, {Narita}, {P{\"a}tzold}, {Rauer}, \& {Ribas}}]{cww89ab1}
{Nowak}, G., {Palle}, E., {Gandolfi}, D., {et~al.} 2017, \aj, 153, 131

\bibitem[{{Oke}(1990)}]{1990AJ.....99.1621O}
{Oke}, J.~B. 1990, \aj, 99, 1621

\bibitem[{{Palle} {et~al.}(2021){Palle}, {Luque}, {Zapatero Osorio}, {Parviainen}, {Ikoma}, {Tabernero}, {Zechmeister}, {Mustill}, {Bejar}, {Narita}, \& {Murgas}}]{Palle_2021AA}
{Palle}, E., {Luque}, R., {Zapatero Osorio}, M.~R., {et~al.} 2021, \aap, 650, A55

\bibitem[{{Parviainen} {et~al.}(2020){Parviainen}, {Palle}, {Zapatero-Osorio}, {Montanes Rodriguez}, {Murgas}, {Narita}, {Hidalgo Soto}, {B{\'e}jar}, {Korth}, {Monelli}, {Casasayas Barris}, {Crouzet}, {de Leon}, {Fukui}, {Hernandez}, {Klagyivik}, {Kusakabe}, {Luque}, {Mori}, {Nishiumi}, {Prieto-Arranz}, {Tamura}, {Watanabe}, {Burke}, {Charbonneau}, {Collins}, {Collins}, {Conti}, {Garcia Soto}, {Jenkins}, {Jenkins}, {Levine}, {Li}, {Rinehart}, {Seager}, {Tenenbaum}, {Ting}, {Vanderspek}, {Vezie}, \& {Winn}}]{toi263b1}
{Parviainen}, H., {Palle}, E., {Zapatero-Osorio}, M.~R., {et~al.} 2020, \aap, 633, A28

\bibitem[{{Pepe, F.} {et~al.}(2021){Pepe, F.}, {Cristiani, S.}, {Rebolo, R.}, {Santos, N. C.}, {Dekker, H.}, {Cabral, A.}, {Di Marcantonio, P.}, {Figueira, P.}, {Lo Curto, G.}, {Lovis, C.}, {Mayor, M.}, {Mégevand, D.}, {Molaro, P.}, {Riva, M.}, {Zapatero Osorio, M. R.}, {Amate, M.}, {Manescau, A.}, {Pasquini, L.}, {Zerbi, F. M.}, {Adibekyan, V.}, {Abreu, M.}, {Affolter, M.}, {Alibert, Y.}, {Aliverti, M.}, {Allart, R.}, {Allende Prieto, C.}, {Álvarez, D.}, {Alves, D.}, {Avila, G.}, {Baldini, V.}, {Bandy, T.}, {Barros, S. C. C.}, {Benz, W.}, {Bianco, A.}, {Borsa, F.}, {Bourrier, V.}, {Bouchy, F.}, {Broeg, C.}, {Calderone, G.}, {Cirami, R.}, {Coelho, J.}, {Conconi, P.}, {Coretti, I.}, {Cumani, C.}, {Cupani, G.}, {D’Odorico, V.}, {Damasso, M.}, {Deiries, S.}, {Delabre, B.}, {Demangeon, O. D. S.}, {Dumusque, X.}, {Ehrenreich, D.}, {Faria, J. P.}, {Fragoso, A.}, {Genolet, L.}, {Genoni, M.}, {Génova Santos, R.}, {González Hernández, J. I.}, {Hughes, I.}, {Iwert, O.}, {Kerber, F.}, {Knudstrup, J.}, {Landoni,
  M.}, {Lavie, B.}, {Lillo-Box, J.}, {Lizon, J.-L.}, {Maire, C.}, {Martins, C. J. A. P.}, {Mehner, A.}, {Micela, G.}, {Modigliani, A.}, {Monteiro, M. A.}, {Monteiro, M. J. P. F. G.}, {Moschetti, M.}, {Murphy, M. T.}, {Nunes, N.}, {Oggioni, L.}, {Oliveira, A.}, {Oshagh, M.}, {Pallé, E.}, {Pariani, G.}, {Poretti, E.}, {Rasilla, J. L.}, {Rebordão, J.}, {Redaelli, E. M.}, {Santana Tschudi, S.}, {Santin, P.}, {Santos, P.}, {Ségransan, D.}, {Schmidt, T. M.}, {Segovia, A.}, {Sosnowska, D.}, {Sozzetti, A.}, {Sousa, S. G.}, {Spanò, P.}, {Suárez Mascareño, A.}, {Tabernero, H.}, {Tenegi, F.}, {Udry, S.}, \& {Zanutta, A.}}]{Pepe2021}
{Pepe, F.}, {Cristiani, S.}, {Rebolo, R.}, {et~al.} 2021, A\&A, 645, A96

\bibitem[{{Persson} {et~al.}(2019){Persson}, {Csizmadia}, {Mustill}, {Fridlund}, {Hatzes}, {Nowak}, {Georgieva}, {Gandolfi}, {Davies}, {Livingston}, {Palle}, {Monta{\~n}es Rodr{\'\i}guez}, {Endl}, {Hirano}, {Prieto-Arranz}, {Korth}, {Grziwa}, {Esposito}, {Albrecht}, {Johnson}, {Barrag{\'a}n}, {Parviainen}, {Van Eylen}, {Alonso Sobrino}, {Beck}, {Cabrera}, {Carleo}, {Cochran}, {Dai}, {Deeg}, {de Leon}, {Eigm{\"u}ller}, {Erikson}, {Fukui}, {Gonz{\'a}lez-Cuesta}, {Guenther}, {Hidalgo}, {Hjorth}, {Kabath}, {Knudstrup}, {Kusakabe}, {Lam}, {Lund}, {Luque}, {Mathur}, {Murgas}, {Narita}, {Nespral}, {Niraula}, {Olofsson}, {P{\"a}tzold}, {Rauer}, {Redfield}, {Ribas}, {Skarka}, {Smith}, {Subjak}, \& {Tamura}}]{epic212036875b1}
{Persson}, C.~M., {Csizmadia}, S., {Mustill}, A.~J., {et~al.} 2019, \aap, 628, A64

\bibitem[{{Phillips} {et~al.}(2020){Phillips}, {Tremblin}, {Baraffe}, {Chabrier}, {Allard}, {Spiegelman}, {Goyal}, {Drummond}, \& {H{\'e}brard}}]{Phillips_2020A&A}
{Phillips}, M.~W., {Tremblin}, P., {Baraffe}, I., {et~al.} 2020, \aap, 637, A38

\bibitem[{{Pont} {et~al.}(2005){Pont}, {Melo}, {Bouchy}, {Udry}, {Queloz}, {Mayor}, \& {Santos}}]{ogle122b}
{Pont}, F., {Melo}, C.~H.~F., {Bouchy}, F., {et~al.} 2005, \aap, 433, L21

\bibitem[{{Pont} {et~al.}(2006){Pont}, {Moutou}, {Bouchy}, {Behrend}, {Mayor}, {Udry}, {Queloz}, {Santos}, \& {Melo}}]{ogle123b}
{Pont}, F., {Moutou}, C., {Bouchy}, F., {et~al.} 2006, \aap, 447, 1035

\bibitem[{{Psaridi} {et~al.}(2022){Psaridi}, {Bouchy}, {Lendl}, {Grieves}, {Stassun}, {Carmichael}, {Gill}, {Pe{\~n}a Rojas}, {Gan}, {Shporer}, {Bieryla}, {Brahm}, {Christiansen}, {Crossfield}, {Galland}, {Hooton}, {Jenkins}, {Jenkins}, {Latham}, {Lund}, {Rodriguez}, {Ting}, {Udry}, {Ulmer-Moll}, {Wittenmyer}, {Zhang}, {Zhou}, {Addison}, {Cointepas}, {Collins}, {Collins}, {Deline}, {Dressing}, {Evans}, {Giacalone}, {Heitzmann}, {Mireles}, {Mounzer}, {Otegi}, {Radford}, {Rudat}, {Schlieder}, {Schwarz}, {Srdoc}, {Stockdale}, {Suarez}, {Wright}, \& {Zhao}}]{psaridi22}
{Psaridi}, A., {Bouchy}, F., {Lendl}, M., {et~al.} 2022, \aap, 664, A94

\bibitem[{{Rayner} {et~al.}(2009){Rayner}, {Cushing}, \& {Vacca}}]{Rayner2009}
{Rayner}, J.~T., {Cushing}, M.~C., \& {Vacca}, W.~D. 2009, \apjs, 185, 289

\bibitem[{{Rayner} {et~al.}(2003){Rayner}, {Toomey}, {Onaka}, {Denault}, {Stahlberger}, {Vacca}, {Cushing}, \& {Wang}}]{Rayner2003}
{Rayner}, J.~T., {Toomey}, D.~W., {Onaka}, P.~M., {et~al.} 2003, \pasp, 115, 362

\bibitem[{{Rebassa-Mansergas} {et~al.}(2023){Rebassa-Mansergas}, {Maldonado}, {Raddi}, {Torres}, {Hoskin}, {Cunningham}, {Hollands}, {Ren}, {G{\"a}nsicke}, {Tremblay}, \& {Camisassa}}]{2023MNRAS.526.4787R}
{Rebassa-Mansergas}, A., {Maldonado}, J., {Raddi}, R., {et~al.} 2023, \mnras, 526, 4787

\bibitem[{{Ricker} {et~al.}(2015){Ricker}, {Winn}, {Vanderspek}, {Latham}, {Bakos}, {Bean}, {Berta-Thompson}, {Brown}, {Buchhave}, {Butler}, {Butler}, {Chaplin}, {Charbonneau}, {Christensen-Dalsgaard}, {Clampin}, {Deming}, {Doty}, {De Lee}, {Dressing}, {Dunham}, {Endl}, {Fressin}, {Ge}, {Henning}, {Holman}, {Howard}, {Ida}, {Jenkins}, {Jernigan}, {Johnson}, {Kaltenegger}, {Kawai}, {Kjeldsen}, {Laughlin}, {Levine}, {Lin}, {Lissauer}, {MacQueen}, {Marcy}, {McCullough}, {Morton}, {Narita}, {Paegert}, {Palle}, {Pepe}, {Pepper}, {Quirrenbach}, {Rinehart}, {Sasselov}, {Sato}, {Seager}, {Sozzetti}, {Stassun}, {Sullivan}, {Szentgyorgyi}, {Torres}, {Udry}, \& {Villasenor}}]{Ricker_2015JATIS_TESS}
{Ricker}, G.~R., {Winn}, J.~N., {Vanderspek}, R., {et~al.} 2015, Journal of Astronomical Telescopes, Instruments, and Systems, 1, 014003

\bibitem[{{Rojas-Ayala} {et~al.}(2012){Rojas-Ayala}, {Covey}, {Muirhead}, \& {Lloyd}}]{Rojas-Ayala2012}
{Rojas-Ayala}, B., {Covey}, K.~R., {Muirhead}, P.~S., \& {Lloyd}, J.~P. 2012, \apj, 748, 93

\bibitem[{{Saumon} \& {Marley}(2008)}]{Saumon_2008ApJ}
{Saumon}, D. \& {Marley}, M.~S. 2008, \apj, 689, 1327

\bibitem[{{Scargle}(1982)}]{Scargle_1982ApJ}
{Scargle}, J.~D. 1982, \apj, 263, 835

\bibitem[{{Schlegel} {et~al.}(1998){Schlegel}, {Finkbeiner}, \& {Davis}}]{Schlegel:1998}
{Schlegel}, D.~J., {Finkbeiner}, D.~P., \& {Davis}, M. 1998, \apj, 500, 525

\bibitem[{Schmidt {et~al.}(2023)Schmidt, Schlaufman, Ding, Grunblatt, Carmichael, Bieryla, Rodriguez, Schulte, Vowell, Zhou, Quinn, Yee, Winn, Hartman, Latham, Caldwell, Fausnaugh, Hedges, Jenkins, Osborn, \& Seager}]{toi1712b}
Schmidt, S.~P., Schlaufman, K.~C., Ding, K., {et~al.} 2023, The Astronomical Journal, 166, 225

\bibitem[{Schwarz(1978)}]{schwarz1978}
Schwarz, G. 1978, Ann. Statist., 6, 461

\bibitem[{{Scott} {et~al.}(2021){Scott}, {Howell}, {Gnilka}, {Stephens}, {Salinas}, {Matson}, {Furlan}, {Horch}, {Everett}, {Ciardi}, {Mills}, \& {Quigley}}]{Scott_2021FrASS}
{Scott}, N.~J., {Howell}, S.~B., {Gnilka}, C.~L., {et~al.} 2021, Frontiers in Astronomy and Space Sciences, 8, 138

\bibitem[{{Sebastian} {et~al.}(2021){Sebastian}, {Gillon}, {Ducrot}, {Pozuelos}, {Garcia}, {G{\"u}nther}, {Delrez}, {Queloz}, {Demory}, {Triaud}, {Burgasser}, {de Wit}, {Burdanov}, {Dransfield}, {Jehin}, {McCormac}, {Murray}, {Niraula}, {Pedersen}, {Rackham}, {Sohy}, {Thompson}, \& {Van Grootel}}]{Sebastian_2021AA}
{Sebastian}, D., {Gillon}, M., {Ducrot}, E., {et~al.} 2021, \aap, 645, A100

\bibitem[{{Sebastian} {et~al.}(2022){Sebastian}, {Guenther}, {Deleuil}, {Dorsch}, {Heber}, {Heuser}, {Gandolfi}, {Grziwa}, {Deeg}, {Alonso}, {Bouchy}, {Csizmadia}, {Cusano}, {Fridlund}, {Geier}, {Irrgang}, {Korth}, {Nespral}, {Rauer}, {Tal-Or}, \& {CoRoT-team}}]{CoRoT34b}
{Sebastian}, D., {Guenther}, E.~W., {Deleuil}, M., {et~al.} 2022, \mnras, 516, 636

\bibitem[{{Shporer} {et~al.}(2017){Shporer}, {Zhou}, {Vanderburg}, {Fulton}, {Isaacson}, {Bieryla}, {Torres}, {Morton}, {Bento}, {Berlind}, {Calkins}, {Esquerdo}, {Howard}, \& {Latham}}]{k276}
{Shporer}, A., {Zhou}, G., {Vanderburg}, A., {et~al.} 2017, \apjl, 847, L18

\bibitem[{{Siverd} {et~al.}(2012){Siverd}, {Beatty}, {Pepper}, {Eastman}, {Collins}, {Bieryla}, {Latham}, {Buchhave}, {Jensen}, {Crepp}, {Street}, {Stassun}, {Gaudi}, {Berlind}, {Calkins}, {DePoy}, {Esquerdo}, {Fulton}, {F{\H{u}}r{\'e}sz}, {Geary}, {Gould}, {Hebb}, {Kielkopf}, {Marshall}, {Pogge}, {Stanek}, {Stefanik}, {Szentgyorgyi}, {Trueblood}, {Trueblood}, {Stutz}, \& {van Saders}}]{kelt1b}
{Siverd}, R.~J., {Beatty}, T.~G., {Pepper}, J., {et~al.} 2012, \apj, 761, 123

\bibitem[{{Skrutskie} {et~al.}(2006){Skrutskie}, {Cutri}, {Stiening}, {Weinberg}, {Schneider}, {Carpenter}, {Beichman}, {Capps}, {Chester}, {Elias}, {Huchra}, {Liebert}, {Lonsdale}, {Monet}, {Price}, {Seitzer}, {Jarrett}, {Kirkpatrick}, {Gizis}, {Howard}, {Evans}, {Fowler}, {Fullmer}, {Hurt}, {Light}, {Kopan}, {Marsh}, {McCallon}, {Tam}, {Van Dyk}, \& {Wheelock}}]{Skrutskie_2006AJ_2MASS}
{Skrutskie}, M.~F., {Cutri}, R.~M., {Stiening}, R., {et~al.} 2006, \aj, 131, 1163

\bibitem[{{Smith} {et~al.}(2012){Smith}, {Stumpe}, {Van Cleve}, {Jenkins}, {Barclay}, {Fanelli}, {Girouard}, {Kolodziejczak}, {McCauliff}, {Morris}, \& {Twicken}}]{Smith_2012PASP}
{Smith}, J.~C., {Stumpe}, M.~C., {Van Cleve}, J.~E., {et~al.} 2012, \pasp, 124, 1000

\bibitem[{{Spiegel} {et~al.}(2011){Spiegel}, {Burrows}, \& {Milsom}}]{Spiegel_2011ApJ}
{Spiegel}, D.~S., {Burrows}, A., \& {Milsom}, J.~A. 2011, \apj, 727, 57

\bibitem[{{Stassun} {et~al.}(2017){Stassun}, {Collins}, \& {Gaudi}}]{Stassun:2017}
{Stassun}, K.~G., {Collins}, K.~A., \& {Gaudi}, B.~S. 2017, \aj, 153, 136

\bibitem[{{Stassun} {et~al.}(2018{\natexlab{a}}){Stassun}, {Corsaro}, {Pepper}, \& {Gaudi}}]{Stassun:2018}
{Stassun}, K.~G., {Corsaro}, E., {Pepper}, J.~A., \& {Gaudi}, B.~S. 2018{\natexlab{a}}, \aj, 155, 22

\bibitem[{{Stassun} {et~al.}(2012){Stassun}, {Kratter}, {Scholz}, \& {Dupuy}}]{Stassun:2012}
{Stassun}, K.~G., {Kratter}, K.~M., {Scholz}, A., \& {Dupuy}, T.~J. 2012, \apj, 756, 47

\bibitem[{{Stassun} {et~al.}(2018{\natexlab{b}}){Stassun}, {Oelkers}, {Pepper}, {Paegert}, {De Lee}, {Torres}, {Latham}, {Charpinet}, {Dressing}, {Huber}, {Kane}, {L{\'e}pine}, {Mann}, {Muirhead}, {Rojas-Ayala}, {Silvotti}, {Fleming}, {Levine}, \& {Plavchan}}]{Stassun_2018AJ_TESS_Catalog}
{Stassun}, K.~G., {Oelkers}, R.~J., {Pepper}, J., {et~al.} 2018{\natexlab{b}}, \aj, 156, 102

\bibitem[{{Stassun} \& {Torres}(2016)}]{Stassun:2016}
{Stassun}, K.~G. \& {Torres}, G. 2016, \aj, 152, 180

\bibitem[{{Stassun} \& {Torres}(2021)}]{StassunTorres:2021}
{Stassun}, K.~G. \& {Torres}, G. 2021, \apjl, 907, L33

\bibitem[{Stumpe {et~al.}(2014)Stumpe, Smith, Catanzarite, Cleve, Jenkins, Twicken, \& Girouard}]{Stumpe_2014}
Stumpe, M.~C., Smith, J.~C., Catanzarite, J.~H., {et~al.} 2014, Publications of the Astronomical Society of the Pacific, 126, 100

\bibitem[{{Stumpe} {et~al.}(2012){Stumpe}, {Smith}, {Van Cleve}, {Twicken}, {Barclay}, {Fanelli}, {Girouard}, {Jenkins}, {Kolodziejczak}, {McCauliff}, \& {Morris}}]{Stumpe_2012PASP}
{Stumpe}, M.~C., {Smith}, J.~C., {Van Cleve}, J.~E., {et~al.} 2012, \pasp, 124, 985

\bibitem[{{Tal-Or} {et~al.}(2013){Tal-Or}, {Mazeh}, {Alonso}, {Bouchy}, {Cabrera}, {Deeg}, {Deleuil}, {Faigler}, {Fridlund}, {H{\'e}brard}, {Moutou}, {Santerne}, \& {Tingley}}]{corot1011}
{Tal-Or}, L., {Mazeh}, T., {Alonso}, R., {et~al.} 2013, \aap, 553, A30

\bibitem[{{Timmermans} {et~al.}(2024){Timmermans}, {Dransfield}, {Gillon}, {Triaud}, {Rackham}, {Aganze}, {Barkaoui}, {Brice{\~n}o}, {Burgasser}, {Collins}, {Cointepas}, {D{\'e}vora-Pajares}, {Ducrot}, {Z{\'u}{\~n}iga-Fern{\'a}ndez}, {Howell}, {Kaltenegger}, {Murray}, {Pass}, {Quinn}, {Raymond}, {Sebastian}, {Stassun}, {Ziegler}, {Almenara}, {Benkhaldoun}, {Bonfils}, {Christiansen}, {Davoudi}, {de Wit}, {Delrez}, {Demory}, {Fong}, {F{\H{u}}r{\'e}sz}, {Ghachoui}, {Garcia}, {G{\'o}mez Maqueo Chew}, {Hooton}, {Horne}, {G{\"u}nther}, {Jehin}, {Jenkins}, {Law}, {Mann}, {Murgas}, {Pozuelos}, {Pedersen}, {Queloz}, {Ricker}, {Rowden}, {Schwarz}, {Seager}, {Smart}, {Srdoc}, {Striegel}, {Thompson}, {Vanderspek}, \& {Winn}}]{Timmermans2024}
{Timmermans}, M., {Dransfield}, G., {Gillon}, M., {et~al.} 2024, \aap, 687, A48

\bibitem[{{Triaud} {et~al.}(2023){Triaud}, {Dransfield}, {Kagetani}, {Timmermans}, {Narita}, {Barkaoui}, {Hirano}, {Rackham}, {Mori}, {Baycroft}, {Benkhaldoun}, {Burgasser}, {Caldwell}, {Collins}, {Davis}, {Delrez}, {Demory}, {Ducrot}, {Fukui}, {Mu{\~n}oz}, {Jehin}, {Garc{\'\i}a}, {Ghachoui}, {Gillon}, {Chew}, {Hooton}, {Ikoma}, {Kawauchi}, {Kotani}, {Levine}, {Pall{\'e}}, {Pedersen}, {Pozuelos}, {Queloz}, {Scutt}, {Seager}, {Sebastian}, {Tamura}, {Thompson}, {Watanabe}, {de Wit}, {Winn}, \& {Z{\'u}{\~n}iga-Fern{\'a}ndez}}]{Triaud2023}
{Triaud}, A. H.~M.~J., {Dransfield}, G., {Kagetani}, T., {et~al.} 2023, \mnras, 525, L98

\bibitem[{{Triaud} {et~al.}(2013){Triaud}, {Hebb}, {Anderson}, {Cargile}, {Collier Cameron}, {Doyle}, {Faedi}, {Gillon}, {Gomez Maqueo Chew}, {Hellier}, {Jehin}, {Maxted}, {Naef}, {Pepe}, {Pollacco}, {Queloz}, {S{\'e}gransan}, {Smalley}, {Stassun}, {Udry}, \& {West}}]{wasp30b2}
{Triaud}, A.~H.~M.~J., {Hebb}, L., {Anderson}, D.~R., {et~al.} 2013, \aap, 549, A18

\bibitem[{{von Boetticher} {et~al.}(2017){von Boetticher}, {Triaud}, {Queloz}, {Gill}, {Lendl}, {Delrez}, {Anderson}, {Collier Cameron}, {Faedi}, {Gillon}, {G{\'o}mez Maqueo Chew}, {Hebb}, {Hellier}, {Jehin}, {Maxted}, {Martin}, {Pepe}, {Pollacco}, {S{\'e}gransan}, {Smalley}, {Udry}, \& {West}}]{j055557ab}
{von Boetticher}, A., {Triaud}, A. H.~M.~J., {Queloz}, D., {et~al.} 2017, \aap, 604, L6

\bibitem[{{von Boetticher} {et~al.}(2019){von Boetticher}, {Triaud}, {Queloz}, {Gill}, {Maxted}, {Almleaky}, {Anderson}, {Bouchy}, {Burdanov}, {Collier Cameron}, {Delrez}, {Ducrot}, {Faedi}, {Gillon}, {G{\'o}mez Maqueo Chew}, {Hebb}, {Hellier}, {Jehin}, {Lendl}, {Marmier}, {Martin}, {McCormac}, {Pepe}, {Pollacco}, {S{\'e}gransan}, {Smalley}, {Thompson}, {Turner}, {Udry}, {Van Grootel}, \& {West}}]{j0954}
{von Boetticher}, A., {Triaud}, A. H.~M.~J., {Queloz}, D., {et~al.} 2019, \aap, 625, A150

\bibitem[{{Vowell} {et~al.}(2025){Vowell}, {Rodriguez}, {Latham}, {Quinn}, {Schulte}, {Eastman}, {Bieryla}, {Barkaoui}, {Ciardi}, {Collins}, {Girardin}, {Heldridge}, {Kotten}, {Mancini}, {Murgas}, {Narita}, {Radford}, {Relles}, {Shporer}, {Soares-Furtado}, {Strakhov}, {Ziegler}, {Brice{\~n}o}, {Calkins}, {Clark}, {Collins}, {Esquerdo}, {Fajardo-Acosta}, {Fukui}, {Watkins}, {He}, {Horne}, {Jenkins}, {Mann}, {Naponiello}, {Palle}, {Schwarz}, {Seager}, {Southworth}, {Srdoc}, {Swift}, \& {Winn}}]{Vowell_2025arXiv250109795V}
{Vowell}, N., {Rodriguez}, J.~E., {Latham}, D.~W., {et~al.} 2025, arXiv e-prints, arXiv:2501.09795

\bibitem[{Vowell {et~al.}(2023)Vowell, Rodriguez, Quinn, Zhou, Vanderburg, Mann, Hooton, Stassun, Howard, Bieryla, Latham, Howell, Guillot, Ziegler, Collins, Carmichael, Jenkins, Shporer, ABE, Bendjoya, Bush, Buttu, Collins, Eastman, Fields, Gasparetto, Günther, Kostov, Kraus, Lester, Levine, Littlefield, Marie-Sainte, Mékarnia, Osborn, Rapetti, Ricker, Seager, Sefako, Srdoc, Suarez, Torres, Triaud, Vanderspek, \& Winn}]{HIP33609b}
Vowell, N., Rodriguez, J.~E., Quinn, S.~N., {et~al.} 2023, The Astronomical Journal, 165, 268

\bibitem[{{{\v{S}}ubjak} {et~al.}(2020){{\v{S}}ubjak}, {Sharma}, {Carmichael}, {Johnson}, {Gonzales}, {Matthews}, {Boffin}, {Brahm}, {Chaturvedi}, {Chakraborty}, {Ciardi}, {Collins}, {Esposito}, {Fridlund}, {Gan}, {Gandolfi}, {Garc{\'\i}a}, {Guenther}, {Hatzes}, {Latham}, {Mathis}, {Mathur}, {Persson}, {Relles}, {Schlieder}, {Barclay}, {Dressing}, {Crossfield}, {Howard}, {Rodler}, {Zhou}, {Quinn}, {Esquerdo}, {Calkins}, {Berlind}, {Stassun}, {Bla{\v{z}}ek}, {Skarka}, {{\v{S}}pokov{\'a}}, {{\v{Z}}{\'a}k}, {Albrecht}, {Sobrino}, {Beck}, {Cabrera}, {Carleo}, {Cochran}, {Csizmadia}, {Dai}, {Deeg}, {de Leon}, {Eigm{\"u}ller}, {Endl}, {Erikson}, {Fukui}, {Georgieva}, {Gonz{\'a}lez-Cuesta}, {Grziwa}, {Hidalgo}, {Hirano}, {Hjorth}, {Knudstrup}, {Korth}, {Lam}, {Livingston}, {Lund}, {Luque}, {Montanes Rodr{\'\i}guez}, {Murgas}, {Narita}, {Nespral}, {Niraula}, {Nowak}, {Pall{\'e}}, {P{\"a}tzold}, {Prieto-Arranz}, {Rauer}, {Redfield}, {Ribas}, {Smith}, {Van Eylen}, \& {Kab{\'a}th}}]{toi503b}
{{\v{S}}ubjak}, J., {Sharma}, R., {Carmichael}, T.~W., {et~al.} 2020, \aj, 159, 151

\bibitem[{{West} {et~al.}(2008){West}, {Hawley}, {Bochanski}, {Covey}, {Reid}, {Dhital}, {Hilton}, \& {Masuda}}]{2008AJ....135..785W}
{West}, A.~A., {Hawley}, S.~L., {Bochanski}, J.~J., {et~al.} 2008, \aj, 135, 785

\bibitem[{{Wildi} {et~al.}(2022){Wildi}, {Bouchy}, {Doyon}, {Blind}, {Genolet}, {Sordet}, {Segovia}, {Grieves}, {Malo}, {Artigau}, {St-Antoine}, {Vall{\'e}e}, {Rasilla}, {Gracia Temich}, {Poulin-Girard}, {Brousseau}, {Sosnowska}, {Reshetov}, {Baron}, {Thibault}, {Bovay}, {Frensch}, {Lo Curto}, {Hubin}, {Zins}, {Peroux}, \& {Cabral}}]{Wildi_NIRPS_2022SPIE12184E}
{Wildi}, F., {Bouchy}, F., {Doyon}, R., {et~al.} 2022, in Society of Photo-Optical Instrumentation Engineers (SPIE) Conference Series, Vol. 12184, Ground-based and Airborne Instrumentation for Astronomy IX, ed. C.~J. {Evans}, J.~J. {Bryant}, \& K.~{Motohara}, 121841H

\bibitem[{{Zacharias} {et~al.}(2012){Zacharias}, {Finch}, {Girard}, {Henden}, {Bartlett}, {Monet}, \& {Zacharias}}]{Zacharias_2012yCat.1322}
{Zacharias}, N., {Finch}, C.~T., {Girard}, T.~M., {et~al.} 2012, VizieR Online Data Catalog, I/322A

\bibitem[{{Zhou} {et~al.}(2019){Zhou}, {Bakos}, {Bayliss}, {Bento}, {Bhatti}, {Brahm}, {Csubry}, {Espinoza}, {Hartman}, {Henning}, {Jord{\'a}n}, {Mancini}, {Penev}, {Rabus}, {Sarkis}, {Suc}, {de Val-Borro}, {Rodriguez}, {Osip}, {Kedziora-Chudczer}, {Bailey}, {Tinney}, {Durkan}, {L{\'a}z{\'a}r}, {Papp}, \& {S{\'a}ri}}]{hats70b}
{Zhou}, G., {Bakos}, G.~{\'A}., {Bayliss}, D., {et~al.} 2019, \aj, 157, 31

\end{thebibliography}

\onecolumn
\begin{appendix}


\section{List of published transiting BDs}
\begin{table*}[h!]
\caption{List of published transiting BDs adapted and updated from \citet{Carmichael_2023} and \citet{Henderson_2024}. Some new objects have been included from   \cite{Vowell_2025arXiv250109795V} .
}             
\label{tab:full_BDs_lists}      
\centering   
{\renewcommand{\arraystretch}{1.2}
\resizebox{0.89\textwidth}{!}{
\begin{tabular}{l l l l l l l l l l l}
\hline\hline                        
Object & P [d] & M$_{2}$ [$M_{\rm Jup}$] & R$_{2}$ [$R_{\rm Jup}$] & $T_{\rm eff}$\ [K]  & M$_{1}$ [$M_\odot$] & R$_{1}$ [$R_\odot$] & ecc & [Fe/H] & $\log g_2$ & Source\\
\hline
TOI-4603b &7.246& $12.90^{+0.58}_{-0.57}$&$1.04\pm0.04$&$6264\pm 95$&$1.77\pm0.06$&$2.74\pm0.05$&$0.325\pm0.02$&$0.34\pm0.04$&  $4.491^{+0.072}_{-0.072}$ &\citet{toi4603b}\\   
HATS-70b&1.89&$12.9^{+1.8}_{-1.6}$&$1.38^{+0.08}_{-0.07}$&$7930^{+630}_{-820}$&$1.78^{+0.12}_{-0.12}$&$1.88^{+0.06}_{-0.07}$&<$0.18$&$0.04^{+0.10}_{-0.11}$& $5.631^{+0.125}_{-0.123}$ &\citet{hats70b}\\
TOI-1278b&14.48&$18.5^{+0.5}_{-0.5}$&$1.09^{+0.24}_{-0.20}$&$3799^{+42}_{-42}$&$0.55^{+0.02}_{-0.02}$&$0.57^{+0.01}_{-0.01}$&$0.013^{+0.004}_{-0.004}$&$-0.01^{+0.28}_{-0.28}$& $4.611^{+0.192}_{-0.160}$ & \citet{toi1278b}\\
GPX-1b&1.74&$19.7^{+1.6}_{-1.6}$&$1.47^{+0.10}_{-0.10}$&$7000^{+200}_{-200}$&$1.68^{+0.10}_{-0.10}$&$1.56^{+0.10}_{-0.10}$&$0$ (fixed)&$0.35^{+0.10}_{-0.10}$& $4.456^{+0.123}_{-0.123}$ &\citet{gpx1b}\\
Kepler-39b&21.09&$20.1^{+1.3}_{-1.2}$&$1.24^{+0.09}_{-0.10}$&$6350^{+100}_{-100}$&$1.29^{+0.06}_{-0.07}$&$1.40^{+0.10}_{-0.10}$&$0.112^{+0.057}_{-0.057}$&$0.10^{+0.14}_{-0.14}$& $4.646^{+0.070}_{-0.083}$ &\citet{kepler39b}\\
CoRoT-3b&4.26&$21.7^{+1.0}_{-1.0}$&$1.01^{+0.07}_{-0.07}$&$6740^{+140}_{-140}$&$1.37^{+0.09}_{-0.09}$&$1.56^{+0.09}_{-0.09}$&$0$ (fixed) &$-0.02^{+0.06}_{-0.06}$& $4.700^{+0.086}_{-0.088}$ &\citet{corot3b}\\
TOI-5882b& 7.1489& $24.36^{+0.85}_{-1.7}$& $0.833^{+0.06}_{-0.059}$& $6000\pm190$& $1.545^{+0.077}_{-0.17}$& $2.303^{+0.066}_{-0.064}$& $0.0347\pm0.0082$& $0.378 \pm0.084$ & $4.959^{+0.092}_{-0.161}$ &  \cite{Vowell_2025arXiv250109795V} \\
KELT-1b&1.22&$27.4^{+0.9}_{-0.9}$&$1.12^{+0.04}_{-0.03}$&$6516^{+49}_{-49}$&$1.34^{+0.06}_{-0.06}$&$1.47^{+0.05}_{-0.04}$&$0.010^{+0.010}_{-0.007}$&$0.05^{+0.08}_{-0.08}$&$4.757^{+0.069}_{-0.071}$ &\citet{kelt1b}\\
NLTT41135b&2.89&$33.7^{+2.8}_{-2.6}$&$1.13^{+0.27}_{-0.17}$&$3230^{+130}_{-130}$&$0.19^{+0.03}_{-0.02}$&$0.21^{+0.02}_{-0.01}$&$< 0.02$&$0$ (fixed)& $4.826^{+0.210}_{-0.134}$ &\citet{NLTTrad}\\
WASP-128b&2.21&$37.2^{+0.8}_{-0.9}$&$0.94^{+0.02}_{-0.02}$&$5950^{+50}_{-50}$&$1.16^{+0.04}_{-0.04}$&$1.15^{+0.02}_{-0.02}$&$<0.007$&$0.01^{+0.12}_{-0.12}$ & $5.045^{+0.042}_{-0.046}$&\citet{wasp128b}\\
CWW89Ab&5.29&$39.2^{+1.1}_{-1.1}$&$0.94^{+0.02}_{-0.02}$&$5755^{+49}_{-49}$&$1.10^{+0.05}_{-0.05}$&$1.03^{+0.02}_{-0.02}$&$0.189^{+0.002}_{-0.002}$&$0.20^{+0.09}_{-0.09}$ & $5.059^{+0.039}_{-0.044}$ &\citet{cww89ab1}\\
KOI-205b&11.72&$39.9^{+1.0}_{-1.0}$&$0.81^{+0.02}_{-0.02}$&$5237^{+60}_{-60}$&$0.93^{+0.03}_{-0.03}$&$0.84^{+0.02}_{-0.02}$&$<0.031$&$0.14^{+0.12}_{-0.12}$&$5.128^{+0.046}_{-0.040}$&\citet{koi205b}\\
TOI-1406b&10.57&$46.0^{+2.6}_{-2.7}$&$0.86^{+0.03}_{-0.03}$&$6290^{+100}_{-100}$&$1.18^{+0.08}_{-0.09}$&$1.35^{+0.03}_{-0.03}$&$0.026^{+0.013}_{-0.010}$&$-0.08^{+0.09}_{-0.09}$& $5.210^{+0.079}_{-0.086}$ & \citet{carmichael20}\\
TOI-3755b& 5.5437& $47.3^{+1.9}_{-2.2}$& $0.876^{+0.05}_{-0.045}$& $5630\pm170$& $1.042^{+0.063}_{-0.073}$& $1.04^{+0.041}_{-0.039}$& $0.005\pm0.0031$& $0.339\pm0.091$& $5.203^{+0.075}_{-0.079}$ & \cite{Vowell_2025arXiv250109795V}  \\
EPIC212036875b&5.17&$52.3^{+1.9}_{-1.9}$&$0.87^{+0.02}_{-0.02}$&$6238^{+59}_{-60}$&$1.29^{+0.07}_{-0.06}$&$1.50^{+0.03}_{-0.03}$&$0.132^{+0.004}_{-0.004}$&$0.01^{+0.10}_{-0.10}$&$5.267^{+0.061}_{-0.060}$&\citet{epic212036875b1}\\
TOI-503b&3.68&$53.7^{+1.2}_{-1.2}$&$1.34^{+0.26}_{-0.15}$&$7650^{+140}_{-160}$&$1.80^{+0.06}_{-0.06}$&$1.70^{+0.05}_{-0.04}$&$0$ (fixed) &$0.30^{+0.08}_{-0.09}$& $4.89^{+0.178}_{-0.113}$ &\citet{toi503b}\\
TOI-852b&4.95&$53.7^{+1.4}_{-1.3}$&$0.83^{+0.04}_{-0.04}$&$5768^{+84}_{-81}$&$1.32^{+0.05}_{-0.04}$&$1.71^{+0.04}_{-0.04}$&$0.004^{+0.004}_{-0.003}$&$0.33^{+0.09}_{-0.09}$& $5.302^{+0.049}_{-0.047}$ &\citet{carmichael21}\\
TOI-2844b& 3.552& $53.8^{+5.0}_{-5.1}$&	$0.775^{+0.048}_{-0.044}$& $6900\pm220$& $1.585^{+0.071}_{-0.073}$&	$1.785^{+0.087}_{-0.081}$& $0.429\pm0.048$& $0.061\pm0.12$ & $5.367^{+0.133}_{-0.133}$ &  \cite{Vowell_2025arXiv250109795V} \\
AD3116b&1.98&$54.2^{+4.3}_{-4.3}$&$1.02^{+0.28}_{-0.28}$&$3184^{+29}_{-29}$&$0.28^{+0.02}_{-0.02}$&$0.29^{+0.08}_{-0.08}$&$0.146^{+0.024}_{-0.016}$&$0$ (fixed)&$5.193^{+0.075}_{-0.072}$&\citet{ad3116b}\\
CoRoT-33b&5.82&$59.2^{+1.8}_{-1.7}$&$1.10^{+0.53}_{-0.53}$&$5225^{+80}_{-80}$&$0.86^{+0.04}_{-0.04}$&$0.94^{+0.14}_{-0.08}$&$0.070^{+0.002}_{-0.002}$&$0.44^{+0.10}_{-0.10}$&$5.102^{+0.420}_{-0.420}$&\citet{corot33b}\\
TOI-3577b& 5.2667& $59.1^{+6.2}_{-8.4}$&	$0.844^{+0.092}_{-0.08}$& $6510\pm870$& $1.29^{+0.21}_{-0.27}$& $1.64^{+0.11}_{-0.11}$& $0.005\pm0.008$& $-0.21\pm0.44$& $5.334^{+0.211}_{-0.256}$ &  \cite{Vowell_2025arXiv250109795V} \\
RIK72b& 97.76& $59.2^{+6.80}_{-6.70}$& $3.10^{+0.31}_{-0.31}$& $3349\pm142$& $0.439^{+0.044}_{-0.044}$& $0.961^{+0.096}_{-0.096}$& $0.1079\pm0.0116$& $0.0\pm0.0$& $4.217^{+0.112}_{-0.106}$ &\citep{rik72b}\\
TOI-811b&25.17&$59.9^{+13.0}_{-8.6}$&$1.26^{+0.06}_{-0.06}$&$6107^{+77}_{-77}$&$1.32^{+0.05}_{-0.07}$&$1.27^{+0.06}_{-0.09}$&$0.509^{+0.075}_{-0.075}$&$0.40^{+0.07}_{-0.09}$ & $5.013^{+0.133}_{-0.105}$ &\citet{carmichael21}\\
TOI-263b&0.56&$61.6^{+4.0}_{-4.0}$&$0.91^{+0.07}_{-0.07}$&$3471^{+33}_{-33}$&$0.44^{+0.04}_{-0.04}$&$0.44^{+0.03}_{-0.03}$&$0.017^{+0.009}_{-0.010}$&$0.00^{+0.10}_{-0.10}$& $5.312^{+0.082}_{-0.082}$ &\citet{toi263b1}\\
KOI-415b&166.79&$62.1^{+2.7}_{-2.7}$&$0.79^{+0.12}_{-0.07}$&$5810^{+80}_{-80}$&$0.94^{+0.06}_{-0.06}$&$1.25^{+0.15}_{-0.10}$&$0.698^{+0.002}_{-0.002}$&$-0.24^{+0.11}_{-0.11}$&$5.353^{+0.129}_{-0.044}$&\citet{koi415b}\\
WASP-30b&4.16&$62.5^{+1.2}_{-1.2}$&$0.95^{+0.03}_{-0.02}$&$6202^{+42}_{-51}$&$1.25^{+0.03}_{-0.04}$&$1.39^{+0.03}_{-0.03}$&$<0.004$&$0.08^{+0.07}_{-0.05}$&$5.236^{+0.050}_{-0.039}$&\citet{wasp30b2}\\
LHS6343c&12.71&$62.7^{+2.4}_{-2.4}$&$0.83^{+0.02}_{-0.02}$&$3130^{+20}_{-20}$&$0.37^{+0.01}_{-0.01}$&$0.38^{+0.01}_{-0.01}$&$0.056^{+0.032}_{-0.032}$&$0.04^{+0.08}_{-0.08}$&$5.369^{+0.027}_{-0.027}$&\citet{lhs6343}\\
CoRoT-15b&3.06&$63.3^{+4.1}_{-4.1}$&$1.12^{+0.30}_{-0.15}$&$6350^{+200}_{-200}$&$1.32^{+0.12}_{-0.12}$&$1.46^{+0.31}_{-0.14}$&$0$ (fixed)&$0.10^{+0.20}_{-0.20}$&$5.273^{+0.169}_{-0.151}$&\citet{corot15b}\\
TOI-569b&6.56&$64.1^{+1.9}_{-1.4}$&$0.75^{+0.02}_{-0.02}$&$5768^{+110}_{-92}$&$1.21^{+0.05}_{-0.05}$&$1.48^{+0.03}_{-0.03}$&$0.002^{+0.002}_{-0.001}$&$0.29^{+0.09}_{-0.08}$& $5.464^{+0.039}_{-0.062}$ &\citet{carmichael20}\\
TOI-2119b&7.20&$64.4^{+2.3}_{-2.2}$&$1.08^{+0.03}_{-0.03}$&$3621^{+48}_{-46}$&$0.53^{+0.02}_{-0.02}$&$0.50^{+0.02}_{-0.02}$&$0.337^{+0.002}_{-0.001}$&$0.06^{+0.08}_{-0.08}$& $5.154^{+0.033}_{-0.031}$ &\citet{carmichael22}\\
TOI-1982b&17.17&$65.9^{+2.8}_{-2.7}$&$1.08^{+0.04}_{-0.04}$&$6325^{+110}_{-110}$&$1.41^{+0.08}_{-0.08}$&$1.51^{+0.05}_{-0.05}$&$0.272^{+0.014}_{-0.014}$&$-0.10^{+0.09}_{-0.09}$& $5.161^{+0.079}_{-0.080}$ &\citet{psaridi22}\\
NGTS-28Ab&1.25&$69.0^{+5.3}_{-4.8}$&$0.95\pm0.05$&$3626^{+47}_{-44}$&$0.56^{+0.02}_{-0.02}$&$0.59^{+0.03}_{-0.03}$&$0.040^{+0.007}_{-0.010}$&$-0.14^{+0.16}_{-0.17}$& $5.313^{+0.051}_{-0.051}$ &\citet{Henderson_2024}\\
EPIC201702477b&40.74&$66.9^{+1.7}_{-1.7}$&$0.76^{+0.07}_{-0.07}$&$5517^{+70}_{-70}$&$0.87^{+0.03}_{-0.03}$&$0.90^{+0.06}_{-0.06}$&$0.228^{+0.003}_{-0.003}$&$-0.16^{+0.05}_{-0.05}$& $5.426^{+0.062}_{-0.059}$&\citet{epic201702477b}\\
TOI-629b&8.72&$67.0^{+3.0}_{-3.0}$&$1.11^{+0.05}_{-0.05}$&$9100^{+200}_{-200}$&$2.16^{+0.13}_{-0.13}$&$2.37^{+0.11}_{-0.11}$&$0.298^{+0.008}_{-0.008}$&$0.10^{+0.15}_{-0.15}$& $5.137^{+0.122}_{-0.122}$ &\citet{psaridi22}\\
TOI-4737b& 9.320& $67.5^{+3.2}_{-3.3}$& $0.816^{+0.052}_{-0.044}$& $6330\pm220$& $1.402^{+0.083}_{-0.085}$& $1.568^{+0.076}_{-0.062}$& $0.018\pm0.022$& $0.25\pm0.11$& $5.4196^{+0.094}_{-0.091}$ & \cite{Vowell_2025arXiv250109795V} \\
TOI-2543b&7.54&$67.6^{+3.5}_{-3.5}$&$0.95^{+0.09}_{-0.09}$&$6060^{+82}_{-82}$&$1.29^{+0.08}_{-0.08}$&$1.86^{+0.15}_{-0.15}$&$0.009^{+0.003}_{-0.002}$&$-0.28^{+0.10}_{-0.10}$& $5.362^{+0.110}_{-0.110}$& \citet{psaridi22}\\
HIP33609b& 39.4718& $ 68.00^{+7.40}_{-7.10}$& $1.580^{+0.074}_{-0.070}$& $10400\pm800$& $2.383^{+0.10}_{-0.095}$& $1.863^{+0.087}_{-0.082}$& $0.560\pm0.031$& $-0.01\pm0.20$& $4.849^{+0.112}_{-0.108}$ &\citet{HIP33609b} \\
LP261-75b&1.88&$68.1^{+2.1}_{-2.1}$&$0.90^{+0.01}_{-0.01}$&$3100^{+50}_{-50}$&$0.30^{+0.02}_{-0.02}$&$0.31^{+0.00}_{-0.00}$&$< 0.007$&0.0&$5.340^{+0.021}_{-0.021}$&\citet{lp261}\\
NGTS-19b&17.84&$69.5^{+5.7}_{-5.4}$&$1.03^{+0.06}_{-0.05}$&$4716^{+39}_{-28}$&$0.81^{+0.04}_{-0.04}$&$0.90^{+0.04}_{-0.04}$&$0.377^{+0.006}_{-0.006}$&$0.11^{+0.07}_{-0.07}$&$5.233^{+0.060}_{-0.061}$&\citet{19b}\\
TOI-2336b& 7.71198& $69.9^{+2.3}_{-2.3}$& $1.05^{+0.04}_{-0.04}$& $6550\pm100$& $1.41^{+0.08}_{-0.08}$& $1.781^{+0.059}_{-0.059}$& $0.010\pm0.006$& $0.0\pm0.03$& $5.218^{+0.080}_{-0.080}$ & \cite{toi1608b} \\ 
CoRoT-34b& 2.1185& $71.40^{+8.90}_{-8.60}$& $1.09^{+0.17}_{-0.16}$& $7820\pm160$& $1.66^{+0.08}_{-0.15}$& $1.85^{+0.29}_{-0.25}$& $0.00\pm0.00$& $-0.20\pm0.20$& $5.193^{+0.215}_{-0.239}$ & \citet{CoRoT34b} \\
TOI-2533b& 6.6847& $72.0^{+3.00}_{-3.00}$& $0.850^{+0.040}_{-0.030}$& $6180\pm80$& $1.020^{+0.060}_{-0.070}$& $1.110^{+0.010}_{-0.010}$& $0.060\pm0.070$& $-0.3\pm0.20$& $5.418^{+0.069}_{-0.072}$ & \cite{toi2533b} \\
\bf TOI-6508b& 18.99& $72.53_{-5.09}^{+7.61}$& $0.985^{+0.031}_{-0.032}$& $3003 \pm 100$& $0.1744^{+0.0293}_{0.0198}$& $0.2041^{+0.0061}_{-0.0061}$& $0.28\pm0.08$& $-0.22\pm 0.08$& $5.255^{+0.054}_{-0.048}$ & This work \\
NGTS-7Ab&0.68&$75.5^{+3.0}_{-13.7}$&$1.38^{+0.13}_{-0.14}$&$3359^{+106}_{-89}$&$0.48^{+0.03}_{-0.12}$&$0.61^{+0.06}_{-0.06}$&$0$ (fixed)&$0$ (fixed) & $4.845^{+0.095}_{-0.092}$ &\citet{ngts7ab}\\
TOI-148b&4.87&$77.1^{+5.8}_{-4.6}$&$0.81^{+0.05}_{-0.06}$&$5990^{+140}_{-140}$&$0.97^{+0.12}_{-0.09}$&$1.20^{+0.07}_{-0.07}$&$0.005^{+0.006}_{-0.004}$&$-0.24^{+0.25}_{-0.25}$& $5.483^{+0.127}_{-0.127}$ &\citet{grieves21}\\
TOI-2521b& 5.5630& $77.5^{+3.3}_{-3.3}$& $1.01^{+0.04}_{-0.04}$& $5600\pm100$& $1.10^{+0.07}_{-0.07}$& $1.770^{+0.068}_{-0.068}$& $0.0\pm1.10$& $-0.3\pm0.3$& $5.333^{+0.090}_{-0.090}$ &\cite{toi1608b}\\
KOI-189b&30.36&$78.0^{+3.4}_{-3.4}$&$1.00^{+0.02}_{-0.02}$&$4952^{+40}_{-40}$&$0.76^{+0.05}_{-0.05}$&$0.73^{+0.02}_{-0.02}$&$0.275^{+0.004}_{-0.004}$&$-0.12^{+0.10}_{-0.10}$ & $5.330^{+0.038}_{-0.036}$&\citet{koi189b}\\
Kepler-503b& 7.2584& $78.6^{+3.1}_{-3.1}$& $0.96^{+0.06}_{-0.04}$& $5670\pm100$& $ 1.154^{+0.047}_{-0.042}$& $1.764^{+0.080}_{-0.068}$& $0.025\pm0.014$& $0.169\pm0.046$& $5.343^{+0.072}_{0.056}$ &\citet{kepler503b}\\
ZTFJ2020+5033& 0.07928& $80.1^{+1.60}_{-1.60}$& $1.050^{+0.010}_{-0.010}$& $2856\pm6$& $0.134^{+0.004}_{-0.004}$& $0.176^{+0.002}_{-0.002}$& $0.0\pm0.0$& $0.0\pm0.0$& $5.278^{+0.031}_{-0.032}$& \citet{ZTFJ2020} \\
TOI-587b&8.04&$81.1^{+7.1}_{-7.0}$&$1.32^{+0.07}_{-0.06}$&$9800^{+200}_{-200}$&$2.33^{+0.12}_{-0.12}$&$2.01^{+0.09}_{-0.09}$&$0.051^{+0.049}_{-0.036}$&$0.08^{+0.11}_{-0.12}$& $4.697^{+0.105}_{-0.096}$ &\citet{grieves21}\\
TOI-1712b& 3.5666& $82.0^{+7.00}_{-7.00}$& $1.740^{+0.080}_{-0.070}$& $6860\pm40$& $1.630^{+0.010}_{-0.020}$& $3.100^{+0.070}_{-0.100}$& $0.090\pm0.072$& $-0.2\pm0.10$& $4.848^{+0.055}_{-0.054}$ & \citet{toi1712b} \\
TOI-746b&10.98&$82.2^{+4.9}_{-4.4}$&$0.95^{+0.09}_{-0.06}$&$5690^{+140}_{-140}$&$0.94^{+0.09}_{-0.08}$&$0.97^{+0.04}_{-0.03}$&$0.199^{+0.003}_{-0.003}$&$-0.02^{+0.23}_{-0.23}$ & $5.372^{+0.143}_{-0.143}$ &\citet{grieves21}\\
TOI-4635b& 12.2769& $84.0^{+2.1}_{-2.0}$& $1.02^{+0.019}_{-0.019}$& $4555\pm67$& $0.698^{+0.027}_{-0.025}$& $0.683^{+0.011}_{-0.011}$& $0.4906\pm0.0015$& $-0.091\pm0.039$& $5.321^{+0.030}_{-0.028}$& \cite{Vowell_2025arXiv250109795V} \\
EBLM J0555-57Ab&7.76&$87.9^{+4.0}_{-4.0}$&$0.82^{+0.13}_{-0.06}$&$6368^{+124}_{-124}$&$1.18^{+0.08}_{-0.08}$&$1.00^{+0.14}_{-0.07}$&$0.090^{+0.004}_{-0.004}$&$-0.04^{+0.14}_{-0.14}$&$5.529^{+0.156}_{-0.098}$&\citet{j055557ab}\\
TOI-681b&15.78&$88.7^{+2.5}_{-2.3}$&$1.52^{+0.25}_{-0.15}$&$7440^{+150}_{-140}$&$1.54^{+0.06}_{-0.05}$&$1.47^{+0.04}_{-0.04}$&$0.093^{+0.022}_{-0.019}$&$-0.08^{+0.05}_{-0.05}$& $4.997^{+0.153}_{-0.097}$&\citet{grieves21}\\
OGLE-TR-123b&1.80&$89.0^{+11.5}_{-11.5}$&$1.29^{+0.09}_{-0.09}$&$6700^{+300}_{-300}$&$1.29^{+0.26}_{-0.26}$&$1.55^{+0.10}_{-0.10}$&$0$ (fixed)&-& $5.140^{+0.245}_{-0.245}$&\citet{ogle123b}\\
TOI-694b&48.05&$89.0^{+5.3}_{-5.3}$&$1.11^{+0.02}_{-0.02}$&$5496^{+87}_{-81}$&$0.97^{+0.05}_{-0.04}$&$1.00^{+0.01}_{-0.01}$&$0.519^{+0.001}_{-0.001}$&$0.21^{+0.08}_{-0.08}$& $5.275^{+0.050}_{-0.042}$&\citet{toi694}\\
TOI-1608b &2.4727 & $90.7^{+3.7}_{-3.7}$& $1.21^{+0.06}_{-0.06}$& $5950\pm100$& $1.38^{+0.08}_{-0.0}$& $2.222^{+0.0076}_{-0.076}$& $0.041^{+0.024}_{-0.019}$& $0.1\pm0.3$& $5.220^{+0.098}_{-0.098}$ &\cite{toi1608b} \\
TOI-5467b & 2.6571& $91.6^{2.8}_{-2.8}$& $1.126^{+0.051}_{-0.049}$& $6740\pm160$& $1.512^{+0.060}_{-0.063}$& $ 1.531^{+0.053}_{-0.049}$& $0.0439\pm0.0006$& $0.28\pm0.09$& $5.272^{+0.070}_{-0.071}$ & \cite{Vowell_2025arXiv250109795V} \\
KOI-607b&5.89&$95.1^{+3.3}_{-3.4}$&$1.09^{+0.09}_{-0.06}$&$5418^{+87}_{-85}$&$0.99^{+0.05}_{-0.05}$&$0.92^{+0.03}_{-0.03}$&$0.395^{+0.009}_{-0.009}$&$0.38^{+0.08}_{-0.09}$& $5.316^{+0.086}_{-0.067}$&\citet{carmichael19}\\
EBLM J1219-39b&6.76&$95.4^{+1.9}_{-2.5}$&$1.14^{+0.07}_{-0.05}$&$5412^{+81}_{-65}$&$0.83^{+0.03}_{-0.03}$&$0.81^{+0.04}_{-0.02}$&$0.055^{+0.000}_{-0.000}$&$-0.21^{+0.07}_{-0.08}$&$5.279^{+0.060}_{-0.047}$&\citet{wasp30b2}\\
TIC-320687387 B&29.77&$96.2^{+1.9}_{-2.0}$&$1.14^{+0.02}_{-0.02}$&$5780^{+80}_{-80}$&$1.08^{+0.03}_{-0.03}$&$1.16^{+0.02}_{-0.02}$&$0.366^{+0.003}_{-0.003}$&$0.30^{+0.08}_{-0.08}$& & \citet{tic320687387b}\\
OGLE-TR-122b&7.27&$96.4^{+9.4}_{-9.4}$&$1.17^{+0.23}_{-0.13}$&$5700^{+300}_{-300}$&$0.98^{+0.14}_{-0.14}$&$1.05^{+0.20}_{-0.09}$&$0.205^{+0.008}_{-0.008}$&$0.15^{+0.36}_{-0.36}$&$5.251^{+0.214}_{-0.162}$&\citet{ogle122b}\\
TOI-1213b&27.22&$97.5^{+4.4}_{-4.2}$&$1.66^{+0.78}_{-0.55}$&$5590^{+150}_{-150}$&$0.99^{+0.07}_{-0.06}$&$0.99^{+0.04}_{-0.04}$&$0.498^{+0.003}_{-0.002}$&$0.25^{+0.13}_{-0.14}$&$4.962^{+0.413}_{-0.293}$&\citet{grieves21}\\
K2-76b&11.99&$98.7^{+2.0}_{-2.0}$&$0.89^{+0.05}_{-0.03}$&$5747^{+70}_{-64}$&$0.96^{+0.03}_{-0.03}$&$1.17^{+0.06}_{-0.03}$&$0.255^{+0.007}_{-0.007}$&$0.01^{+0.04}_{-0.04}$&$3.528^{+0.061}_{-0.0422}$&\citet{k276}\\
CoRoT-101186644&20.68&$100.6^{+11.5}_{-11.5}$&$1.01^{+0.25}_{-0.06}$&$6090^{+200}_{-200}$&$1.20^{+0.20}_{-0.20}$&$1.07^{+0.07}_{-0.07}$&$0.402^{+0.006}_{-0.006}$&$0.20^{+0.20}_{-0.20}$&$5.406^{+0.282}_{-0.190}$&\citet{corot1011}\\
TOI-3122b & 6.1836& $101.5^{+4.1}_{-4.8}$& $1.235^{+0.075}_{-0.057}$& $6120^{+180}_{-220}$& $1.247^{+0.074}_{-0.091}$& $1.336^{+0.062}_{-0.045}$& $0.4704 \pm 0.008$ & $ 0.29\pm0.11$& $5.237^{+0.091}_{-0.098}$&\cite{Vowell_2025arXiv250109795V}  \\
TOI-4759b & 9.65779& $ 102^{+7.8}_{-5.3}$& $0.993^{+0.082}_{-0.075}$& $5680\pm150$& $1.242^{+0.15}_{-0.097}$& $ 1.927^{+0.11}_{-0.098}$& $0.2424\pm0.0032$& $0.28\pm0.17$& $5.428^{+0.155}_{-0.111}$ &\cite{Vowell_2025arXiv250109795V}  \\
J2343+29Ab&16.95&$102.7^{+7.3}_{-7.3}$&$1.24^{+0.07}_{-0.07}$&$5150^{+90}_{-60}$&$0.86^{+0.10}_{-0.10}$&$0.85^{+0.05}_{-0.06}$&$0.161^{+0.002}_{-0.003}$&$0.07^{+0.01}_{-0.17}$&$5.235^{+0.105}_{-0.105}$&\citet{j2343}\\
EBLM J0954-23Ab&7.57&$102.8^{+5.9}_{-6.0}$&$0.98^{+0.17}_{-0.17}$&$6406^{+124}_{-124}$&$1.17^{+0.08}_{-0.08}$&$1.23^{+0.17}_{-0.17}$&$0.042^{+0.001}_{-0.001}$&$-0.01^{+0.14}_{-0.14}$&$5.424^{+0.173}_{-0.173}$&\citet{j0954}\\
KOI-686b&52.51&$103.4^{+4.8}_{-4.8}$&$1.22^{+0.04}_{-0.04}$&$5834^{+100}_{-100}$&$0.98^{+0.07}_{-0.07}$&$1.04^{+0.03}_{-0.03}$&$0.556^{+0.004}_{-0.004}$&$-0.06^{+0.13}_{-0.13}$&$5.254^{+0.071}_{-0.071}$&\citet{koi189b}\\
TIC-220568520b&18.56&$107.2^{+5.2}_{-5.2}$&$1.25^{+0.02}_{-0.02}$&$5589^{+81}_{-81}$&$1.03^{+0.04}_{-0.04}$&$1.01^{+0.01}_{-0.01}$&$0.096^{+0.003}_{-0.003}$&$0.26^{+0.07}_{-0.07}$&$5.239^{+0.042}_{-0.042}$&\citet{toi694}\\
TOI-4462b & 4.9133& $107.6^{6.6}_{-5.8}$& $1.286^{+0.071}_{-0.058}$& $5960\pm120$& $1.36^{+0.13}_{-0.11}$& $2.149^{+0.085}_{-0.064}$& $ 0.0202\pm0.0037$& $0.09\pm0.16$& $5.226^{+0.128}_{-0.109}$ & \cite{Vowell_2025arXiv250109795V}  \\
TOI-5240b & 4.17930& $ 127.6^{+5}_{-5}$& $1.7^{+0.11}_{-0.11}$& $7340\pm360$& $ 1.744^{+0.096}_{-0.096}$& $2.38^{+0.12}_{-0.12}$& $ 0.0109\pm0.012$& $-0.13\pm0.19$& $5.058^{+0.111}_{-0.111}$ & \cite{Vowell_2025arXiv250109795V}   \\
\hline
\multicolumn{10}{l}{}
\end{tabular}}}
\end{table*}

\section{Posterior probability distribution for the system TOI-6508.}

\begin{figure*}[h!]
	\centering
	\includegraphics[scale=0.2]{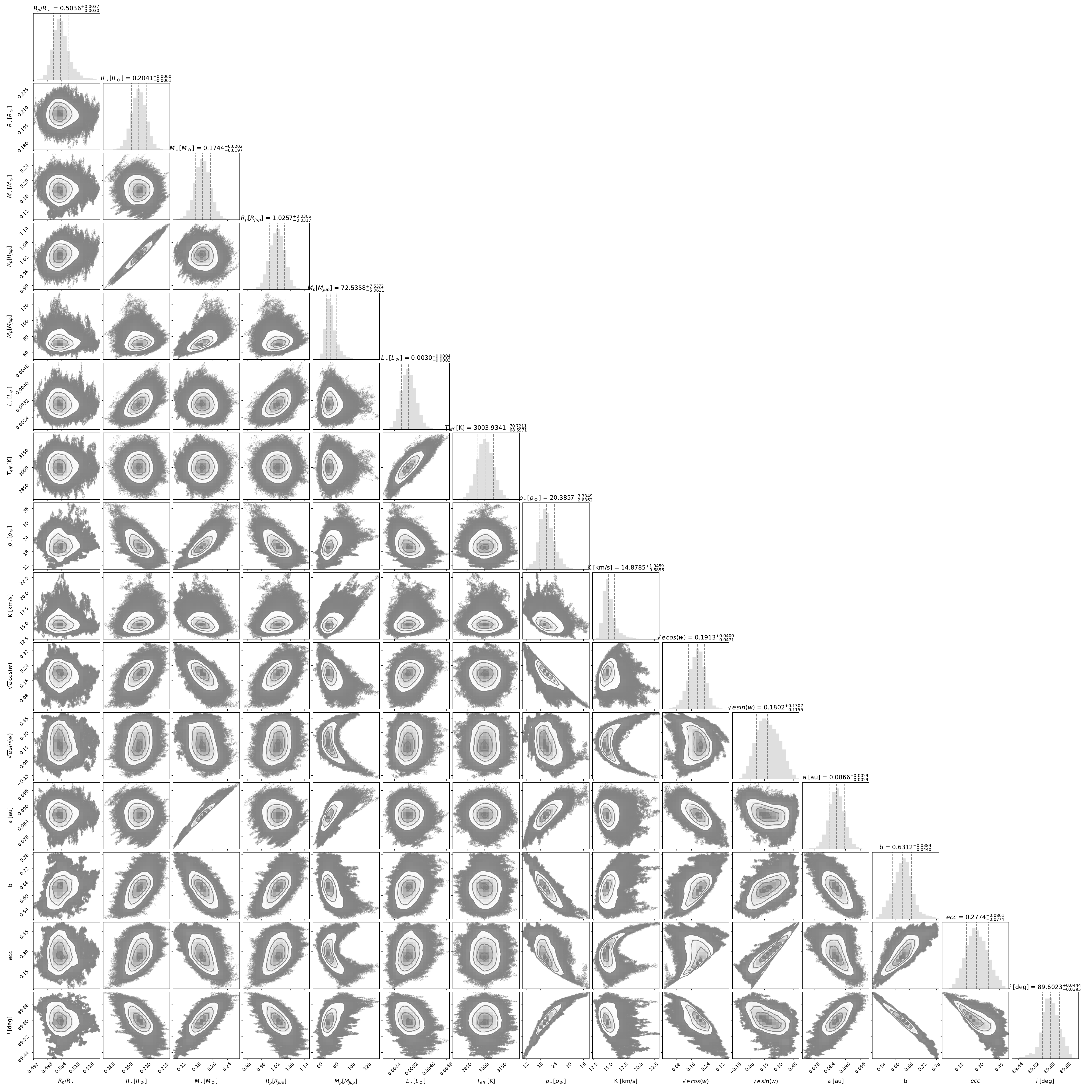}
	\caption{ Posterior probability distribution for the TOI-6508 system parameters derived from our global MCMC analysis.  The median value for each parameter is represented by the vertical dashed lines.}
	\label{corner_TOI6508}
\end{figure*}

\end{appendix}

\end{document}